\documentclass[%
 aip,
 amsmath,amssymb,
 reprint,%
]{revtex4-1}

\usepackage{graphicx}% Include figure files
\usepackage{dcolumn}% Align table columns on decimal point
\usepackage{bm}% bold math
\usepackage[utf8]{inputenc}
\usepackage[T1]{fontenc}
\usepackage{mathptmx}

\newcommand{\medno}{\medskip\noindent}

\begin{document}

\preprint{AIP/123-QED}

%\pagenumbering{roman}
%\setcounter{page}{0}

\title{An Extended-MHD Model for Handling Low-density Plasmas with Tabular Material Models}

\author{N. D. Hamlin}%
\email{ndhamli@sandia.gov.}
\affiliation{ 
Sandia National Laboratories, P.O. Box 5800, Albuquerque, New Mexico 87185, USA%\\This line break forced with \textbackslash\textbackslash
}%

\author{M. R. Martin}%
\affiliation{ 
Sandia National Laboratories, P.O. Box 5800, Albuquerque, New Mexico 87185, USA%\\This line break forced with \textbackslash\textbackslash
}%

\author{J. M. Woolstrum}%
\affiliation{ 
Sandia National Laboratories, P.O. Box 5800, Albuquerque, New Mexico 87185, USA%\\This line break forced with \textbackslash\textbackslash
}%

\date{\today}% It is always \today, today, but any date may be explicitly specified

\begin{abstract}
An extended-MHD model, interfaced with tabular equation-of-state and conductivity models, has been developed in PERSEUS (Physics as an Extended-MHD Relaxation System with an Efficient Upwind Scheme) for simulating a plasma-vacuum interface under experimentally-relevant conditions for a pulsed-power system, and with minimal sensitivity to parameters characterizing the numerical vacuum.  For several test problems, we demonstrate convergence of this model for sufficiently low density floors and with respect to certain vacuum parameters.  This capability is crucial for predictively modeling the coupling of energy and current onto a target in a pulsed-power system.
\end{abstract}

\maketitle

\section{Introduction}

Accurate modeling of the plasma-vacuum interface is a problem that has plagued simulations of pulsed-power systems for over 20 years.  In a pulsed-power device, current is delivered along magnetically insulated transmission lines, radially converging on a small target.  In inertial confinement fusion (ICF), the resulting magnetic forces drive a cylindrical implosion of the target, and the fuel inside it, generating fusion reactions.  One ICF experiment, Magnetized Liner Inertial Fusion (MagLIF), is performed on the Z Accelerator at Sandia National Laboratories, which is one of the most powerful pulsed-power devices in the world.  In MagLIF \cite{gome14}, the target is a cylindrical canister known as a liner, and contains fuel (Deuterium or a mixture of Deuterium and Tritium) which is preheated by a laser and then magnetized by an applied axial magnetic field, so that as the fuel implodes, the thermal energy is magnetically insulated.  To predict the efficiency of such an implosion requires predictive models of the efficiency with which energy and current in the transmission lines are coupled onto the target.

In turn, the main challenge faced by these models is their ability to model the interaction of a solid target with experimental vacuum conditions.  An experimental vacuum pressure on a pulsed-power facility, such as the Z Accelerator at Sandia National Laboratories, is about $10^{-5}$ Torr, corresponding to densities which are 12 or 13 orders of magnitude below solid density.  Moreover, plasma at these densities is known to influence target performance in several ways.  Namely, this low-density plasma (i) carries a significant amount of current which is then coupled onto the target, and (ii) substantially affects the nature of plasma ablation from the target and electrode surfaces.  A predictive model of target performance therefore has to simulate across 12-13 orders of magnitude in density variation in order to capture plasma at these densities.  This has, for more than 2 decades, been a numerically intractable problem, mainly because of the unbounded nature of the phase velocities of the magnetohydrodynamic (MHD) waves at low densities, resulting in prohibitively small timesteps.  Most resistive MHD codes are restricted to vacuum floors on the order of $10^{-6}$ g/cc, well above the experimental vacuum density.  These challenges of extended-MHD modeling have been discussed in a number of pioneering works on this subject, including Refs.~\onlinecite{huba03tut, shum03, schn09}.

This is the problem that motivated the development of PERSEUS (Plasma as an Extended-MHD Relaxation System with an Efficient Upwind Scheme) at Cornell University in 2010 by Charles Seyler and Matt Martin \cite{seyl11}.  The extended-MHD model in PERSEUS includes displacement current, which bounds the Alfven speed by the numerically reduced speed of light (see Sec. \ref{ssec:model} for discussion of numerical speed of light), enabling timesteps which are no longer computationally prohibitive.  This feature, together with modeling of extended-MHD phenomena, in particular Hall physics, enable the PERSEUS model to capture current coupling by and ablation of plasma at densities close to the experimental vacuum, which in turn enables a more predictive model of target performance in experimental vacuum conditions.

In addition to a more predictive model of the experimental vacuum, the use of a sufficiently low-density vacuum floor enables convergence of simulation results with respect to numerical vacuum parameters, including the density floor and other adjustable parameters, which are much more influential in resistive MHD codes, due to their use of higher density floors.  These include resets to floor and ceiling values of various quantities including pressure, temperature, and conductivity, along with floor multipliers defining ``buffer" intervals just above a floor (see Sec. \ref{ssec:limiting} for further discussion).  Because resistive MHD codes are restricted to densities well above vacuum levels, significant tweaking of these adjustable parameters is often required in order to achieve approximate agreement with experimental data, and the settings of these parameters often vary significantly between different experiments being simulated.  

Therefore, by minimizing sensitivity to vacuum modeling, the PERSEUS extended-MHD model enables significantly improved portability between validations against different experiments, and enabling validation against a broader range of pulsed-power experiments extending to higher current densities over longer timescales, and at smaller spatial scales.

Note that in this context, vacuum convergence is relative to a numerical solution at a given timestep and grid resolution, and is separate from convergence with respect to grid resolution (which has also been verified in PERSEUS for numerous problems).  As will be demonstrated, both types of convergence have comparable importance in assessing the reliability of simulation results.  %Resistive MHD codes are often converged with respect to grid resolution, but rarely converged with respect to vacuum modeling.

It should be noted that PERSEUS is a single-material code, and therefore does not capture the multi-material physics needed for a high-fidelity simulation of actual experimental conditions.  The goal of this paper is to demonstrate the existence of numerical solutions which are minimally sensitive to vacuum parameters in a single-material model, which constitutes an important step toward making such solutions available in a multi-material model.  A multi-material capability is being developed in FLEXO (Flux-Limited Extended-MHD Ohm's Law), an extended-MHD code under development at Sandia whose extended-MHD formulation is similar to the PERSEUS formulation. % TODO Ref?

In addition, the minimally-vacuum-sensitive solution is inevitably dependent on the equation-of-state (EOS) and conductivity tables being used, as well as additional physics for modeling various other transport phenomena (thermal conduction, radiation, thermoelectric transport, etc.).

Another challenge of modeling low-density plasma is modeling the extended-MHD physics that becomes more important at low densities, in particular the Hall term in the Generalized Ohms Law (GOL), which dominates at ion inertial length scales or in strongly magnetized plasmas, and has been shown to have a number of important consequences for target performance \cite{seyl11, haml18, haml18b, haml19, seyl18, zhao15}.  Hall physics, in turn, introduces small electron length and time scales.  Due to the need to resolve these scales in order for an explicit numerical scheme to remain stable, prohibitively small timesteps are required at small length scales relative to the ion inertial length.  This is why resistive MHD codes have only recently begun efforts to model the Hall term.

As Sec. \ref{ssec:model} discusses, PERSEUS surmounts this obstacle with a semi-implicit relaxation formulation of the GOL and Ampere's law, implicit only in the source terms and not flux terms, which overcomes the need to resolve electron time scales.

These features of PERSEUS were mostly incorporated at Cornell University.  In the version of PERSEUS being developed at Sandia, which is the subject of this study, the main additional feature is an interface to utilize  SESAME tables for equation-of-state (EOS) and conductivity \cite{lyon92}.  As discussed further in Sec. \ref{ssec:tabular}, certain modifications needed to be made to the tables in order to interface them properly with the extended-MHD model in PERSEUS, as these tables had mostly been used with resistive MHD codes using higher density floors.  

A central goal of this study, therefore, is to demonstrate the use of the PERSEUS extended-MHD model in identifying numerical solutions, for a given grid resolution and for given EOS and conductivity tables, which are minimally sensitive to parameters characterizing the numerical vacuum.

The tabular models of EOS and conductivity impose certain constraints, e.g. the need to avoid extrapolation, which would be needed by any numerical model that uses these tables.  Apart from these, the main forms of limiting needed by PERSEUS are current limiting to prevent superluminal Hall velocities (since the model is non-relativistic) and positivity preservation of density.  The other adjustable parameters mentioned in this study, e.g. density buffer, pressure floor, etc., are included as examples of modeling parameters needed by resistive MHD codes, much more than by PERSEUS, which (i) become influential at the high ($\sim 10^{-6}$ g/cm$^3$) density floors used by many of these codes, and (ii) can alter the numerical set of equations and initial conditions from the physical system being addressed.  It will be shown that at the lower density floors ($\le 10^{-10}$ g/cm$^3$) used by PERSEUS, these parameters become much less influential. 

For ICF problems, the domain can be roughly divided into three regions of interest: the solid material, the vacuum, and the low-density plasma that serves as a transition between vacuum and solid.  The tabular EOS is most useful for addressing the solid and low-density regions, while extended-MHD physics is needed for addressing the low-density and vacuum regions, in particular for correctly computing the current that is coupled from the vacuum onto the target.

The structure of this paper is as follows.  Section \ref{sec:methods} discusses the numerical model used in PERSEUS, including the extended-MHD model in Sec. \ref{ssec:model}, the tabular interface in Sec. \ref{ssec:tabular}, and the implementation of limiting in Sec. \ref{ssec:limiting}.  Section \ref{sec:results} then discusses results from the various tests performed with this model.  The vacuum region can be thought of as the region for which the electron inertial length has become large enough that the electromagnetic fields are decoupled from the plasma.

%Broadly speaking, the tests are arranged in order of increasing coupling of Hall physics into the behavior of the $E$ and $B$-fields, which is found to be correlated with an increase in vacuum sensitivity of the bulk implosion dynamics.

Before proceeding, we present a brief discussion of the need for vacuum resistivity models used in resistive MHD codes, which is one central factor motivating the need to reduce vacuum-modeling sensitivity.

\subsection{Models of Vacuum Resistivity}
\label{ssec:vacres}

A fully self-consistent model of vacuum resistivity requires extended-MHD physics, in particular Hall physics, to capture the behavior of the conductivity tensor as the density transitions to vacuum levels.  This Hall conductivity tensor formulation, described in Appendix \ref{app:XMHD}, is missing in resistive MHD codes.  Therefore, in order for the resistivity to remain finite in vacuum, resistive MHD codes have often invoked various sources of anomalous resistivity to justify vacuum resistivity models.  Reference~\onlinecite{step04} invokes anomalous resistivity due to plasma turbulence.  One frequently used source is the ion acoustic instability, described in Ref.~\onlinecite{bisk71}; another is the two-stream instability, described in Ref.~\onlinecite{chod71}.

The lower-hybrid drift instability (LHDI) has been described in the context of HED plasmas \cite{chit95}, and the associated LHDI microturbulence has been invoked as another basis for MHD vacuum resistivity models \cite{chit97}.

As described in Ref.~\onlinecite{subr18}, in models of plasma expansion into a vacuum, which should only produce a rarefaction wave, spurious shocks can arise when the vacuum is modeled as a low-density material with finite resistivity.

Reference~\onlinecite{mast21} compares several vacuum resistivity settings in resistive MHD simulations of nonlinear electrothermal instability growth, showing pronounced sensitivity if a sufficiently large vacuum-to-liner resistivity ratio is not used.

These vacuum resistivity models often involve numerical parameters (conductivity floors, density thresholds, etc.), which are not prescribed by physical considerations, and whose values often need to be adjusted depending on the experimental setup against which validation is being performed.  These parameters therefore also introduce sensitivity into the numerical results.  Moreover, if the parameters exceed the minimum values required for numerical stability, the simulation often becomes overly dissipative, damping out high-frequency modes which may have physical importance.

The extended-MHD model in PERSEUS provides a more physical description of the behavior of current at vacuum densities, including a continuous transition to zero current perpendicular to the $B$-field (see Appendix \ref{app:XMHD}), which substantially reduces its reliance on a vacuum resistivity model.  As is demonstrated in what follows, the PERSEUS results are minimally sensitive to vacuum resistivity modeling parameters. 

A fully self-consistent treatment of the vacuum requires an anisotropic resistivity model, which, as noted in the Conclusion, is not yet implemented but is under development.

\section{Numerical Methods}
\label{sec:methods}

\subsection{Extended-MHD Model}
\label{ssec:model}

We begin with the extended-MHD model used in PERSEUS, further discussion of which is presented in Refs.~\onlinecite{seyl11, zhao15,  haml18, seyl18, haml19}. 
The extended-MHD equations are solved in the following hyperbolic semi-conservative form, for a plasma with ion density $n$, ion mass $m_i$, electron mass $m_e$, electron density $n_e$, velocity ${\bf u}$, pressure $P$, current density ${\bf J}$, and resistivity $\eta$ in a magnetic field ${\bf B}$ and electric field ${\bf E}$:  
 
\begin{align}
\frac{\partial}{\partial t}(m_in) + \nabla\cdot(m_in{\bf u}) &\ =\ 0\label{density}\\
\frac{\partial}{\partial t}(m_in{\bf u}) + \nabla\cdot \left(m_in{\bf u}{\bf u} + P{\bf I}\right) &\ =\ {\bf J}\times {\bf B}\label{momentum}\\
\frac{\partial\epsilon}{\partial t}+ \nabla\cdot \left[{\bf u}\left(\epsilon + P\right)\right] &\ =\ ({\bf J}\times {\bf B})\cdot {\bf u} + \eta J^2\label{energy}
\end{align} 

\begin{align}
\frac{\partial{\bf B}}{\partial t} &\ =\ -{\nabla}\times {\bf E}\label{faraday}\\
\frac{\partial{\bf E}}{\partial t} &\ =\ {c^2}\left({\nabla}\times {\bf B} - \mu_0{\bf J}\right)\label{ampere}\\
\frac{\partial{\bf J}}{\partial t} &\ =\ \frac{n_ee^2}{m_e}\left({\bf E} + {\bf u}\times {\bf B} - \frac{{\bf J}}{n_ee}\times {\bf B} - \eta {\bf J}\right)\label{gol}
\end{align}   
 
 \medno Equation \eqref{density} describes continuity of the fluid, while Eq. \eqref{momentum} and Eq. \eqref{energy} describe conservation of momentum and energy, respectively.  Equation  \eqref{faraday} is Faraday's Law, Eq. \eqref{ampere} is Ampere's Law with displacement current, while Eq. \eqref{gol} is the GOL.  The GOL is obtained from the two-fluid model by taking a charge-to-mass superposition of the ion and electron momentum equations.
 
References such as Ref.~\onlinecite{kimu14, croc23, mart10} present the GOL with terms modeling non-quasineutral and electron inertial effects.  The derivation of the GOL \eqref{gol}, along with relevant approximations for an extended-MHD model, are presented in detail in the Appendix of Ref.~\onlinecite{mart10} .  The present work invokes the assumption of quasineutrality, i.e. $Zn \approx n_e$, where $Z$ is the ionization state, and retains only the electron inertial $\partial J/\partial t$ term for purposes of formulating the semi-implicit advance described below, dropping other electron inertial terms, e.g. electron advection, since the electron inertial length is under-resolved; see the discussion, below, of the relaxation density used in the GOL.  The electron pressure $e\nabla P_e /m_e$ term is also neglected.  This term is associated with the Biermann battery mechanism, whereby magnetic field is generated from differences in the direction of electron density and temperature gradients.  This study examines problems similar to those studied in Refs.~\onlinecite{seyl18,haml18,haml19}, for which non-quasineutral and Biermann battery effects are generally dominated by Hall physics.  In particular, for strongly magnetized systems such as these, the $B$-field generated by the Biermann battery mechanism is not expected to be a significant fraction of the driven and applied $B$-fields.  A more quantitative analysis will become possible once the Biermann battery mechanism has been incorporated into PERSEUS or FLEXO.  
 
To improve computational efficiency, the speed of light $c$ is numerically reduced by a factor $\le 30$, the approximate range for which displacement current remains negligible.  See Sec. 2D of Ref.~\onlinecite{seyl11} for further discussion.  The energy $\epsilon$ is the sum of a kinetic energy $(1/2)m_inu^2$ and an internal energy $\epsilon_{\rm int}(n,T)$, where, as will be discussed further in Sec. \ref{ssec:tabular}, the temperature $T$ is obtained from reverse interpolation on internal energy and density $n$.
 
\begin{align}
\epsilon &\ =\ \epsilon_{\rm int}(n,T) + \frac{1}{2}m_in{u}^2
\end{align}  
   
Equations \eqref{density} - \eqref{ampere} are consistent with a resistive MHD model with displacement current.  The extended-MHD terms enter through the GOL \eqref{gol}, in particular the electron inertial term $\partial J/\partial t$ on the left, and the Hall term on the right, given by ${\bf J}/(n_ee)\times {\bf B}$.  The Hall term becomes important relative to the MHD inductive ${\bf u}\times{\bf B}$ term at length scales comparable to or smaller than the ion inertial length $\lambda_i$, and becomes important relative to the resistive $\eta {\bf J}$ term in strongly magnetized plasmas for which the electron cyclotron frequency is large compared to the electron-ion collision frequency.  The displacement current and electron inertial term enable a relaxation formulation that, in a dense plasma, enforces an equilibrium for which the right-hand sides of \eqref{ampere} and \eqref{gol} are both zero.

We now elaborate on how the extended-MHD formulation in PERSEUS has enabled modeling of the plasma-vacuum interface.  The challenges of Hall MHD modeling discussed here have also been examined in Refs. such as ~\onlinecite{huba03tut,shum03,schn09}.   Much of what limits MHD codes pertains to prohibitively small timesteps at densities typical of an experimental vacuum.  In particular, when the code omits displacement current, referred to as the magneto-quasistatic approximation, the Alfven speed $\sim B/\sqrt{\mu_0\rho}$ is essentially unbounded as the density $\rho$ decreases, making the timestep prohibitively small at density floors consistent with experimental measurements of initial pump-down vacuum pressure (about 12 orders of magnitude below solid density) .  PERSEUS, on the other hand, includes displacement current in Ampere's law, so the Alfven speed is bounded by the speed of light \cite{huba03tut,shum03}:

\begin{align}
v_A &\ =\ \frac{v_{A0}\,c}{\sqrt{v_{A0}^2 + c^2}},\qquad {\rm where}\\
v_{A0} &\ =\ \frac{B}{\sqrt{\mu_0\rho}}\nonumber
\end{align}   

\medno The second complication involves modeling the Hall term, which in an explicit code requires resolving high-frequency electron cyclotron and electron plasma waves.   
In PERSEUS, the semi-implicit formulation of the GOL and Ampere's law circumvents the need to resolve these waves.  
 
 The semi-implicit advance occurs with Ampere's law \eqref{ampere} and the GOL \eqref{gol}, after all quantities have been advanced explicitly in their fluxes. Let $Q^{n+1/2}$ denote quantities advanced explicitly in their fluxes after timestep $n$.  For example, in what appears below, we have ${\bf B}^{n+1/2} = {\bf B}^n + \Delta t \nabla\times {\bf E}^n$ from an explicit advance of Faraday's law.  Then the semi-implicit advance of \eqref{ampere} and \eqref{gol} is given, in dimensional form, by
 
\begin{align}
&\frac{{\bf E}^{n+1} - {\bf E}^n}{\Delta t} \ =\ \mu_0c^2\left(\nabla\times {\bf B}^n - {\bf J}^{n+1}\right)\nonumber\\
&\frac{{\bf J}^{n+1} - {\bf J}^n}{\Delta t} \ =\ \frac{n_e^{n+1/2}e^2}{m_e}\times\\
&\left({\bf E}^{n+1} + {\bf u}^{n+1/2}\times {\bf B}^{n+1/2} - \frac{1}{n_e^{n+1/2}e}{\bf J}^{n+1}\times {\bf B}^{n+1/2} - {\eta}^n {\bf J}^{n+1}\right)\label{semiimpl0}
\end{align}
 
 \medno To convert Eqs. \eqref{semiimpl0} to the dimensionless form used by PERSEUS, a scale factor $X_0$ is introduced for each quantity in \eqref{semiimpl0}, namely $n_e$, ${\bf u}$, ${\bf E}$, ${\bf B}$, ${\bf J}$, and $\eta$.  The derived scale factors are as follows:
 
\begin{align}
u_0 &\ =\ \frac{L_0}{t_0},\qquad 
P_0 \ =\ m_in_0u_0^2,\qquad
B_0 \ =\ \sqrt{\mu_0 P_0},\nonumber\\
E_0 &\ =\ B_0u_0,\qquad
J_0 \ =\ \frac{B_0}{\mu_0L_0},\qquad \eta_0 \ =\ \mu_0u_0L_0\label{qscales}
\end{align}
 
 \medno We also introduce an ion inertial length scale defined as $\lambda_{i0} = \sqrt{m_i/(Zn_0e^2 \mu_0)}$, along with the electron inertial length $\lambda_e = c/\omega_{pe} = \sqrt{m_e/(n_ee^2\mu_0)}$.  Using these definitions, along with Eqs. \eqref{qscales}, and letting $\tilde{X} = X/X_0$ denote $X$ in the dimensionless units used by PERSEUS, the dimensionless form of Eqs. \eqref{semiimpl0} becomes

\begin{align}
&\frac{\tilde{\bf E}^{n+1} - \tilde{\bf E}^n}{\Delta \tilde{t}} \ =\ \frac{c^2}{u_0^2}\left(\nabla\times \tilde{\bf B}^n  - \tilde{\bf J}^{n+1}\right)\nonumber\\
&\frac{\tilde{\bf J}^{n+1} - \tilde{\bf J}^n}{\Delta \tilde{t}} \ =\ \nonumber\\
&\qquad \frac{L_0^2}{\lambda_e^2}\left(\tilde{\bf E}^{n+1} + \tilde{\bf u}^{n+1/2}\times \tilde{\bf B}^{n+1/2} - H - \tilde{\eta}^n \tilde{\bf J}^{n+1}\right), \  {\rm where}\label{semiimpl}\\
H &\ \equiv\ \frac{\lambda_{i0}}{\tilde{n}_e^{n+1/2}L_0}\tilde{\bf J}^{n+1}\times \tilde{\bf B}^{n+1/2}\nonumber
\end{align}  

 \medno In the simulations in this study, electron inertia does not play a significant role.  Accordingly, in order for the GOL relaxation parameter $L_0^2/\lambda_e^2$ to maintain accurate relaxation conditions, a sufficiently large reference electron density is used for $\lambda_e$  to ensure that $\lambda_e$ is small relative to the grid scale.  In the present simulations, a reference density of $10^{-3}$ of solid density is used, though the results are insensitive to this value as long as it is large enough to satisfy $\lambda_e\Delta x \ll 1$.  This is only done in the GOL relaxation parameter $L_0^2/\lambda_e^2$, and no other terms.  See Sec. II of Ref.~\onlinecite{seyl18}, along with Appendix \ref{app:XMHD}, for further discussion.
 
 \medno When the relaxation parameters $c^2/u_0^2$ and $L_0^2/\lambda_e^2$ are large, as occurs with the latter parameter in a dense plasma, the system \eqref{semiimpl} enforces the following steady-state Hall equilibrium:
 
\begin{align}
{\nabla}\times {\bf B} &\ =\ \mu_0{\bf J}\\
0 &\ =\ {\bf E} + {\bf u}\times {\bf B} - \frac{{\bf J}}{n_ee}\times {\bf B} - \eta {\bf J}\label{gol_rlx}
\end{align}  

The semi-implicit formulation \eqref{semiimpl} is implicit only in the sources, not the fluxes, so the solve is completely local, which saves the CPU expense of a global solve of a spatially coupled system.   

Because the time-advance of current density introduces an electron inertial term, PERSEUS can in principle model an electron plasma wave, which, owing to the semi-implicit time advance, does not need to be resolved. 

The system \eqref{density}-\eqref{gol} is solved using Discontinuous Galerkin (DG) spatial discretization \cite{zhao14, zhao15} in linear basis (2 nodes per cell in each direction), for which the 3-D basis elements are $\{1,x,y,z,yz,zx,xy,xyz\}$, and more generally, at arbitrary basis order, are products of Legendre polynomials $P_i(x)P_j(y)P_k(z)$.  In a 2-D axisymmetric geometry, the linear DG basis elements $\{1,r,z,rz\}$ have the form $P_i(r)P_j(z)$.  To compute fluxes through cell interfaces, a local Lax-Friedrichs scheme is used for fluxes of the $E$ and $B$-fields, and unless otherwise stated, an HLLC approximate Riemann solver \cite{batt97} is used for the hydrodynamic fluxes, namely density, momentum, and energy.  The implementation of these schemes is described in Refs.~\onlinecite{haml16, haml18, haml18b}.

$\nabla\cdot {\bf B} = 0$ is enforced locally in each cell using the linear slopes in the DG basis.  For the problems in this study, $\nabla\cdot {\bf B}$ is negligible.

PERSEUS uses Message-Passing-Interface (MPI) parallelization by sub-dividing the computational domain into a grid of subdomains, where each subdomain is assigned its own processor.

If the temperature exceeds a certain value, a damping term $(0.2/\Delta t)\epsilon_{\rm int}$ is subtracted from the energy source in order to mitigate numerical instabilities associated with thermal runaway heating of current-carrying low-density plasma (discussed, e.g., in Ref.~\onlinecite{seyl18}).  This energy damping was also used in PERSEUS simulations in Refs. such as~\onlinecite{haml18}, ~\onlinecite{haml18b}, and ~\onlinecite{seyl18}, and is found to have a minimal impact on the present results apart from providing numerical stability.

\subsection{Interfacing with Tabular EOS and Conductivity}
\label{ssec:tabular}

The tabular interface consists of Equation-of-State (EOS) SESAME tables which tabulate internal energy and pressure, and conductivity tables for electrical conductivity, ionization state, and thermal conductivity.  See Ref.~\onlinecite{lyon92}.  Each quantity is tabulated over an array of densities and temperatures.  At the start of the simulation, these tables are loaded for the material being simulated.  Then, during the simulation, interpolation routines are used for computing any of these quantities at a given simulated density and temperature.  This is typically done in the following sequence.  

\begin{enumerate}
\item First, the internal energy is recovered by subtracting the kinetic energy from the conserved energy.  
\item Then, reverse interpolation is performed on the internal energy and density to recover the temperature.  
\item Finally, forward interpolation is performed on the density and temperature to interpolate the desired quantity (pressure, electrical conductivity, ionization state, etc.) onto tabulated values.
\end{enumerate}

%\begin{figure*} %[!h]
%\centering
%\includegraphics[scale=0.5]{figs/tabular2.png}
%\caption{Tabular interface in PERSEUS}
%\label{fig:tabular}
%\end{figure*}

\medno The interpolation routines are described in more detail in Ref.~\onlinecite{kerl77}.  To avoid extrapolation, the inputs to the tables are clipped, as needed, to remain within tabulated ranges.  Clipping of density and temperature is only actually applied at the upper tabulated bounds, not the lower bounds, as density and temperature floors are already above the tabular minima.  A  pressure floor, if used, is computed by interpolating the tabulated pressure onto the density and temperature floors at the start of the simulation.  %As we shall see in Secs. \ref{ssec:vac1d} to \ref{sec:vac2dmrt}, at sufficiently low density floors, the simulation is minimally influenced by imposing a pressure floor.

\begin{figure*} %[!h]
\centering
\includegraphics[scale=0.5]{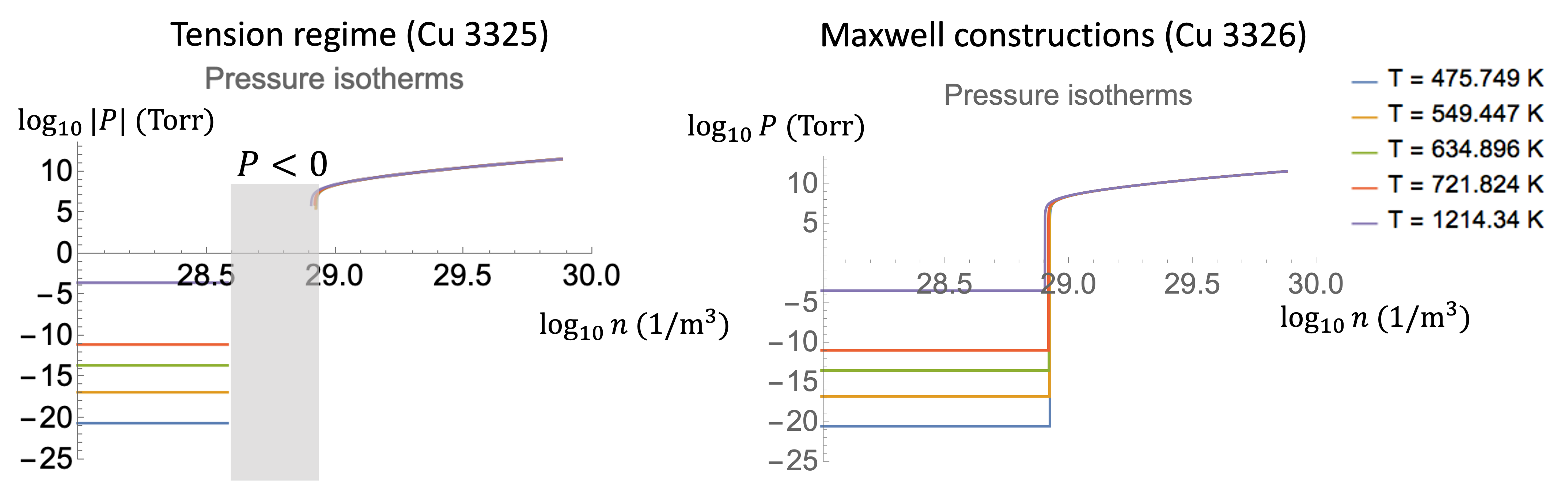}
\caption{Copper SESAME EOS: Tension regime replaced by Maxwell constructions}
\label{fig:tensionCu}
\end{figure*}

\begin{figure} %[!h]
\includegraphics[scale=0.45]{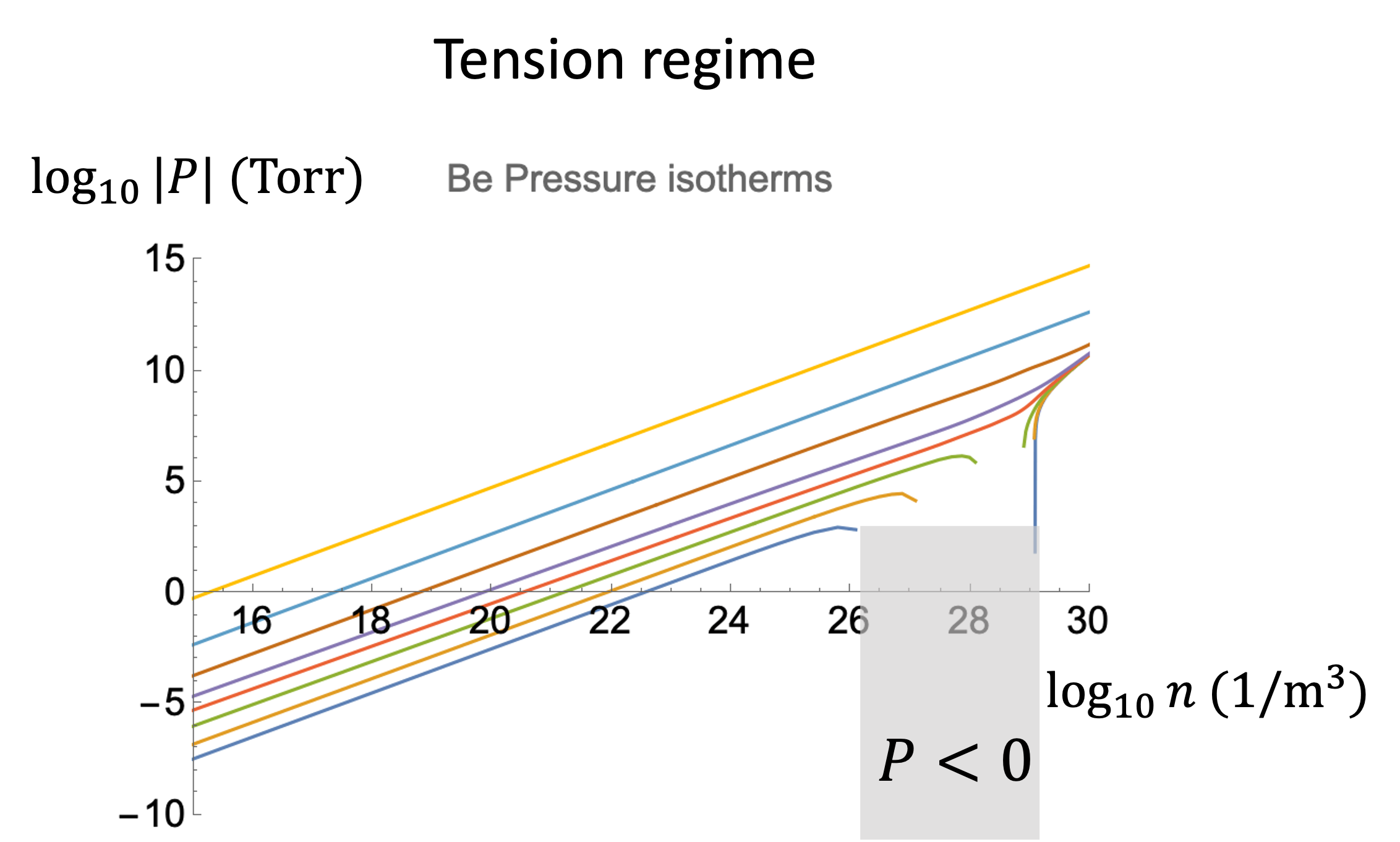}
\includegraphics[scale=0.45]{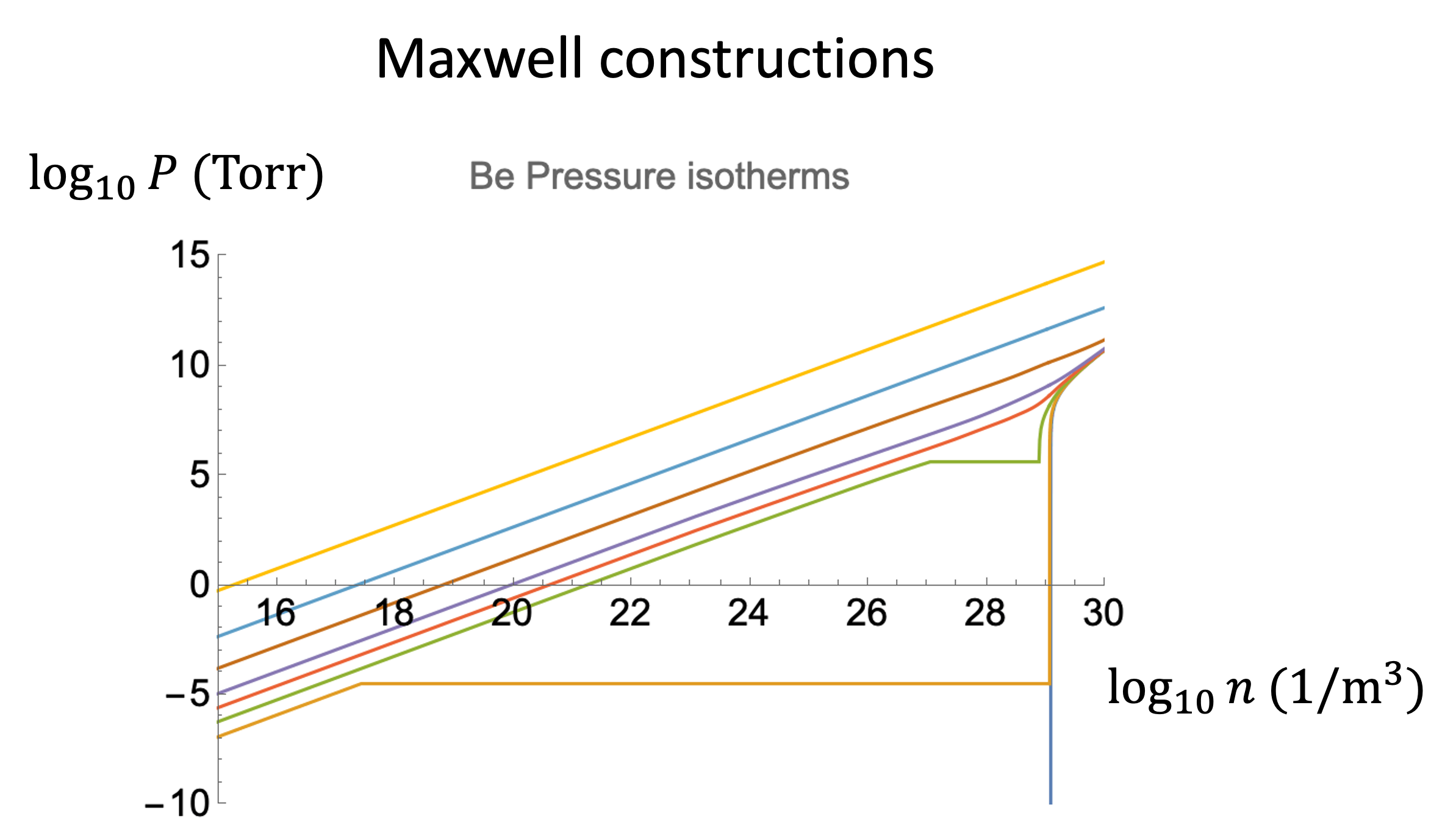}
\includegraphics[scale=0.45]{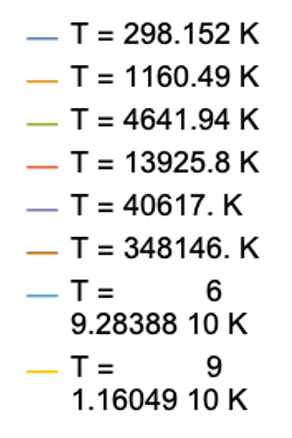}
\caption{Beryllium SESAME EOS: Tension regime replaced by Maxwell constructions}
\label{fig:tensionBe}
\end{figure}

Several improvements were made to the tables in order to interface them with PERSEUS.  The first was to extend the range of tabulated densities downward, to below the physical vacuum, so that physical quantities could be interpolated properly at or near vacuum densities.  The second was to impose Maxwell constructions in tables for which the vapor dome (a region of phase space in which liquid and vapor phases can coexist) contained negative pressures or regions of undefined sound speed due to $\partial P/\partial\rho < 0$.  Specifically, for densities along pressure isotherms for which either $P < 0$ or oscillatory regions where $\partial P/\partial\rho$ changes sign, typically in the vapor dome, a Maxwell construction applies a constant pressure across these densities along the isotherm.  In the case of positive oscillatory pressures, this constant pressure is typically the average of the pressures being replaced.  Because PERSEUS does not yet have a stable model for addressing negative pressures, we found it desirable, for this stage of testing, to restrict our attention to the compression regime.  Figures \ref{fig:tensionCu} show pressure isotherms for Copper with a tension regime versus with Maxwell constructions, and Figs. \ref{fig:tensionBe} show the same for the Beryllium EOS.  These are the two materials modeled in this study.

\subsection{Limiting}
\label{ssec:limiting}

The limiting in PERSEUS is focused on positivity preservation of density, and bounds preservation of internal energy, both of which are needed by DG algorithms to address oscillations near discontinuities, particularly at the plasma-vacuum interface \cite{zhao14}.  To bound the density, the cell-centered density $\rho_{cell,0}$ is first compared to the density floor $\rho_{floor}$.  If $\rho_{cell,0} > \rho_{floor}$, an evaluation is performed of the densities at all nodes on the inside and surface of the cell.  If any of these fall below the density floor, the density slopes are adjusted such that the minimum of these densities is now equal to $\rho_{floor}$.  That is, the density slopes are multiplied by a factor $\theta$ given by

\begin{align}
\theta &\ =\ \frac{\rho_{cell,0} - \rho_{floor}}{\rho_{cell,0} - {\rm min}(\rho_{nodes})}\label{theta}
\end{align} 

\medno If $\rho_{cell,0} < \rho_{floor}$, the slopes of density, momentum, energy, and current are zeroed out, while the cell-centered values of these quantities are reset to floor values.

Internal energy is handled in much the same way, where the minimum tabulated internal energy serves as the lower bound, and the limiting in this case acts to prevent tabular extrapolation.   Limiting on internal energy was found to provide much more numerical stability than limiting on pressure, likely due to the fact that internal energy at each node can be computed directly from the conserved quantities, while computation of nodal pressures requires interpolation over densities and temperatures.  The slopes of each conserved fluid quantity, namely density, momentum, and energy, are then multiplied by the factor $\theta$ found for internal energy.  

Moreover, the positivity properties noted in Ref.~\onlinecite{zhao14} for an ideal gas pressure can readily be extended to bounds preservation of internal energy in an arbitrary EOS.  To see this, note that Ref.~\onlinecite{zhao14} discusses pressure-floor limiting in the context of an ideal gas EOS for which internal energy is $P/(\gamma - 1)$ for a constant $\gamma$.  For an arbitrary EOS, the bounds preservation analysis for internal energy $e_{\rm int}$ involves merely (i) replacing $P/(\gamma - 1)$ with $e_{\rm int}$ in the expression for total energy, and (ii) using, as the lower bound, the minimum tabulated internal energy, which in this case can be negative.  What is being preserved about internal energy, therefore, is not ``positivity" but rather the internal energy remaining within tabulated ranges, i.e. above the cold curve, so as not to require extrapolation.  Therefore, the analysis of Ref.~\onlinecite{zhao14} can readily be applied to internal energy limiting, partly because for $X = $ internal energy and for $X = $ ideal gas pressure, the total energy can be written as

\begin{align}
\epsilon &\ =\ CX + \frac{1}{2}mnu^2
\end{align}

\medno where $C = 1$ for internal energy and $C = 1/(\gamma-1)$ for ideal gas pressure.  However, no such analytic form exists for writing total energy in terms of pressure computed from a tabular EOS, and thus tabular pressure should not be used in bounds preservation.

%Moreover, the internal energy, energy, momenta, and density form a convex set in a linear DG basis, which is to say, if the internal energy is above some minimum threshold at the nodes on a cell boundary, The same is true of the nodes internal to the cell.  This follows from a simple extension of the argument presented in Ref.~\onlinecite{zhao14}.  Reference \cite{zhao14} notes that the pressure-floor limiting forms a convex set in the energy, momenta, and density when assuming an ideal gas EOS for which internal energy is $P/(\gamma - 1)$ for a constant $\gamma$.  The convexity analysis for internal energy $e_{\rm int}$ involves merely replacing $P/(\gamma - 1)$ with $e_{\rm int}$ in the expression for total energy.  Therefore, internal energy must have the same convexity properties.

A density buffer interval can be optionally applied just above the density floor, characterized by a floor multiplier $f_{mult}$, such that at cell-centered densities below $f_{mult}\rho_{floor}$, energy is reset to a floor value, and momentum, current density, and higher-order modes of each of these quantities are all zeroed out.  Such an interval is frequently used in resistive MHD codes \cite{seyl18, haml19}, and is included here for purposes of demonstrating (i) that this buffer interval often suppresses physical behavior (see Sec. \ref{ssec:rzbuffer}), and, as we shall see in Sec. \ref{ssec:vac1d}, (ii) that this interval is much more influential at the higher density floors typical of resistive MHD codes than at the lower floors enabled by the PERSEUS extended-MHD model.  Because resistive MHD codes lack extended-MHD physics, the transition from plasma to vacuum is artificially abrupt, which can introduce numerical instability associated with suddenly switching off hydrodynamics and electromagnetics at the density floor.  The density buffer thus provides a transitional interval for resistive MHD that mimics the more physical transition to vacuum modeled by extended-MHD, discussed in Appendix \ref{app:XMHD}.

%\begin{figure*}
%\centering
%\includegraphics[scale=0.5]{figs/positivity.png}
%\caption{Bounds preservation in PERSEUS}
%\label{fig:limiting}
%\end{figure*}

We have now described the essential features of the PERSEUS model of a plasma-vacuum interface, namely the extended-MHD model, the limiting, and the interface with tabular EOS and conductivity.  We now turn to numerical results of several representative 1-D and 2-D problems, and demonstrate the ways in which the PERSEUS extended-MHD model has reduced sensitivity of the results to various aspects of vacuum modeling.

\section{Numerical Results}
\label{sec:results}

This section examines the implosion of a liner first in a 1-D radially convergent geometry, and then in a 2-D $r-z$ axisymmetric geometry.  Compared to the 1-D problem, the 2-D problem will be shown to have (i) stronger coupling of Hall physics to the electromagnetics, and (ii) stronger coupling of vacuum sensitivity into the bulk implosion dynamics if the density floor is too high.

The vacuum sensitivity of the bulk implosion dynamics will be quantified, in both problems, by graphing the time-dependence of the liner centroid, and observing that the sensitivity, to vacuum modeling parameters, of the centroid position increases from the 1-D radially convergent geometry to 2-D $r-z$ geometry.  

In both problems, vacuum modeling sensitivity is observed in the low-density plasma, and is examined because this sensitivity is often coupled into the bulk dynamics, particularly in 2-D $r-z$ or 3-D geometries.

As will be shown in Secs. \ref{ssec:vac1d} and \ref{ssec:vac2drz}, sensitivity to vacuum modeling is minimized when (i) a sufficiently low density floor is used, and (ii) Hall physics is modeled.  

These results do not include the effects of anisotropic transport such as thermal conduction and viscosity.  As discussed in the Conclusion, this transport is under development and will be incorporated in future publications.  While these transport terms may affect details of the results, we do not expect the salient conclusions to be altered.

\subsection{1-D liner implosion in radially convergent geometry}
\label{ssec:vac1d}

As our first test of convergence with respect to vacuum parameters, we model a Copper liner imploding in a 1-D radially convergent geometry with driven $B_{\theta}$.  SESAME EOS and conductivity tables are used for Copper, where the EOS table uses Maxwell constructions in the vapor dome.  The liner has an outer radius of 1.92 mm and inner radius of 1.6 mm, corresponding to an aspect ratio $R_{\rm outer}/(R_{\rm outer} - R_{\rm inner}) = 6$.  The domain extends from $r = 0$ to $r = 8$ mm, spanning 400 cells, corresponding to a cell size of 20 $\mu$m.  An azimuthal $B$-field is driven at the outer radial boundary using the following analytic current profile:

\begin{align}
I(t) &\ =\ I_{\rm peak}\sin^2\left(\frac{t}{\tau}\right)\label{icurr}
\end{align}

\medno where $I_{\rm peak} = 26$ MA is the peak current, and $\tau = 100$ ns is the rise time of the pulse.  Inside the liner is a 1 mg/cc gas fill (modeled as a Copper vapor, as only a single material can be modeled), which models the fuel and is typical of fill densities used in MagLIF experiments at Sandia.

For a one-dimensional problem such as this, for which the magnetic field is perpendicular to the simulation direction, the influence of the Hall term enters through the coupling to hydrodynamics, without which the Hall term is not predicted to have any effect in this geometry (apart from a negligible effect due to displacement current). 

To see this, consider the equations in 1-D axisymmetric geometry (no axial variation, only radial) for driven $B_{\theta}$.  The momentum equation \eqref{momentum}, Faraday's law \eqref{faraday}, Ampere's law \eqref{ampere}, and the GOL \eqref{gol} become

\begin{align}
\frac{\partial}{\partial t}(\rho u_r) &\ =\ -\frac{\partial P}{\partial r} - \frac{\partial}{\partial r}(\rho u_r^2) - J_zB_{\theta}\label{rmomtm}\\
\frac{\partial}{\partial t}(\rho u_z) &\ =\ -\frac{\partial}{\partial r}(\rho u_ru_z) + J_rB_{\theta}\label{zmomtm}\\
\frac{\partial B_{\theta}}{\partial t} &\ =\ \frac{\partial E_z}{\partial r}\label{dbdt}\\
\frac{\partial E_r}{\partial t} &\ =\ -\mu_0c^2 J_r\label{derdt}\\
\frac{\partial E_z}{\partial t} &\ =\ \frac{1}{r}\frac{\partial}{\partial r}(rB_{\theta}) - \mu_0c^2 J_z\label{dezdt}\\
\frac{\partial J_r}{\partial t} &\ =\ \frac{n_ee^2}{m_e}\left(E_r - u_zB_{\theta} + \frac{J_z}{n_ee}B_{\theta} - \eta J_r\right)\label{djrdt}\\
\frac{\partial J_z}{\partial t} &\ =\ \frac{n_ee^2}{m_e}\left(E_z + u_rB_{\theta} - \frac{J_r}{n_ee}B_{\theta} - \eta J_z\right)\label{djzdt}
\end{align} 

This system describes the time-evolution of $u_r$, $u_z$, $B_{\theta}$, $E_r$, $E_z$, $J_r$, and $J_z$.  Note that the complementary $B_r$, $B_z$, $E_{\theta}$, and $J_{\theta}$ components are not generated, as none of these appear in Eqs. \eqref{rmomtm} - \eqref{djzdt}.

Equations \eqref{rmomtm} - \eqref{zmomtm} and \eqref{derdt} - \eqref{djzdt} constitute a system $u_r$, $u_z$, $E_r$, $E_z$, $J_r$, and $J_z$ for which no subset of these variables can be decoupled from the rest, and for which the Hall term increases this coupling by introducing the perpendicular $J$-components into the GOL.  Therefore, a nonzero $B_{\theta}$, $E_z$, $J_z$, and therefore $u_r$ (Eq. \eqref{rmomtm}) mean that $J_r$ must also be nonzero.  $E_r$ and $J_r$ are in fact both close to zero in vacuum, and are both nonzero in the vicinity of the imploding liner, where the $E$ and $B$-fields are strongly coupled to hydrodynamics.

This nonzero $J_r$ introduces a nonzero Hall velocity $J_r/(n_ee)$ multiplying $B_{\theta}$ in Eq. \eqref{djzdt}, which in turn modifies the $E_z$ computed from Eq. \eqref{djzdt}.  This results in a modification to the $\partial_rE_z$ that governs the time-evolution of $B_{\theta}$ in Faraday's law \eqref{dbdt}.

The nonzero $E_r$ introduces a non-negligible $\nabla \cdot {\bf E}$ in high-density current-carrying regions.  However,  enforcing $\nabla\cdot {\bf E} = 0$ would not be consistent for a quasi-neutral model.  To see this, note that in resistive MHD, we have $\nabla\cdot {\bf E} = \nabla \cdot (-{\bf u}\times {\bf B} + \eta {\bf J})$, which is generally nonzero.  It has been verified that enforcing $\nabla \cdot {\bf J} = 0$ (in the same manner as described for $\nabla \cdot {\bf B}$ in Sec. \ref{ssec:model}) has little impact on test results.

%TODO: Did we verify the above?

Having characterized the manner in which extended-MHD physics is expected to influence this problem, we are now in a position to discuss the resulting impact of extended-MHD on the numerical results, regarding first the density floor, and then other aspects of vacuum modeling.

\subsubsection{Convergence with respect to density floor}
\label{ssec:1Dfloor}

We begin by examining the impact of Hall and electron inertial physics on convergence with respect to the density floor.  That is, at a given mesh resolution, we wish to show that at sufficiently low density floors, the numerical solution becomes minimally sensitive to the density floor.

\begin{figure}[!h]
\includegraphics[scale=0.5]{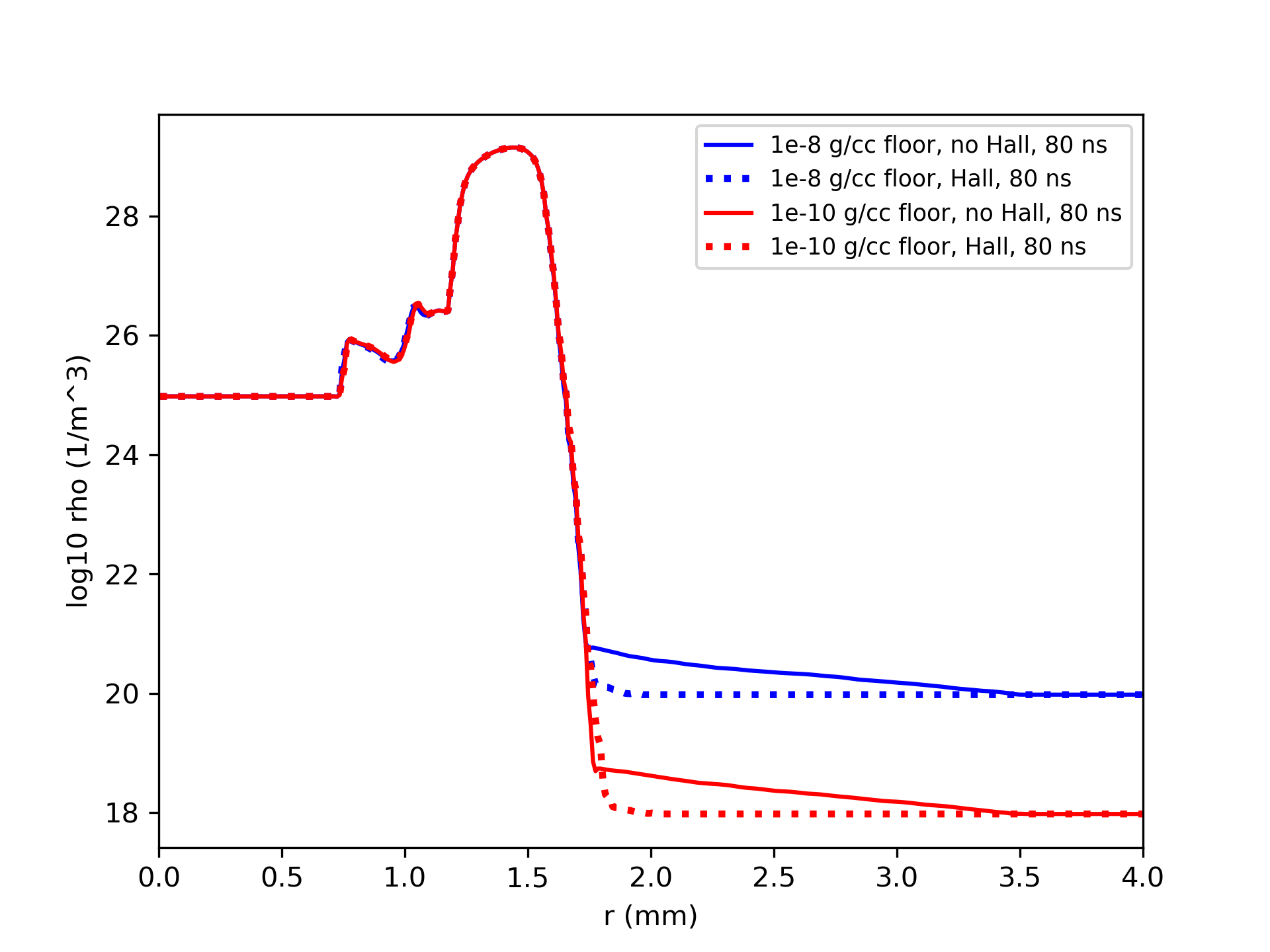}
\caption{Base-10 logarithm of density during liner implosion phase using different density floors.  Bulk dynamics are approximately converged with respect to floor, while low-density dynamics are not converged.}
\label{fig:1Dliner_impl}
\end{figure}

The implosion phase of the liner is shown in Fig. \ref{fig:1Dliner_impl} at two values of the density floor, with and without the Hall term being modeled.  In this phase, the bulk dynamics show minimal sensitivity to the density floor and to the Hall term; sensitivity to either of these emerges mainly in the low-density dynamics.   For a system with axial variation, e.g. a 2-D $r-z$ axisymmetric geometry for which the Hall term generates significant $E$ and $B$-field components (see Sec. \ref{ssec:vac2drz}), the implosion phase shows much more sensitivity both to the Hall term and density floor.

\begin{figure}[!h]
\includegraphics[scale=0.5]{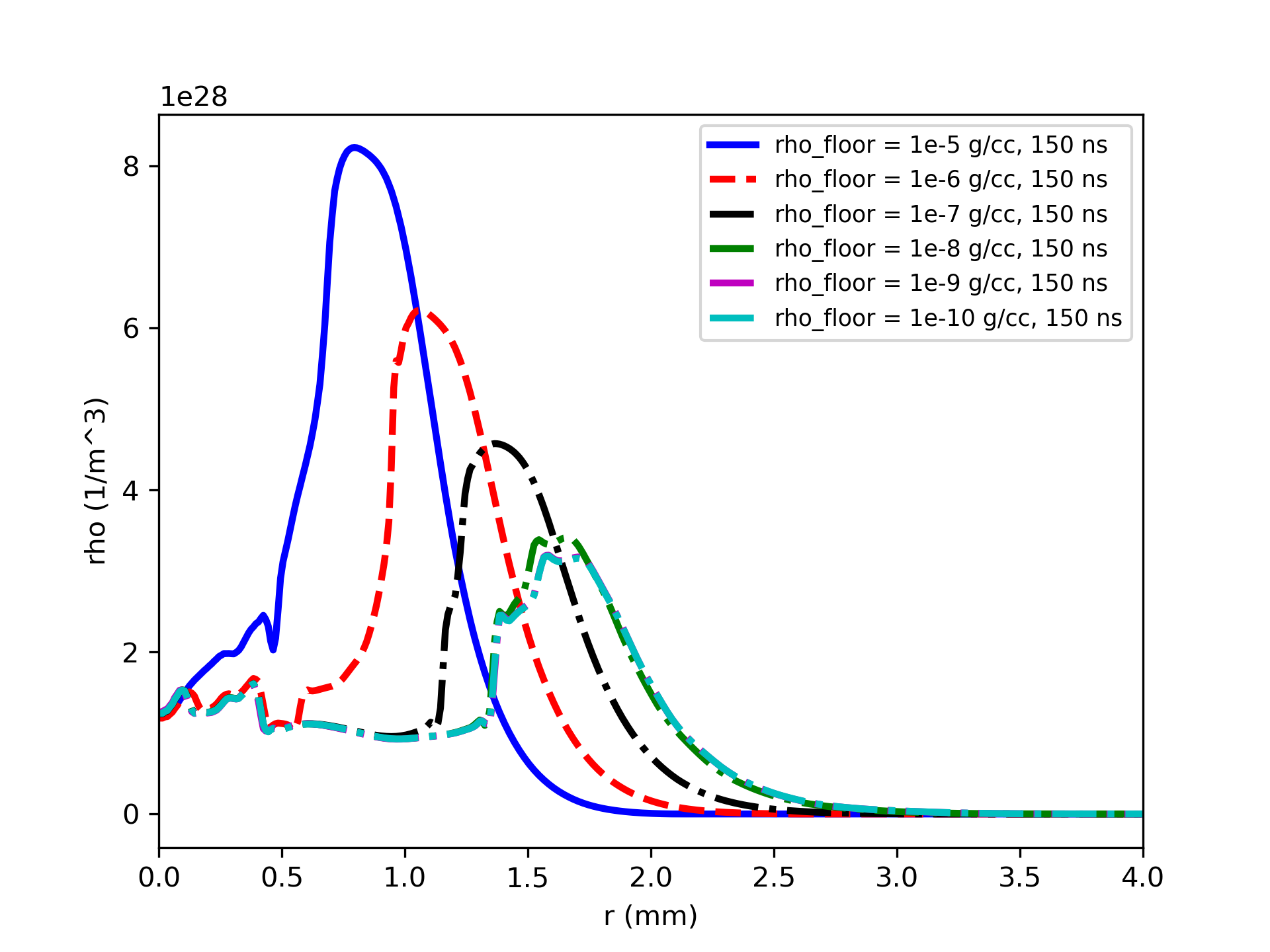}\\
\includegraphics[scale=0.5]{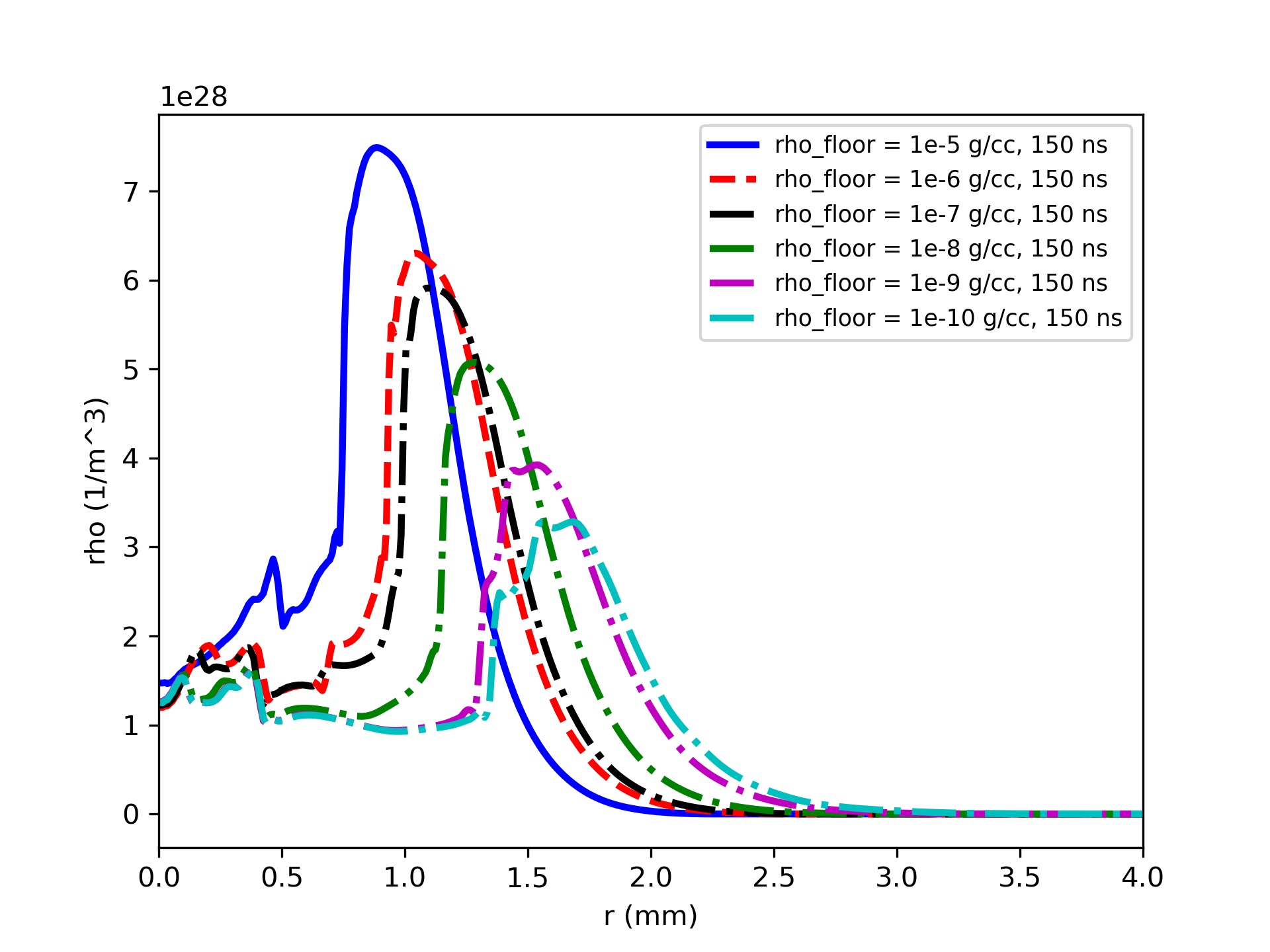}
\caption{Density displays at 150 ns showing the liner post-stagnation modeled at different density floors, with the Hall term modeled (top) and without the Hall term (bottom).  Convergence at higher floors is observed when the Hall term is included.}
\label{fig:1Dliner2}
\end{figure}

Sensitivity to Hall term and density floor are also much stronger in the post-stagnation phase.  This is seen in Figs. \ref{fig:1Dliner2}, which demonstrate (i) convergence of bulk dynamics at sufficiently low density floors both with and without the Hall term being modeled, and (ii) that the results converge at higher density floors when the Hall term is modeled, seen in particular by comparing the curves at or below the $10^{-7}$ g/cc floor between the two graphs.  

Importantly, the post-stagnation is not converged (shows high vacuum sensitivity) at the floors of $10^{-5}$ and $10^{-6}$ g/cc typical of many MHD codes (recall the MHD timestep restrictions noted in Sec. \ref{ssec:model}).  This is consequential for ICF targets; in order to accurately predict the confinement times of the fuel in an imploded target, the dynamics during and after stagnation need to be minimally sensitive to vacuum modeling \cite{knap17}.  The use of a thick metal tamper, which for this problem would be the liner, assumes a long confinement time $\tau$ to enable values of $P\tau$ sufficient for fusion reactions while reducing the requirement on the stagnation pressure $P$.  Testing this assumption in turn requires a predictive model of confinement.  The post-stagnation results in this study provide an important step toward such a model in that, for a given mesh, we have reduced the sensitivity of this phase to parameters characterizing the numerical vacuum, as we demonstrate in this section.  %Future work will focus on improving convergence of the post-stagnation with respect to mesh refinement, which will require the use of adaptive mesh refinement (AMR) to adequately resolve the stagnation.  

To quantify floor convergence before, during, and after the implosion, we examine the time-dependence of the liner centroid, which is computed as 

\begin{align}
\rho_c &\ =\ \int_0^{r_{\rm max}}\rho r^2dr \left(\int_0^{r_{\rm max}}\rho rdr\right)^{-1}\label{centroid}
\end{align}

\begin{figure}[!h]
(i) \includegraphics[scale=0.5]{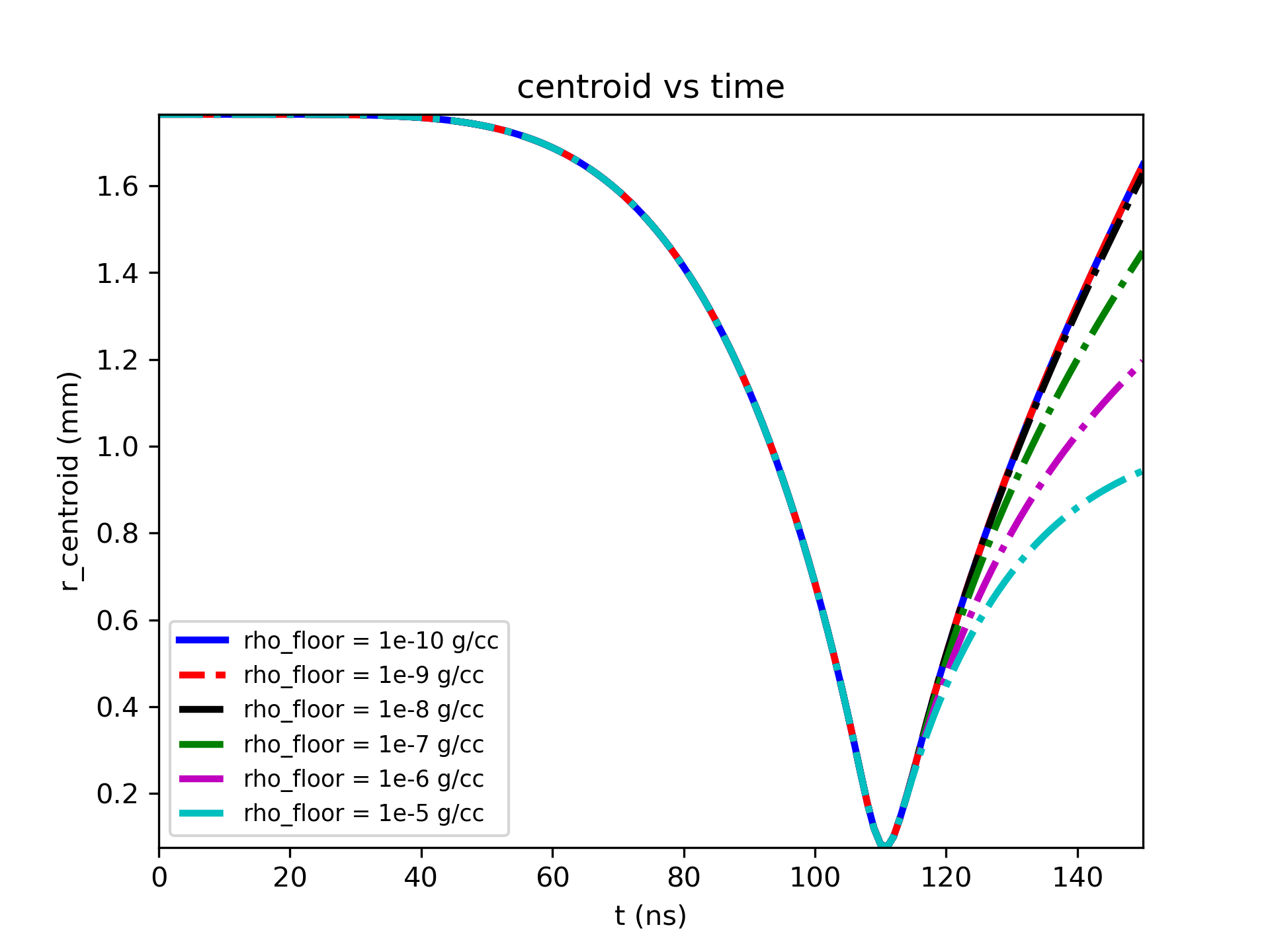}\\
(ii) \includegraphics[scale=0.5]{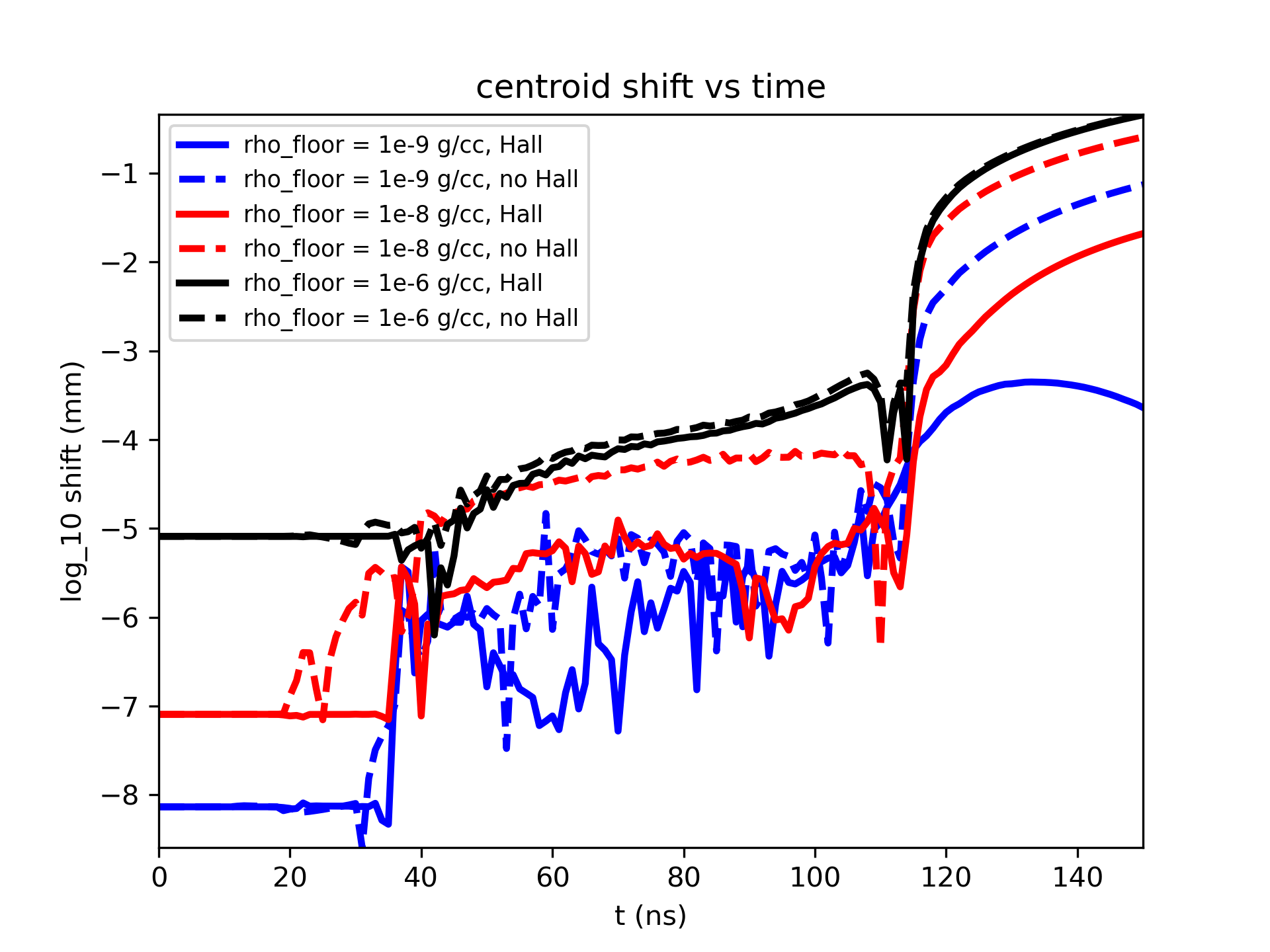}
\caption{Time-dependence of (i) liner centroid with Hall included, and (ii) shift in liner centroid relative to corresponding $10^{-10}$-g/cc-floor case with Hall included (solid curves) and with Hall omitted (dashed curves).}
\label{fig:1Dliner_cent_hall}
\end{figure}

\medno using $r_{\rm max} = 6$ mm.  Figures \ref{fig:1Dliner_cent_hall} plot the time-dependence of the (i) liner centroid (with Hall included) and (ii) centroid shift relative to the corresponding $10^{-10}$ g/cc floor case, where the Hall term is included in the solid curves, and omitted in the dashed curves.  Note that during the implosion, the centroid shift is minimal, indicating minimal vacuum modeling sensitivity of the bulk implosion dynamics for this 1-D problem.  By contrast, in Sec. \ref{ssec:vac2drz}, which discusses a liner imploding in 2-D $r-z$ axisymmetric geometry, the centroid shifts are larger during the implosion phase.  

However, even with very small shifts in centroid position, several observations are worth noting.  First, with or without Hall included, the centroid shift decreases with decreasing floor, which is to say, the results converge with respect to density floor.  Second, Hall physics provides improved convergence, i.e. smaller centroid shift, not only in post-stagnation but also during the implosion.  That is, at each floor, the solid curve is below the corresponding dashed curve of the same color.

These observations are consistent with the analysis, below, of $L_2$ differences versus time.  The $L_2$ difference between discretely sampled spatial profiles $f_2(r)$ and $f_1(r)$, normalized relative to $f_1(r)$, is computed as

\begin{align}
L_2 &\ =\ \sqrt{\frac{\sum_{i=1}^N [f_2(r_i) - f_1(r_i)]^2}{\sum_{i=1}^N f_1(r_i)^2}}\label{L2diff}
\end{align}

\begin{figure}[!h]
\includegraphics[scale=0.5]{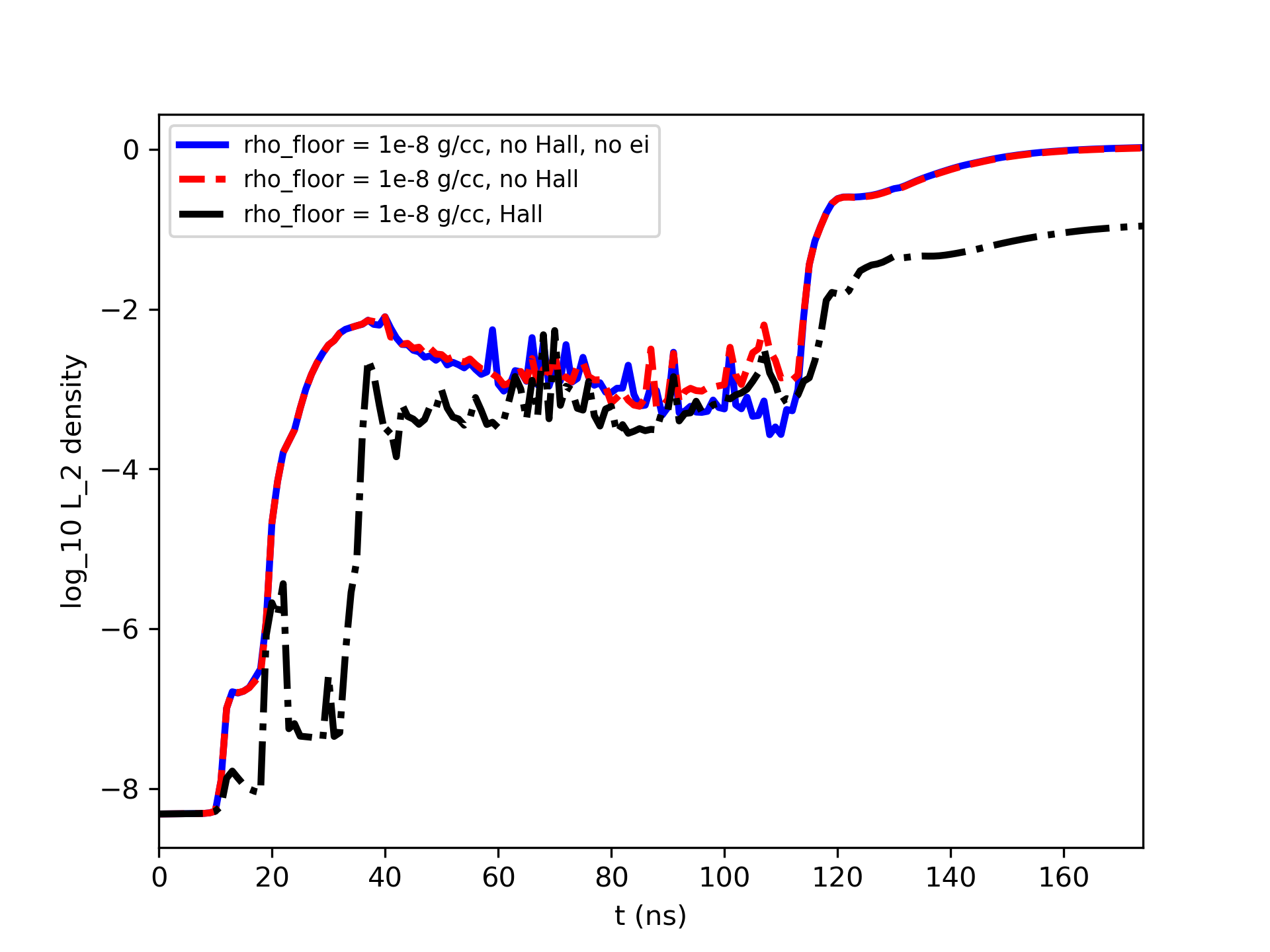}\\
\caption{Base-10 logarithm of normalized $L_2$ difference, computed using Eq. \eqref{L2diff}, between spatial profiles of density at a $10^{-8}$ g/cc floor versus a $10^{-10}$ g/cc floor, normalized to the $10^{-10}$ g/cc-floor result.  The three cases compared are (i) Hall physics modeled, (ii) Hall physics omitted, and (iii) Hall physics and electron inertial term omitted.}
\label{fig:1Dliner_L2norms}
\end{figure}

\medno where the numerical profiles $f_1(r)$ and $f_2(r)$ have been mapped from a linear DG basis representation onto a uniform sub-grid of 2 cells per computational cell.  Figure \ref{fig:1Dliner_L2norms} shows the time-dependence of the normalized $L_2$ difference between spatial profiles of density obtained using a $10^{-8}$ g/cc density floor and a $10^{-10}$ g/cc floor.  Three cases are compared, namely (i) no Hall term or electron inertial $\partial_tJ$ term, (ii) electron inertial term included but no Hall term, and (iii) both Hall and electron inertial terms included.  While the electron inertial term has a minimal effect, the Hall term has a much more pronounced effect in accelerating convergence with respect to the density floor, particularly during the post-stagnation and prior to the implosion, and to a lesser extent during the implosion.

This result is expected, and is consistent with the intended purpose of the electron inertial term described in Sec. \ref{ssec:model}.  Recall that this term manifests as an artificially large electron plasma frequency included not for modeling electron inertial physics, but rather for modeling relaxation of the GOL to a Hall MHD equilibrium.  

\begin{figure}[!h]
\includegraphics[scale=0.5]{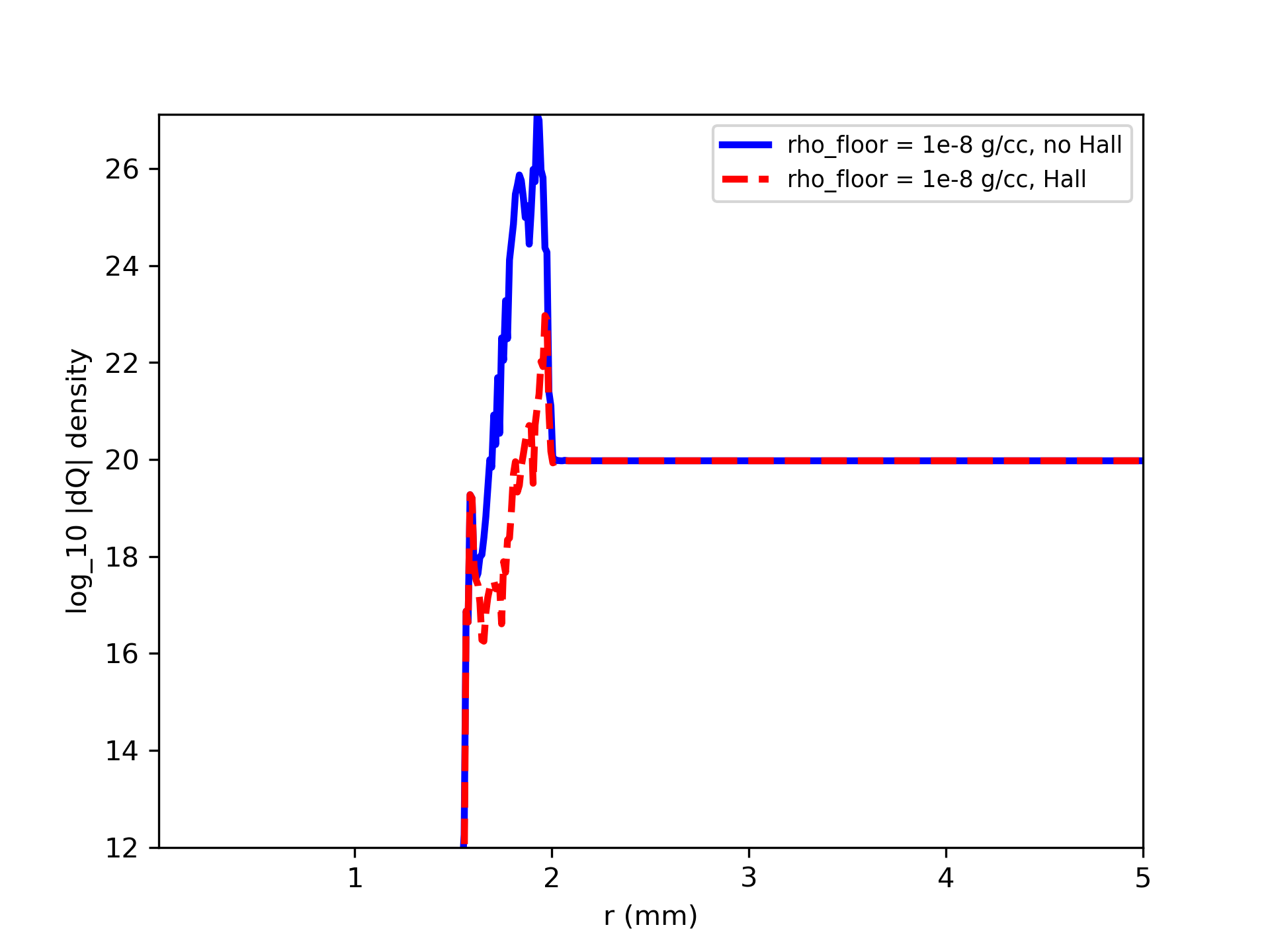}\\
\caption{Base-10 logarithm of absolute difference in density profiles between $10^{-8}$ g/cc and $10^{-10}$ g/cc floors at 30 ns, showing that Hall physics improves floor convergence of the mass distribution in the liner prior to the implosion.}  %  Bottom graph: base-10 logarithm of density profiles at 30 ns at  $10^{-8}$ g/cc and $10^{-10}$ g/cc floors.}
\label{fig:Qdiff30}
\end{figure}

\begin{figure}[!h]
\includegraphics[scale=0.5]{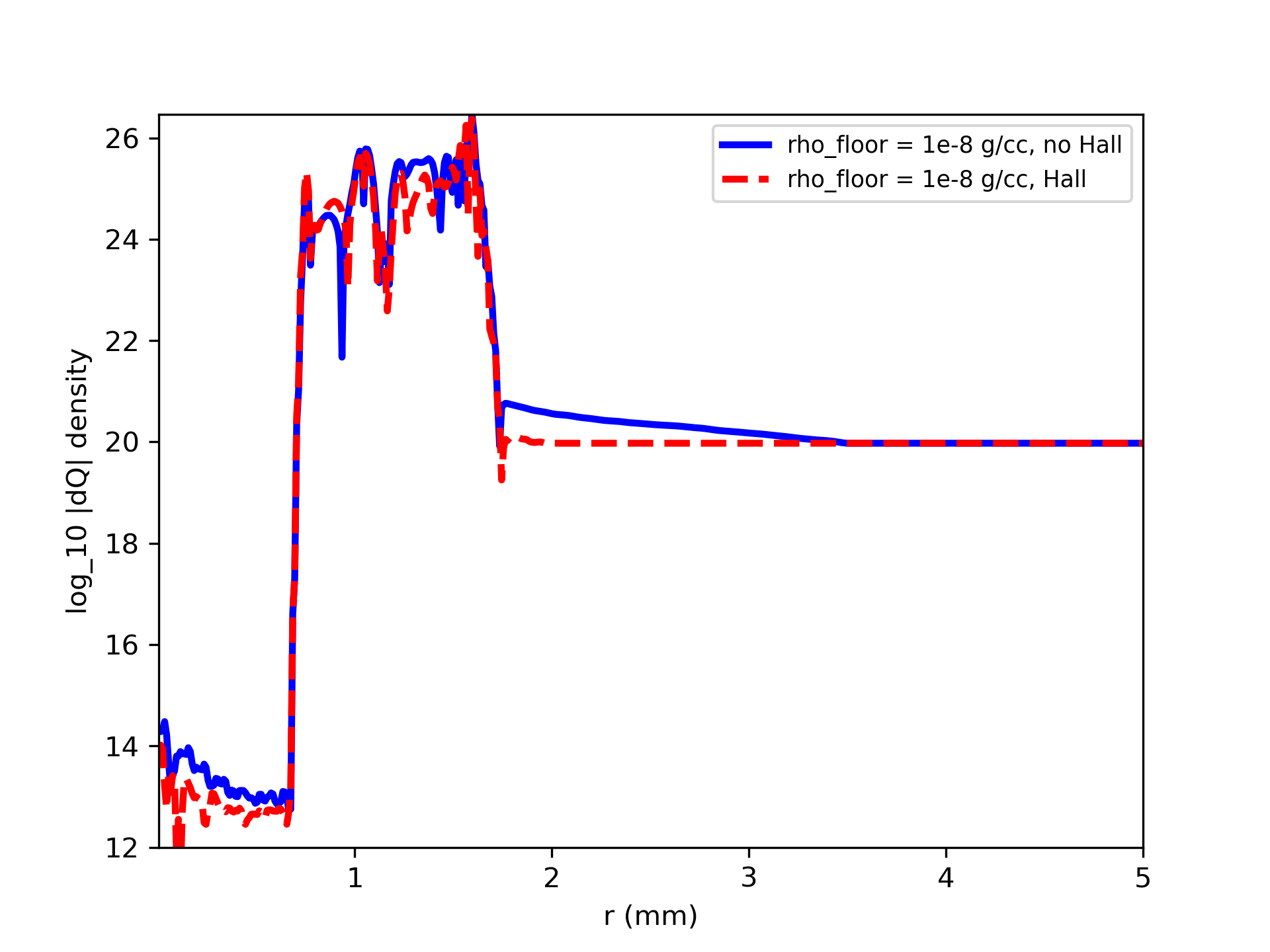}\\
\caption{Base-10 logarithm of absolute difference in density profiles between $10^{-8}$ g/cc and $10^{-10}$ g/cc floors at 80 ns, showing that, during the implosion, Hall physics has less impact on floor convergence of the mass distribution in the liner.}  %  Bottom graph: base-10 logarithm of density profiles at 80 ns at  $10^{-8}$ g/cc and $10^{-10}$ g/cc floors.}
\label{fig:Qdiff80}
\end{figure}

To understand the improvement provided by Hall in floor convergence prior to the implosion shown in Fig. \ref{fig:1Dliner_L2norms}, we examine Fig. \ref{fig:Qdiff30}, which is a snapshot at 30 ns of the spatial dependence of the base-10 logarithm of the absolute difference in densities between $10^{-8}$ g/cc and $10^{-10}$ g/cc floors, comparing the cases with and without the Hall term.  Evidently, most of the improved floor convergence due to Hall is in the modeling of the liner material.  On the other hand, Fig. \ref{fig:Qdiff80} makes this same comparison at 80 ns, and shows that during the liner implosion, once more current has been coupled into the liner, the Hall and no-Hall cases exhibit more comparable floor sensitivity in the liner material.  These figures compare the spatial profiles of density, however the same observations are true of the profiles of thermal pressure, magnetic pressure, and current density.

More generally, the accelerated density-floor convergence provided by Hall can be understood in terms of conductivity.  Hall physics models a full 3-by-3 conductivity tensor, given in Appendix \ref{app:XMHD}, which, compared to a resistive MHD model, more self-consistently captures the transition to zero vacuum current density perpendicular to the $B$-field.  Further discussion of this mechanism is provided in Appendix \ref{app:XMHD} and Ref.~\onlinecite{seyl18}.   Compared to resistive MHD, Hall physics is therefore less susceptible to numerical instabilities (e.g. thermal runaway instabilities; see Ref.~\onlinecite{seyl18}) associated with unphysical vacuum currents.

%TODO: how does this relate to plasma frequency in two-fluid model?

\subsubsection{Sensitivity to density buffer}
\label{ssec:1Dvacparams}

\begin{figure}
\includegraphics[scale=0.5]{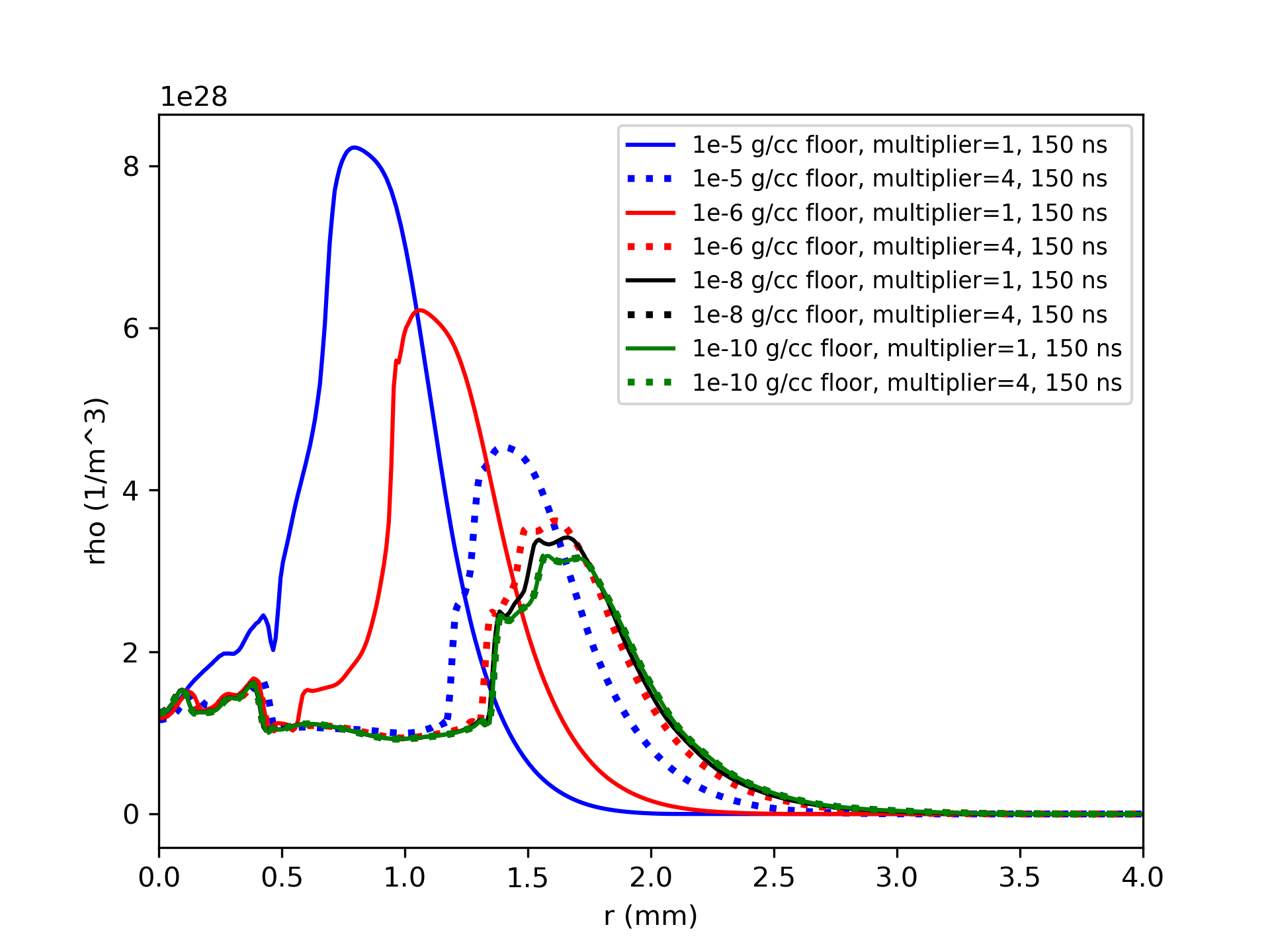}
\caption{Density display at 150 ns showing the liner post-stagnation modeled at different density floors with versus without the use of a density ``buffer".  This buffer has minimal influence at a sufficiently low density floor.}
\label{fig:1Dliner_buffer}
\end{figure}

With Hall physics turned on, we now examine convergence with respect to the density ``buffer" discussed in Sec. \ref{ssec:limiting}, as an example of an adjustable parameter frequently used in vacuum modeling.  Figure \ref{fig:1Dliner_buffer} shows the impact, on post-stagnation density, of the density ``buffer" .  Recall that a floor multiplier of 1 means that no buffer is used.  A comparison of the black curves shows that this buffer significantly impacts the results at a floor of $10^{-5}$ g/cc, and a comparison of the blue curves shows the same at a floor of $10^{-6}$ g/cc.  However, a comparison of the red curves, which overlay one another, shows that the influence of the buffer is no longer visible once the density floor is lowered to $10^{-10}$ g/cc.  Moreover, the use of a moderate buffer causes the result to converge at higher density floors, thus showing that for this 1-D problem, the floor convergence in post-stagnation bulk dynamics is associated with a reduction in numerically introduced current just above vacuum density, which is filtered out by the density buffer.  ``Numerical" in this context refers to a quantity introduced by pressure gradients and magnetic forces acting on vacuum cells.  The buffer also resets numerical momentum and energy, but these were found to have minimal impact on the results.

\begin{figure}
\includegraphics[scale=0.5]{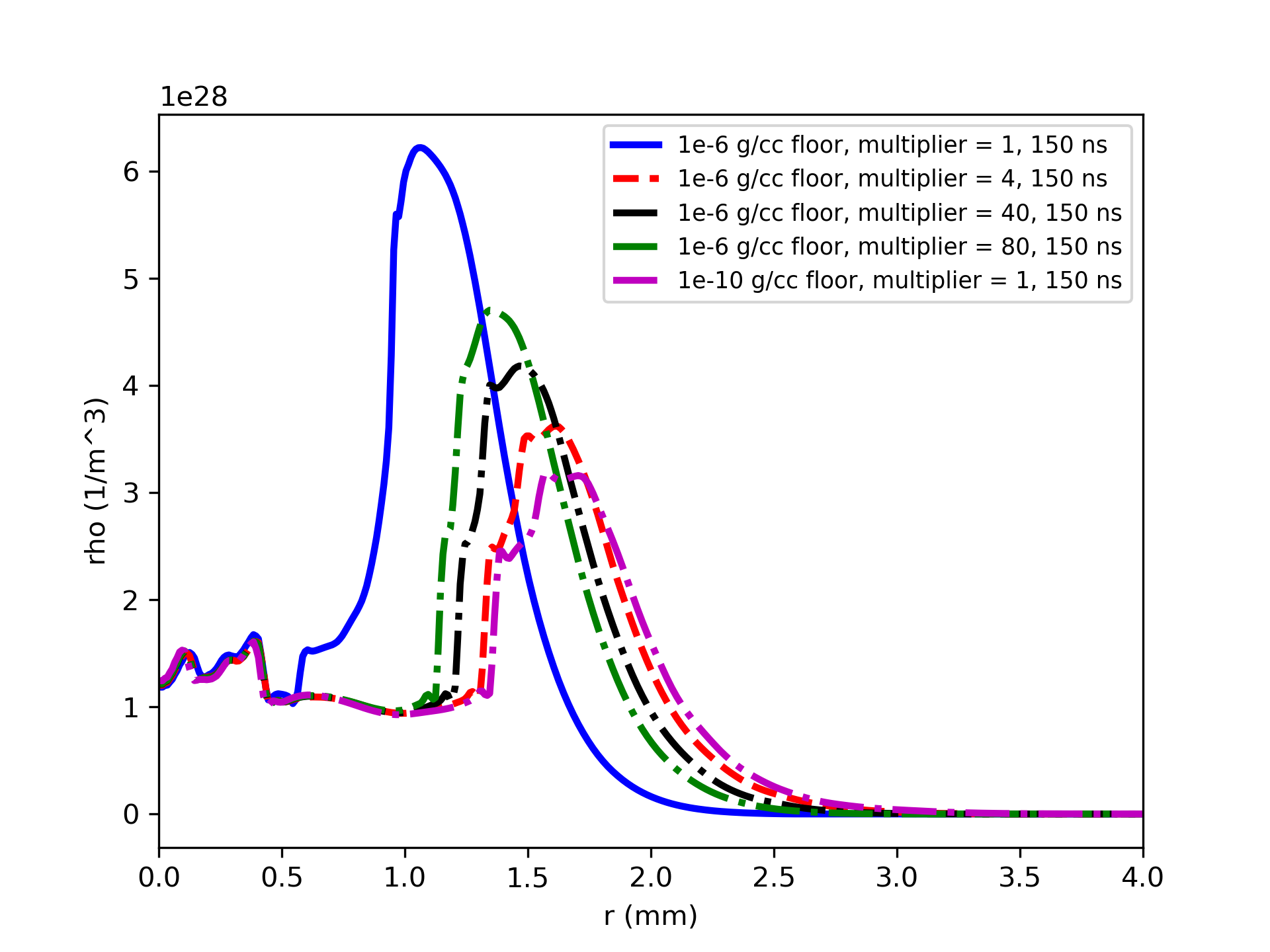}
\caption{Density display at 150 ns showing the liner post-stagnation at a $10^{-6}$ g/cc density floor and different values of the floor multiplier characterizing the density buffer, compared with the solution at a $10^{-10}$ g/cc density floor.  Behavior at a lower density floor is not reproduced by simply increasing the floor multiplier.}
\label{fig:1Dliner_buffer2}
\end{figure}

However, the solution at a low density floor cannot be recovered with a higher floor by simply increasing the floor multiplier in the buffer.  As this floor multiplier continues to increase, the current removed from the system is physical (generated in cells which were never part of the vacuum), and not just numerical, causing the solution to diverge from the approximately-floor-converged solution.  This is illustrated in Fig. \ref{fig:1Dliner_buffer2}, which compares the post-stagnation at a $10^{-6}$ g/cc density floor at several values of the floor multiplier, demonstrating that as the floor multiplier increases, the solution converges toward, and then diverges from, the solution at a $10^{-10}$ g/cc floor.

\begin{figure}
\includegraphics[scale=0.5]{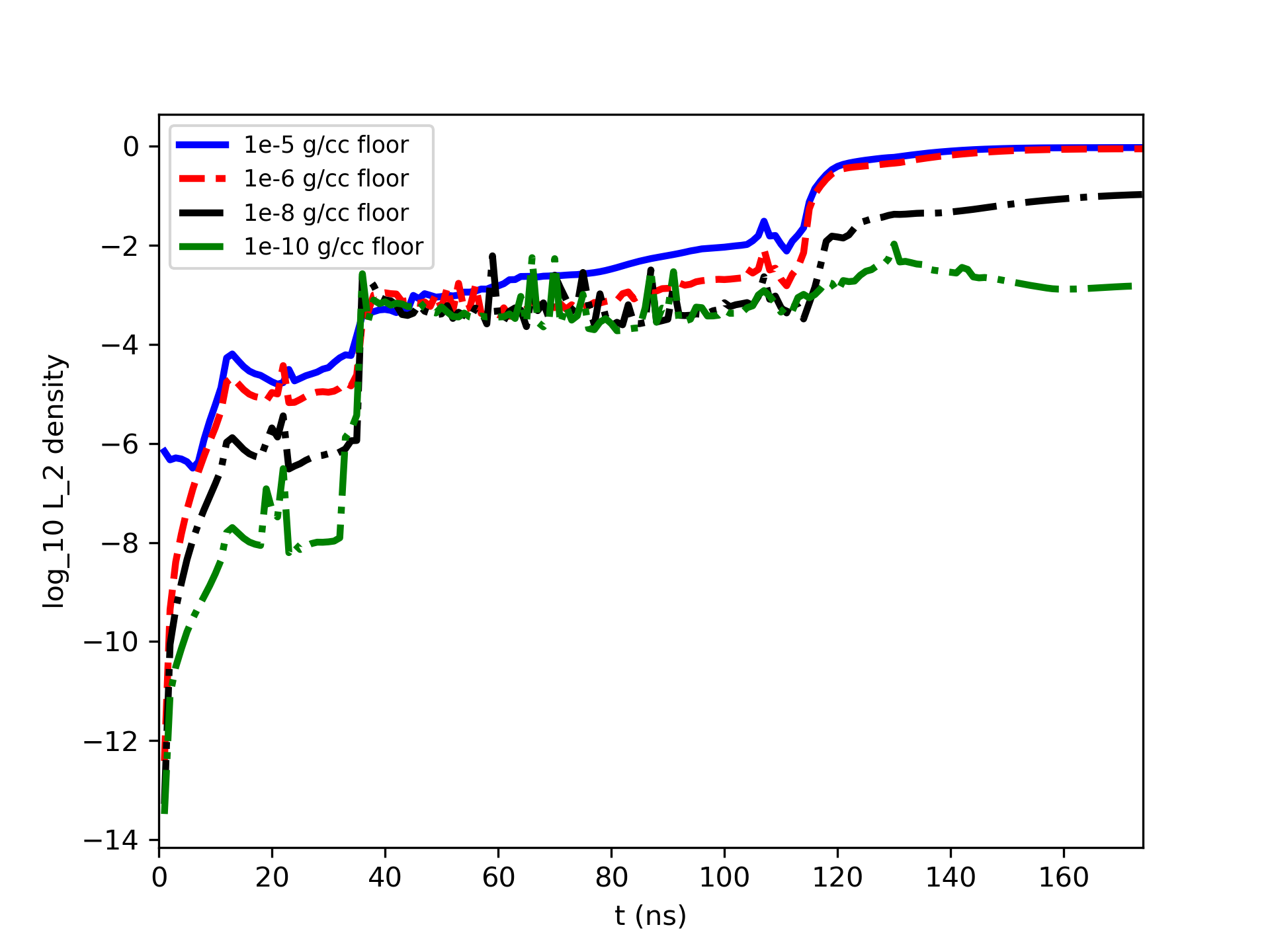}
\caption{Base-10 logarithm of normalized $L_2$ difference, computed using Eq. \eqref{L2diff}, between density profiles with versus without the use of a density buffer with a floor multiplier of 4.  Particularly before the implosion and in the post-stagnation, but also during the implosion, the influence of the density buffer decreases with decreasing density floor.}
\label{fig:1Dliner_L2_buffer}
\end{figure}

We now examine the influence of the density buffer throughout the simulation by plotting, in Fig. \ref{fig:1Dliner_L2_buffer}, the time-dependence of the $L_2$ norm of the difference between density profiles at several density floors.  This graph shows that (i) buffer sensitivity at a given floor generally increases throughout the simulation, and (ii) the reduced buffer sensitivity at lower floors occurs not just in the post-stagnation, but also prior to the implosion, and even during the implosion.  Other quantities, e.g. pressure, current density, $E$-field, and $B$-field, show qualitatively this same behavior in buffer sensitivity vs time, and the floor-dependence of this sensitivity.  Insensitivity at sufficiently low floors is also observed with the use of pressure floor resets (reset of the pressure to the value corresponding to the density and temperature floor, if less than that value).

Finally, a diagnostic was added that outputs the limiting slopes ($\theta$ values discussed in the Sec. \ref{ssec:limiting}) applied to density, momentum, and energy to keep density equal to or above the floor, and to keep internal energy equal to or above the tabular minimum.  It was verified that for the 1-D liner section, this limiting occurs in a small percentage of the domain localized around the liner.  In particular, the following metric was calculated to represent, at a given output time, the fractional limiting of density or internal energy applied to the entire domain, where $r_{max} = 6$ mm:

\begin{align}
f &\ =\ \int_0^{r_{max}} (1-\theta) r\,dr \left( \int_0^{r_{max}} r\,dr\right)^{-1}\label{ftheta}
\end{align}

\begin{figure}
(i) \includegraphics[scale=0.5]{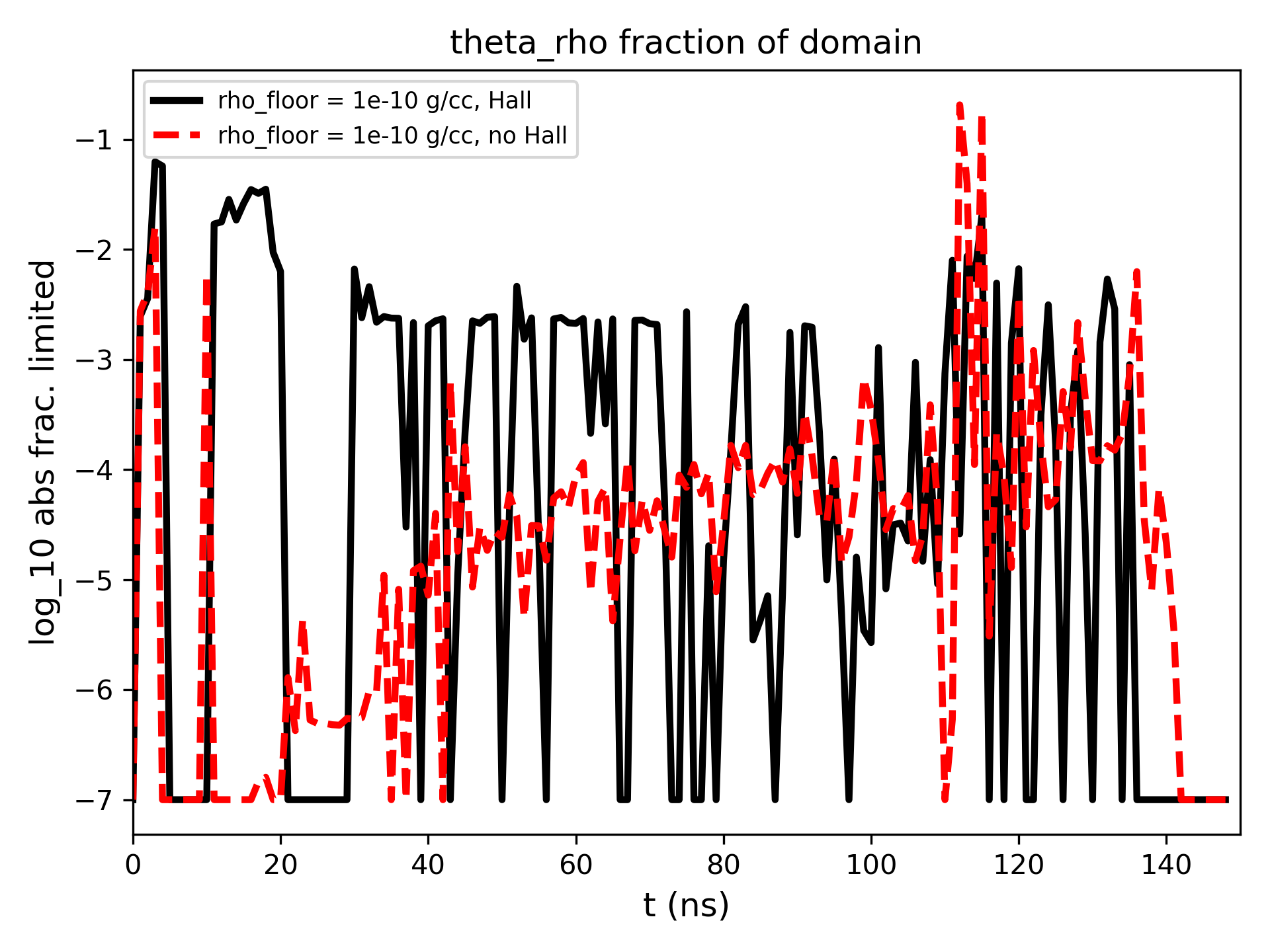}
(ii) \includegraphics[scale=0.5]{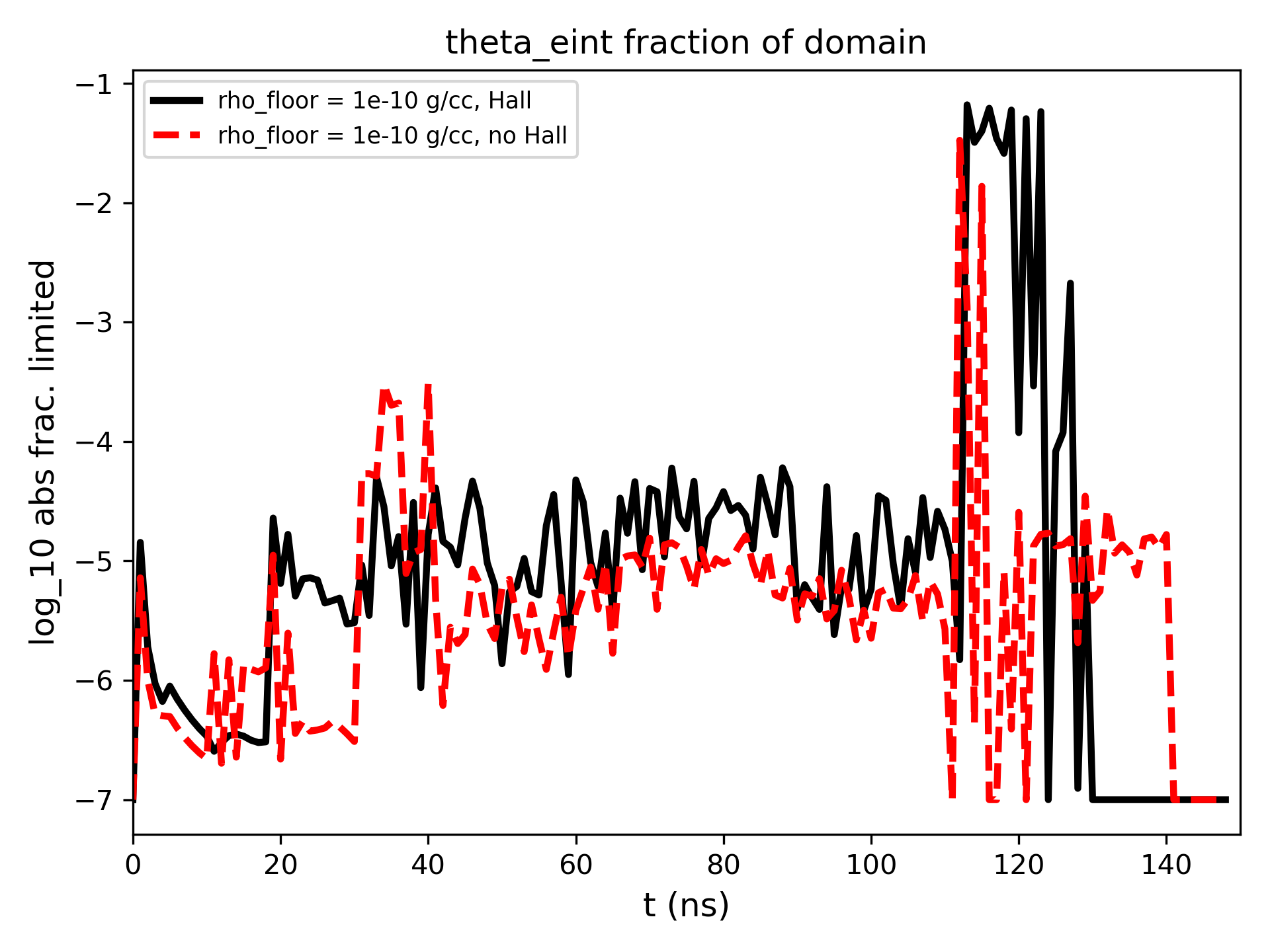}
\caption{Base-10 logarithm of time-dependence of fraction of domain for which limiting occurs (i) to maintain density above the density floor, and (ii) to maintain internal energy above the tabular minimum.  This domain fraction is computed using Eq. \eqref{ftheta}.  Cases with Hall (solid) and without Hall (dashed) are compared. Density floor is $10^{-10}$ g/cm$^3$.}
\label{fig:1Dliner_theta}
\end{figure}

\begin{figure}
(i) \includegraphics[scale=0.5]{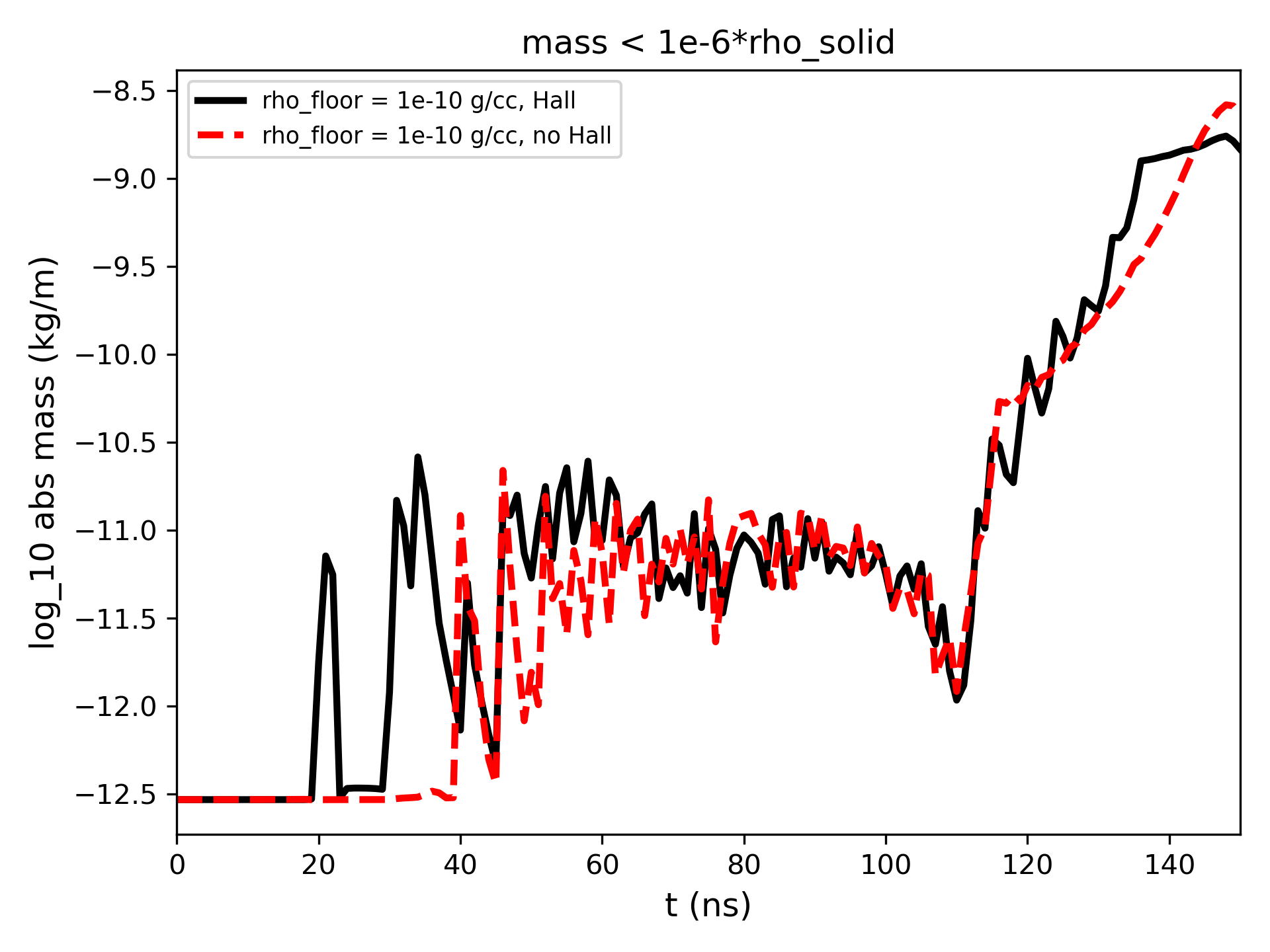}
(ii) \includegraphics[scale=0.5]{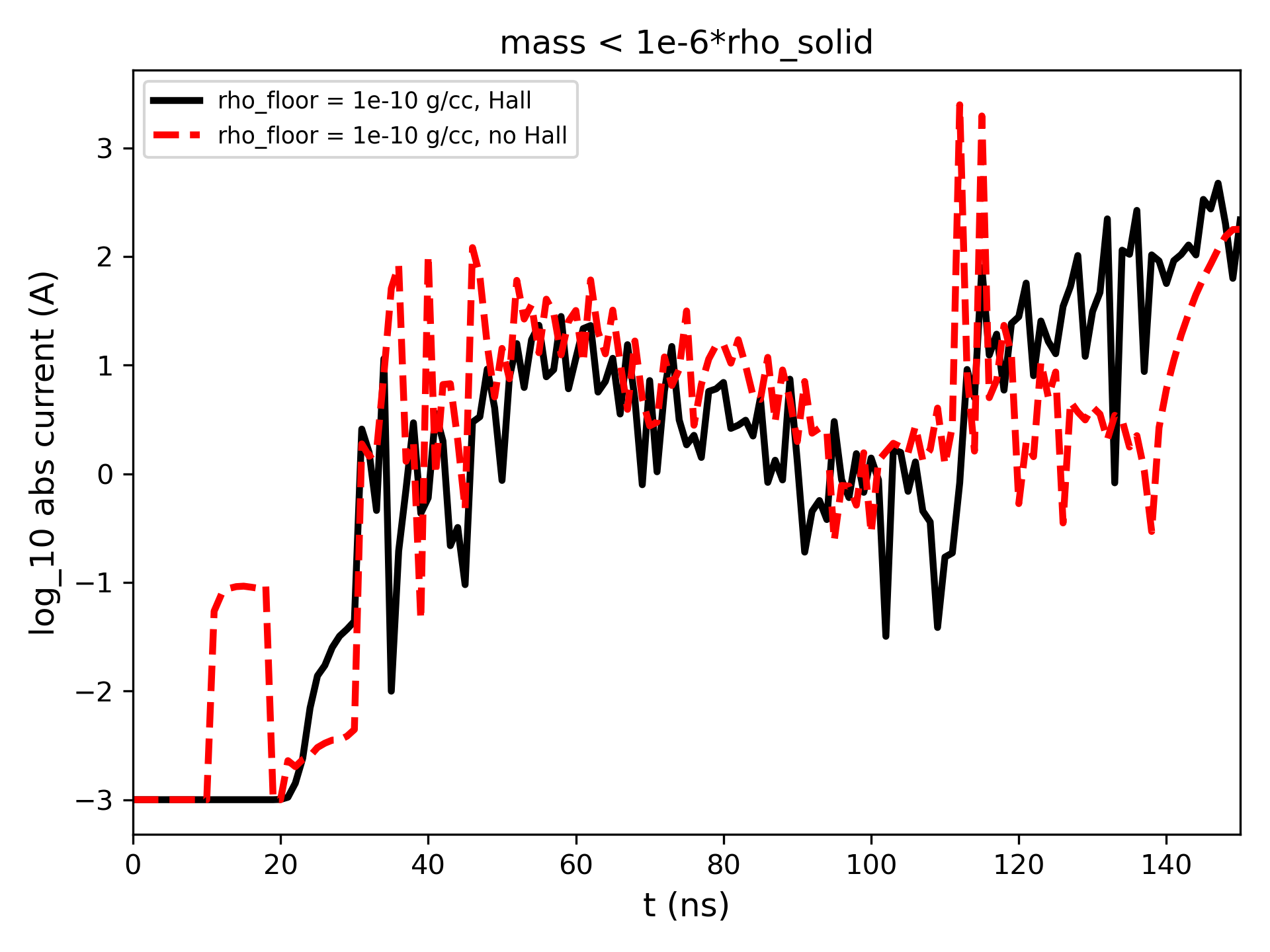}
\caption{Base-10 logarithm of time-dependence of (i) total linear mass density below $10^{-6}\rho_{\rm solid}$, and (ii) total axial current carried by densities below $10^{-6}\rho_{\rm solid}$, each integrated over a cross-section in the $r-\theta$ plane. Cases with Hall (solid) and without Hall (dashed) are compared. Density floor is $10^{-10}$ g/cm$^3$.}
\label{fig:1Dliner_low_density}
\end{figure}

The time dependence of this quantity is shown in Figs. \ref{fig:1Dliner_theta} for limiting on density and internal energy.  During the implosion, we almost always have $f < 0.01$, i.e. less than 1$\%$ of the domain is limited, while in the post-stagnation, there are isolated times at which more than 1$\%$, but less than 10$\%$, of the domain is limited.  The amount of limiting does not show a clear correlation with density floor.  Somewhat more frequent limiting occurs in the presence of the Hall term (solid curves), which could be partly due to the Hall term modeling a somewhat larger mass of current-carrying low-density plasma in the vacuum.  See Fig. \ref{fig:1Dliner_low_density} (i), which shows the time-dependence of the total linear mass density below $10^{-6}\rho_{\rm solid}$, computed (with $r_{\rm max} = 8$ mm) as 

\begin{align}
\lambda &\ =\ \int_0^{r_{max}} dr\,2\pi r [\rho(r) \le 10^{-6}\rho_{\rm solid}]
\end{align}

During the early implosion, and for much of the post-stagnation, the Hall term (solid curve) models more mass in this density range than without Hall (dashed curve).  Prior to the implosion and in its early stages, this observation is independent of the density threshold used, though at later simulation times and other thresholds, the impact of the Hall term on mass is not as clear-cut.  This impact of the Hall term is also observed in the 2D liner implosion examined in Sec. \ref{ssec:vac2drz}.

Figure \ref{fig:1Dliner_low_density} (ii) shows total axial current carried by plasma with density below $10^{-6}\rho_{\rm solid}$.  Prior to and during the implosion, the Hall term generally reduces the amount of current carried by low-density plasma, at all density thresholds that were tried.  This is consistent with the observation, noted earlier in this section, that the Hall term more self-consistently models a transition to zero current density as density drops to vacuum levels.  The Hall term no longer has this effect in the post-stagnation, likely due to the rapid filling of the domain with plasma expanding at low densities.

The salient results of this section are as follows:
\begin{enumerate}
\item For this 1-D geometry with perpendicular $B_{\theta}$, the influence of Hall physics enters through coupling to hydrodynamics.
\item Sensitivity to Hall physics and density floor is primarily confined to low-density plasma during the implosion, and is most visible in the bulk dynamics after stagnation.  This can be quantified by examining the position of the liner centroid at different floors.
\item Hall physics, more than electron inertia, reduces sensitivity to density floor and other aspects of vacuum modeling, including the use of a density buffer.
\item The reduced floor sensitivity provided by Hall is greater prior to the implosion than during the implosion, which is evidently associated with floor sensitivity of current coupling onto the liner, and consequent mass distribution.
\item Vacuum modeling sensitivity of the post-stagnation is reduced by modeling Hall physics, and using sufficiently low density floors, which is relevant to developing more predictive models of confinement times.
\item For most of the implosion, bounds-preserving limiting occurs in a small percentage of the domain which, as expected, is generally localized around the plasma-vacuum interface just outside the imploding liner.
\end{enumerate}

We have shown that in one dimension, vacuum modeling sensitivity is significantly reduced by (i) using extended-MHD, in particular Hall physics, and (ii) using lower density floors.  We now examine the extent to which these observations hold in two dimensions.

\subsection{2-D liner implosion in $r-z$ geometry}
\label{ssec:vac2drz}

We now examine a liner implosion in an axisymmetric $r-z$ cylindrical geometry, which is based on Sandia's Magnetized Liner Inertial Fusion (MagLIF) program.   In particular, the liner sits on top of an annular feed section with a central cathode post, with a conducting wall at the outer radial and upper axial boundaries, so that the system is driven with axial Poynting inflow at the base of the feeds.  In the presence of an in-plane $B$-field, in this case an applied $B_z$, and in particular when the axial direction is modeled with Poynting inflow at an axial boundary, the Hall term has been found to generate in-plane $E$ and out-of-plane $B$-fields which, compared to other geometries (2-D $r-\theta$, 1-D radial, no axial Poynting inflow, no in-plane $B$-field, etc.), are much larger fractions of the driven $E$ and $B$-fields.  These Hall-generated $E$ and $B$-field components therefore exert a stronger influence not only on the behavior of low-density plasma but on the dynamics of the liner implosion itself.  See Ref.~\onlinecite{seyl18} for further discussion.

The material modeled throughout the domain is Beryllium, and Beryllium SESAME tables are used for EOS and conductivity.

This problem was examined in Ref.~\onlinecite{wool23} in the context of the Hall instability; the present work uses the setup described in Ref.~\onlinecite{wool23} for 20 MA peak current.  This test problem is also similar to the one discussed in Refs.~\onlinecite{seyl18, haml19}.  The present work elaborates on the vacuum modeling sensitivity discussion in Ref.~\onlinecite{haml19}, and adapts it to the use of tabular material models.

The cell size is 101 $\mu$m in $r$ and $z$.  At the annular base of the feeds, between the cathode at $r = 10$ mm and anode at $r = 13$ mm, Poynting inflow of $B_{\theta}$ and $E_r$ is driven with a current pulse that follows the sine-squared profile described in Sec. \ref{ssec:vac1d}, rising to a peak current of 20 MA in 100 ns.  A uniform axial $B$-field of 10 T is applied throughout the system, and is assumed to have completely penetrated the conductors, which is consistent with experiment.  On top of the cathode post is a Beryllium liner with parameters typical of MagLIF experiments, and similar to those in Refs.~\onlinecite{seyl18,haml19}.  The liner is 10 mm tall, with inner radius 2.325 mm and outer radius 2.79 mm.  A 1$\%$ density perturbation is applied on the liner surface, designed to model instabilities, but which does not significantly affect the discussion in this section.  Surrounding the liner is a coronal plasma layer of uniform initial density $10^{16}$ /cm$^3$ extending from $r = 2.79$ mm to $r = 3.255$ mm, and representing plasma ablating from the liner surface.  A 1-cell-thick, $10^{19}$ /cm$^3$  layer lines the cathode and anode in the feeds, which represents plasma that ${\bf E}\times {\bf B}$ drifts inward from the feeds in a device powerful enough to drive 20 MA of current, such as the Z accelerator.  A vacuum temperature of 116 K is used, with the electrodes and liner initialized at 300 K.  A vacuum temperature less than 300 K was found in certain cases to address stability issues likely associated with the vapor dome of the EOS table.  The salient conclusions of this section are unaffected by initializing at 116 K instead of 300 K.

In order to prevent Hall velocities from exceeding the numerical speed of light, limiting is imposed on the current density.  That is, for some Hall velocity $v_H$ (the value used in this test was $2\times 10^6$ m/s), and for each component $s = x,y,z$ of current density $J$, the magnitude of $J_s$ is reset to $n_eev_H$ if it exceeds this value.  

If current-limiting is not applied to prevent superluminal Hall velocities, a common bifurcation in low-density behavior is associated with excessive plasma filling the vacuum and causing premature liner implosion.  This is associated with a numerical thermal runaway effect that begins in the first few ns with the formation of isolated highly conducting hot-spots in the 10s of keV in a small amount of low-density plasma.   Thermal runaway is somewhat more common in simulations with no Hall term, which is consistent with the explanation in Appendix \ref{app:XMHD} of the Hall term modeling more physically consistent behavior of the current at near-vacuum densities.  Reference~\onlinecite{seyl18} also discusses thermal runaway in the absence of Hall physics. 

In addition to current-limiting, temperature-limiting was also applied.  This can be thought of as an inverse of the positivity-preserving limiting discussed in Sec. \ref{ssec:limiting}, where in this case the adjustment of density, momentum, and energy slopes acts to prevent temperature from exceeding a certain threshold.  The limiting value of 5 keV is chosen so that plasma and vacuum temperatures are consistent with the average of the values reported in Ref.~\onlinecite{benn21}.  Apart from maintaining temperatures at or below this value, temperature-limiting has little effect on the dynamics, since most of the stabilization is provided by current-limiting.

With current-limiting and temperature-limiting in place, pressure floor resets were found to have very little impact on the results, with or without the Hall term being modeled.  The same is true of the numerical speed of light, for values $\ge 10^7$ m/s.  A value of $10^7$ m/s is used in the displayed results.

For this simulation, as with the 1-D simulation, the limiting $\theta$ values discussed in Sec. \ref{ssec:limiting} were displayed at various times.  It was found that this limiting occurs in a small percentage of the vacuum, along with internal energy limiting on the inner surface of the outer return can, which is consistent with the expectation that this limiting should be confined primarily to the plasma-vacuum interface.

We now discuss several factors that influence the vacuum sensitivity of the liner implosion results, namely (i) convergence with respect to density floor and (ii) sensitivity to the density buffer.

\subsubsection{Floor convergence with Hall and without Hall}
 
\begin{figure*}
\centering
(i) \includegraphics[scale=0.5]{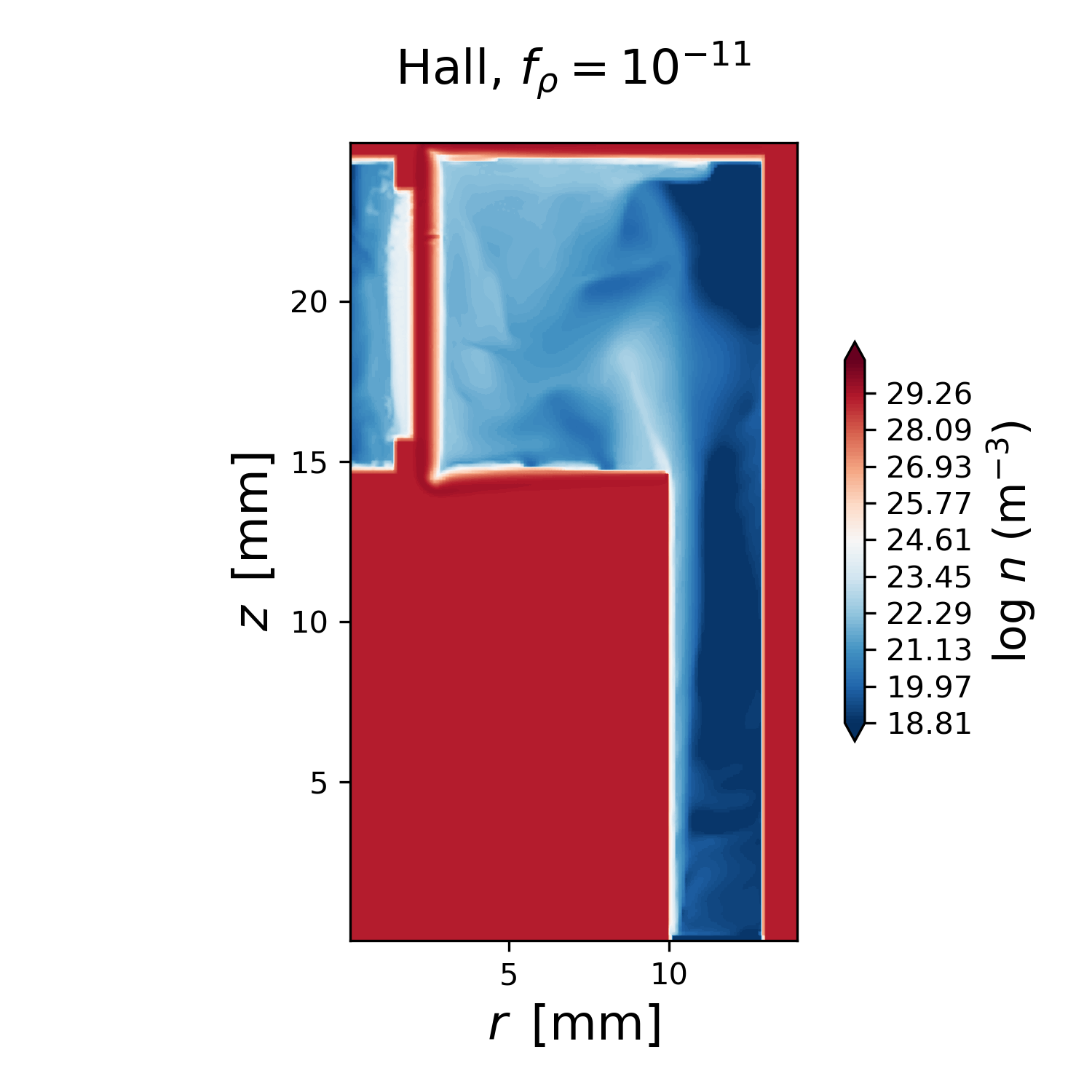}
(ii) \includegraphics[scale=0.5]{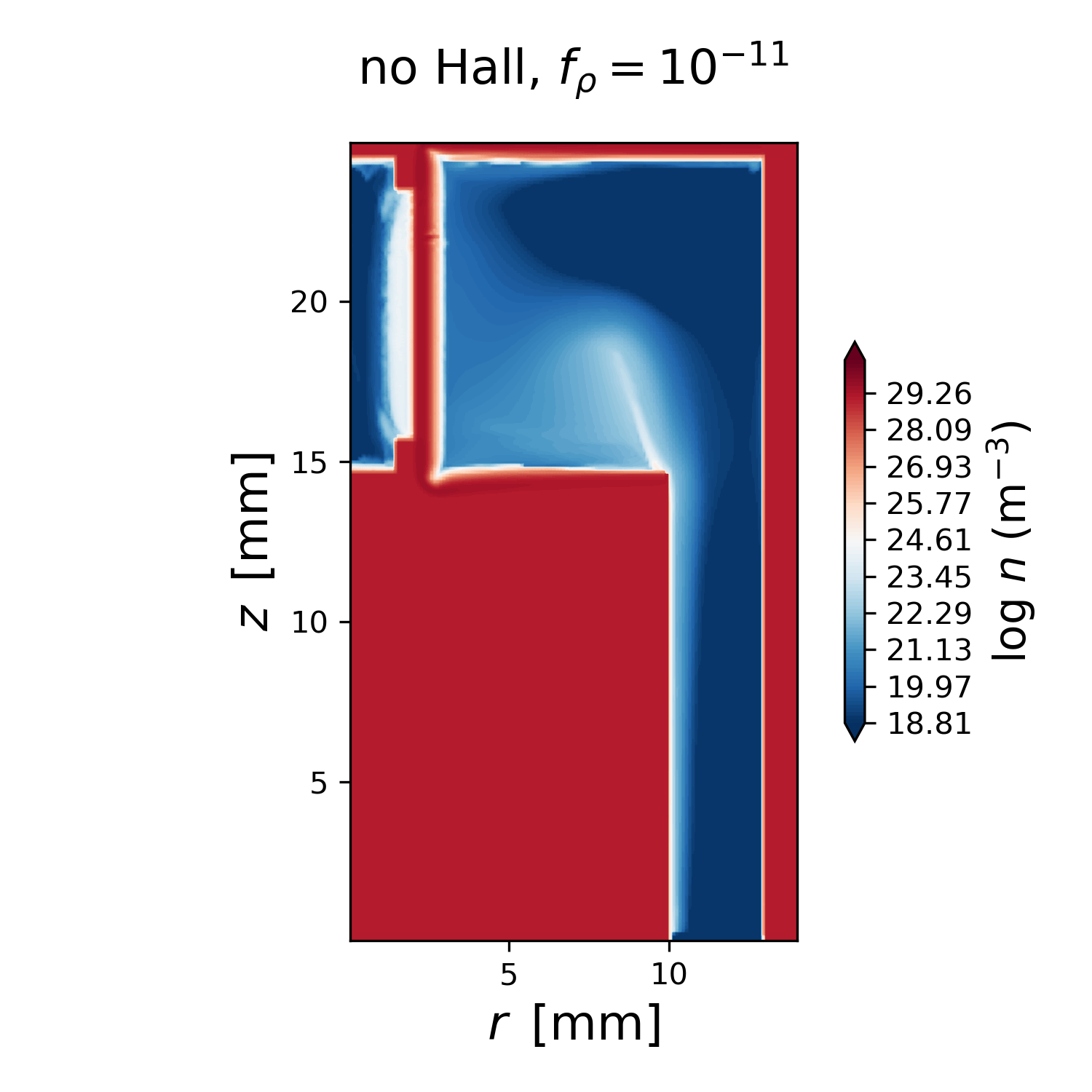}\\
(iii) \includegraphics[scale=0.5]{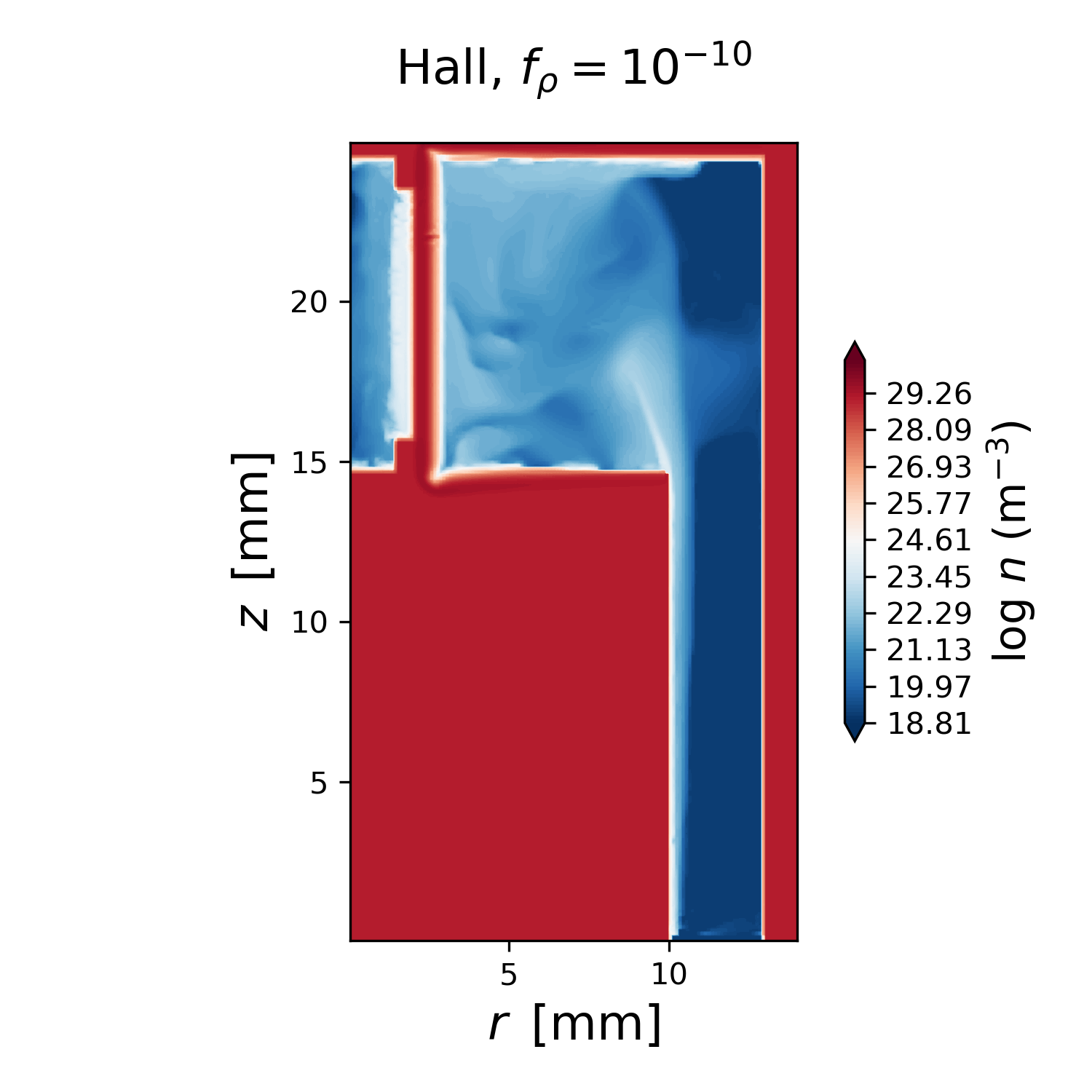}
(iv) \includegraphics[scale=0.5]{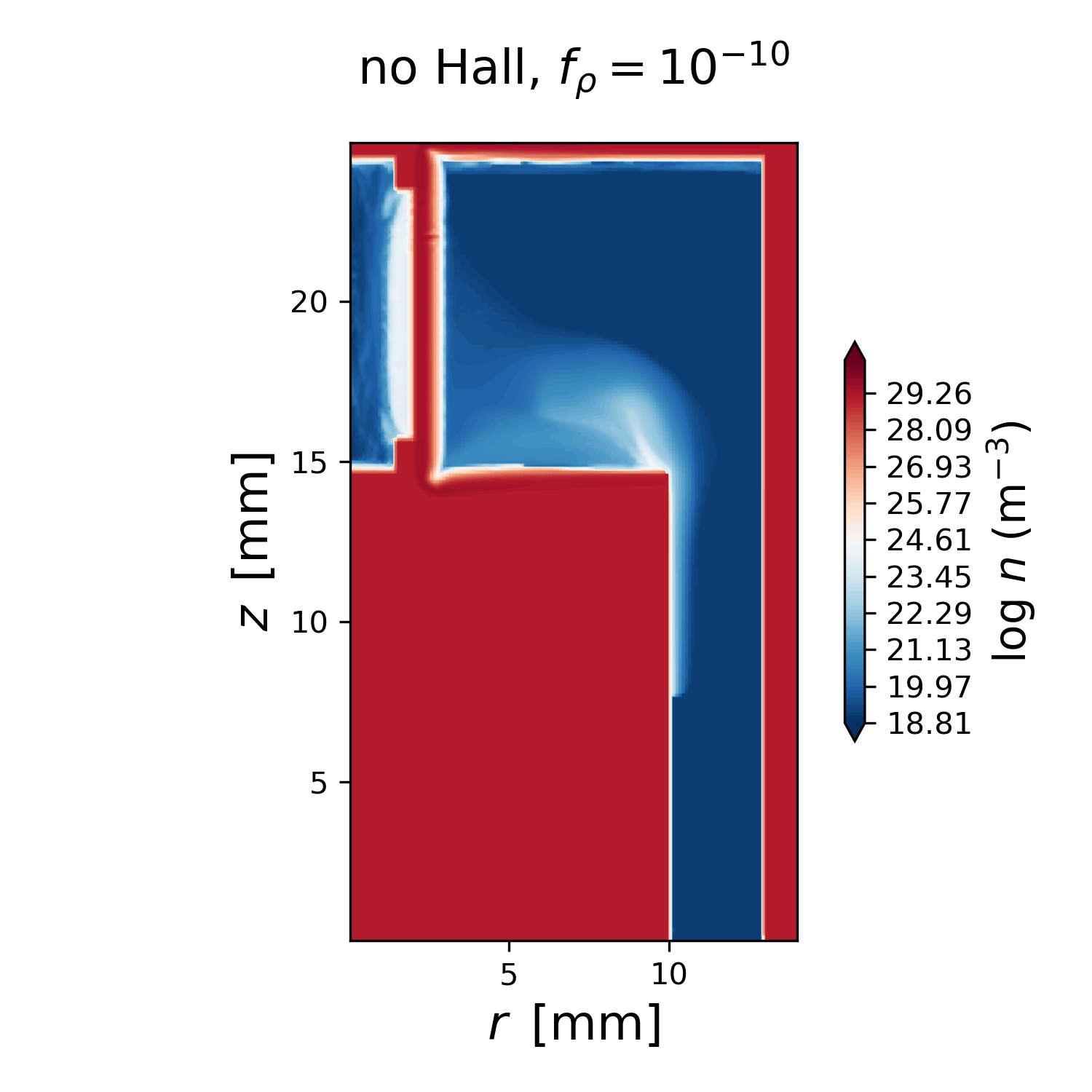}\\
(v) \includegraphics[scale=0.5]{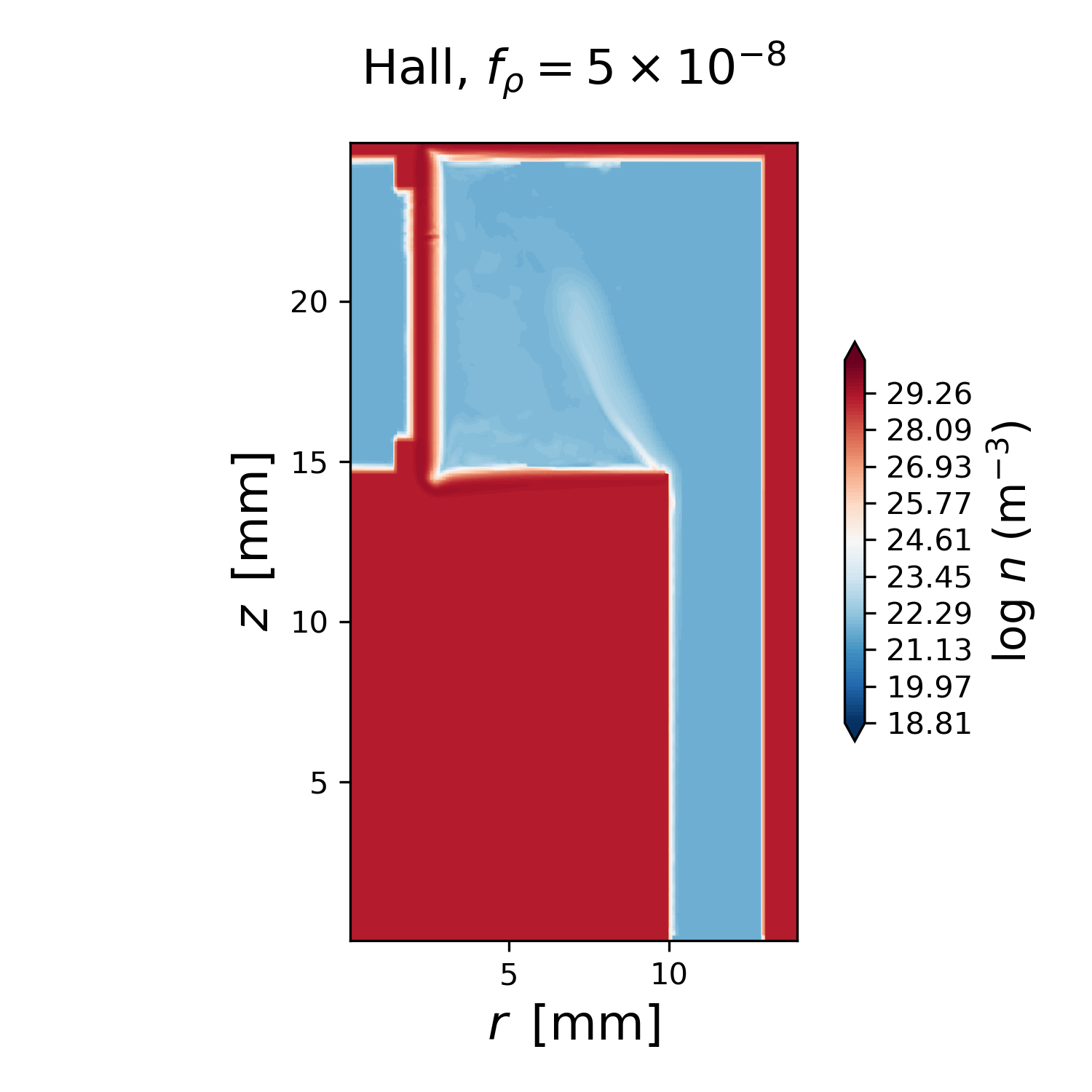}
(vi) \includegraphics[scale=0.5]{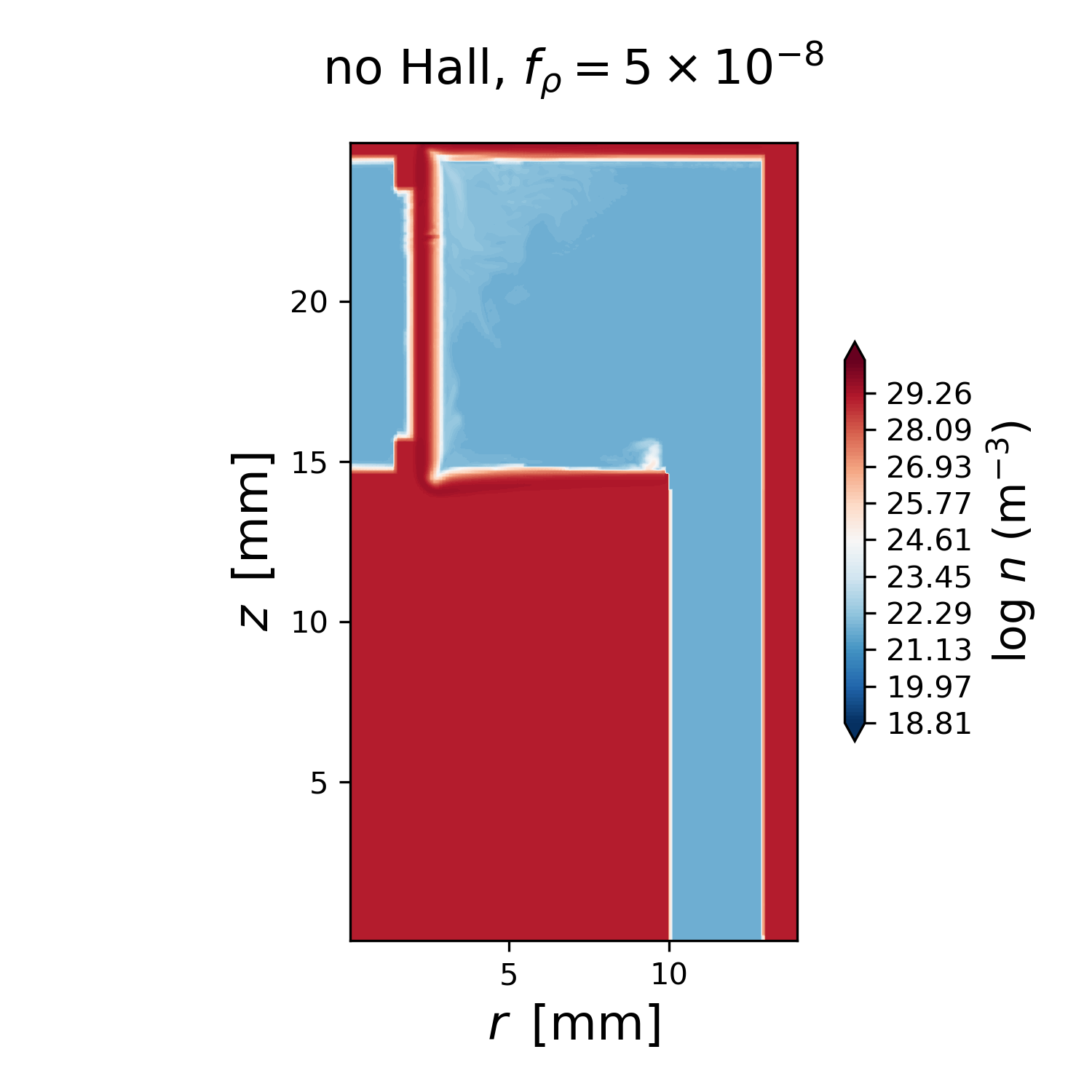}\\
\caption{Displays of base-10 log density at 80 ns at $\rho_{\rm floor} = f_{\rho}\rho_{\rm solid}$ with (i) Hall included, $f_{\rho} = 10^{-11}$, (ii) Hall omitted, $f_{\rho} = 10^{-11}$, (iii) Hall included, $f_{\rho} = 10^{-10}$, (iv) Hall omitted, $f_{\rho} = 10^{-10}$, (v) Hall included, $f_{\rho} = 5\times 10^{-8}$, (vi) Hall omitted, $f_{\rho} = 5\times 10^{-8}$.  $\rho_{\rm  multiplier} = 1.01$.  Hall physics reduces sensitivity, to density floor, of the dynamics of the low-density plasma, provided a sufficiently low density floor is used.}
\label{fig:rz_rho_hall_0hall_80}
\end{figure*}

\begin{figure*}
\centering
(i) \includegraphics[scale=0.45]{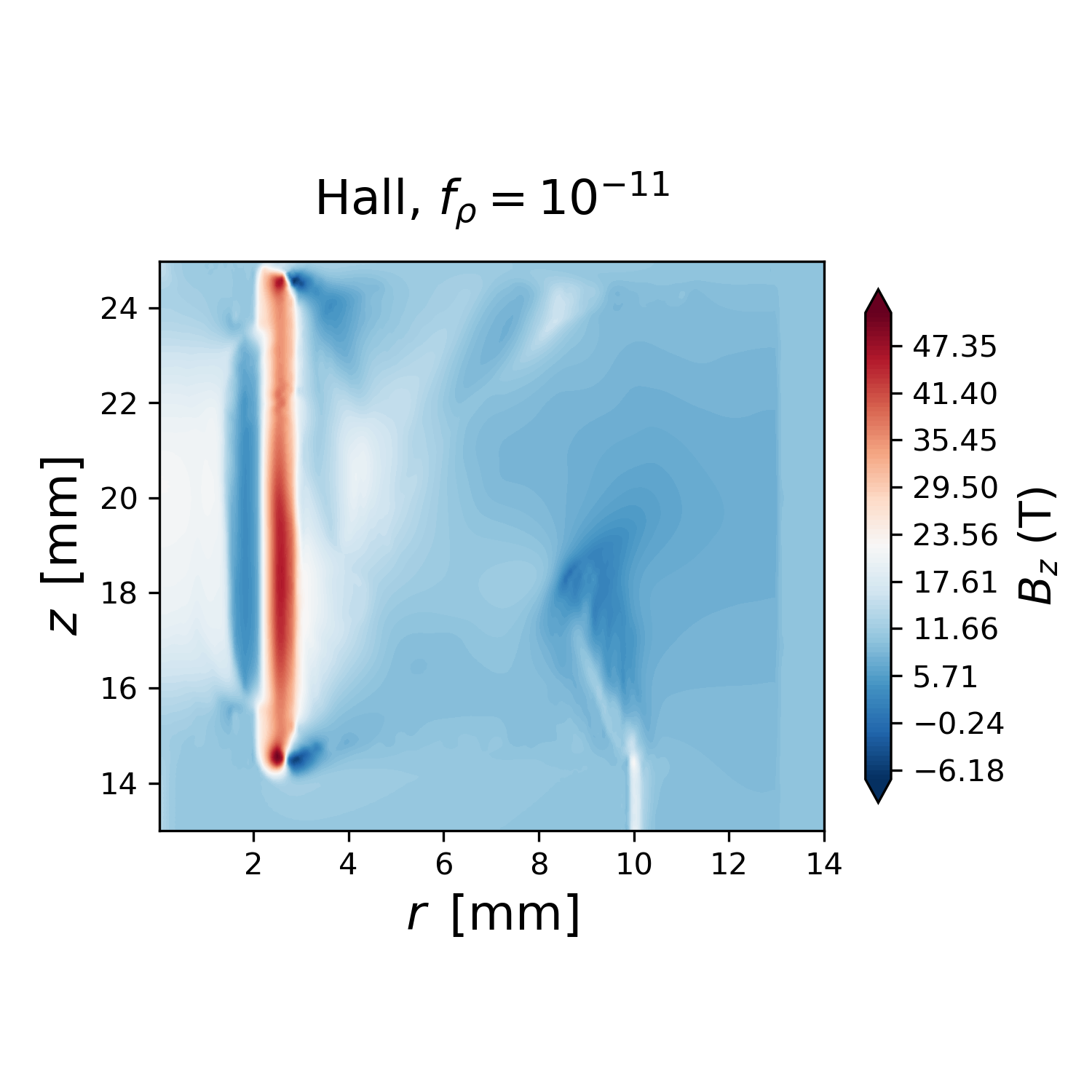}
%(iii) \includegraphics[scale=0.35]{figs2/Hall4_0hall_jmax_2e6__h9_Bz_80ns.png}
(ii) \includegraphics[scale=0.45]{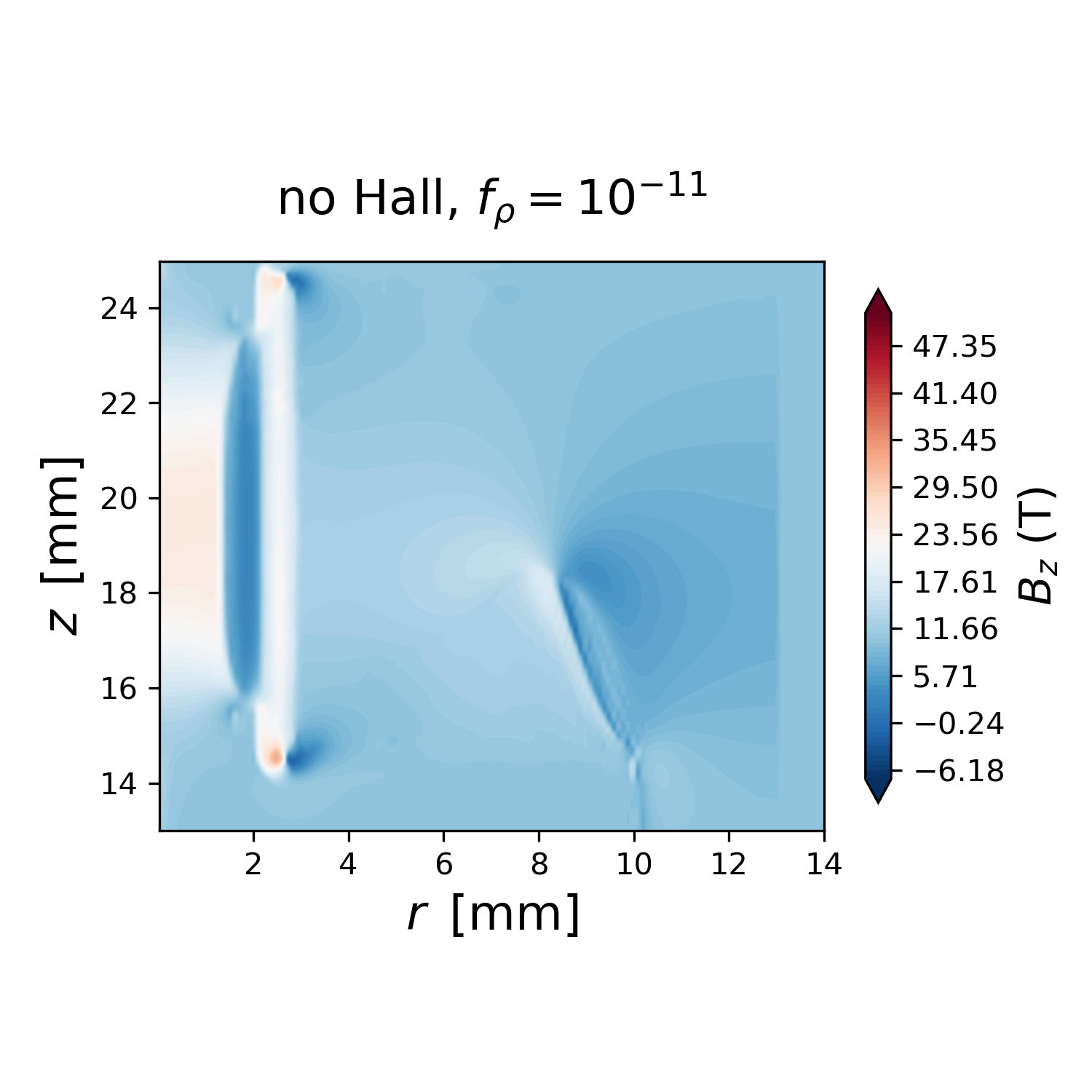}\\
(iii) \includegraphics[scale=0.45]{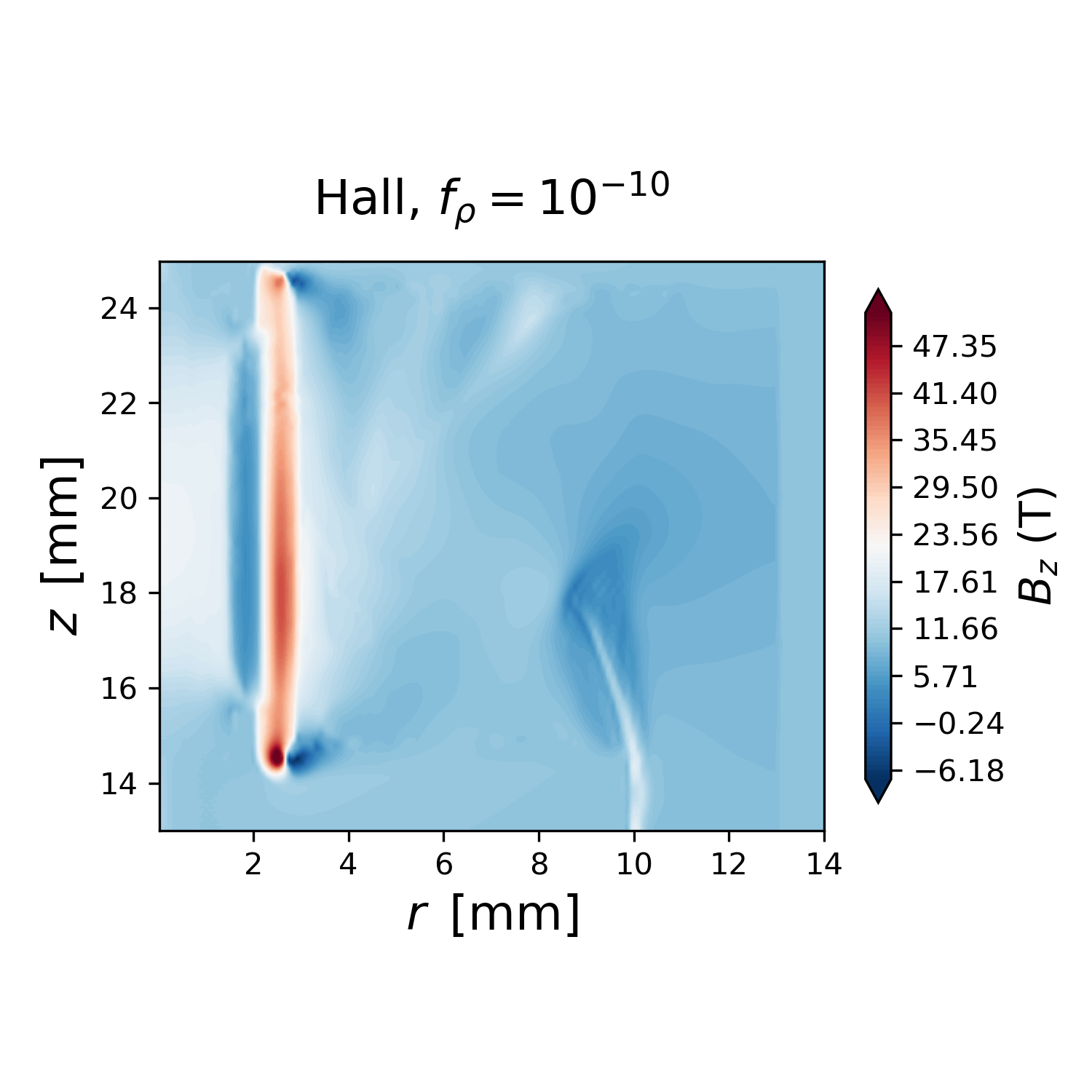}
(iv) \includegraphics[scale=0.45]{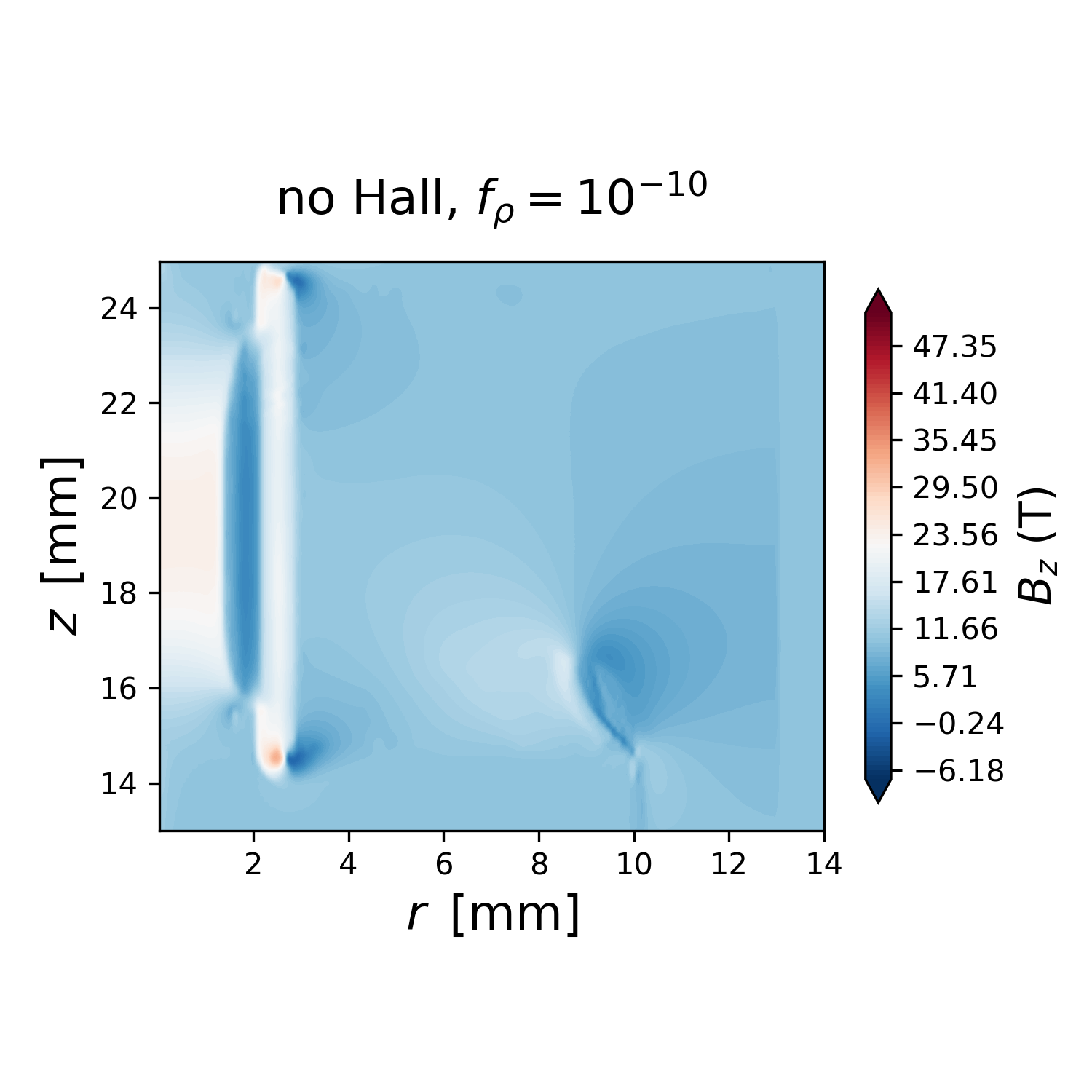}\\
%(vi) \includegraphics[scale=0.35]{figs2/Hall4_0hall_jmax_2e6__0h9_Bz_80ns.png}
(v) \includegraphics[scale=0.45]{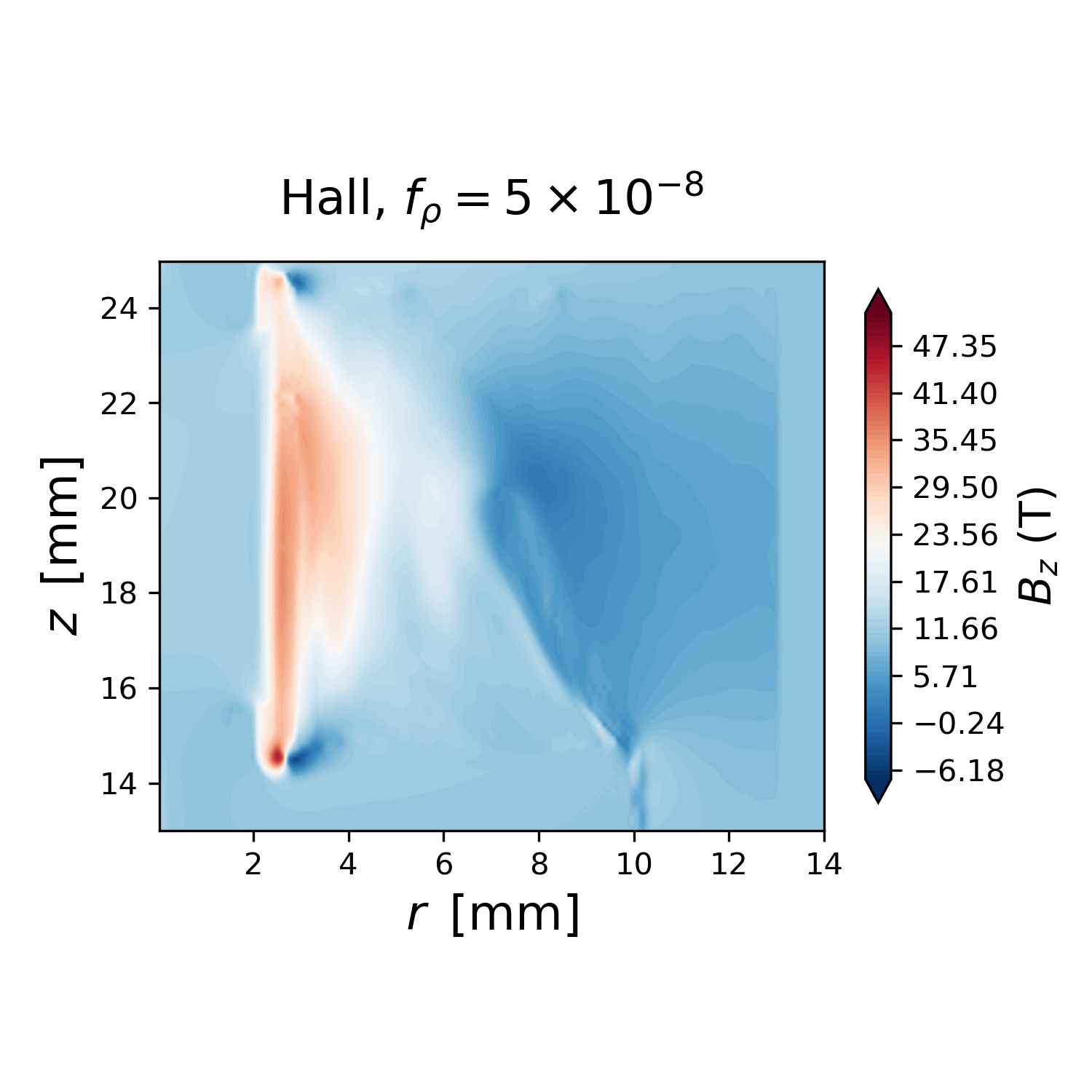}
(vi) \includegraphics[scale=0.45]{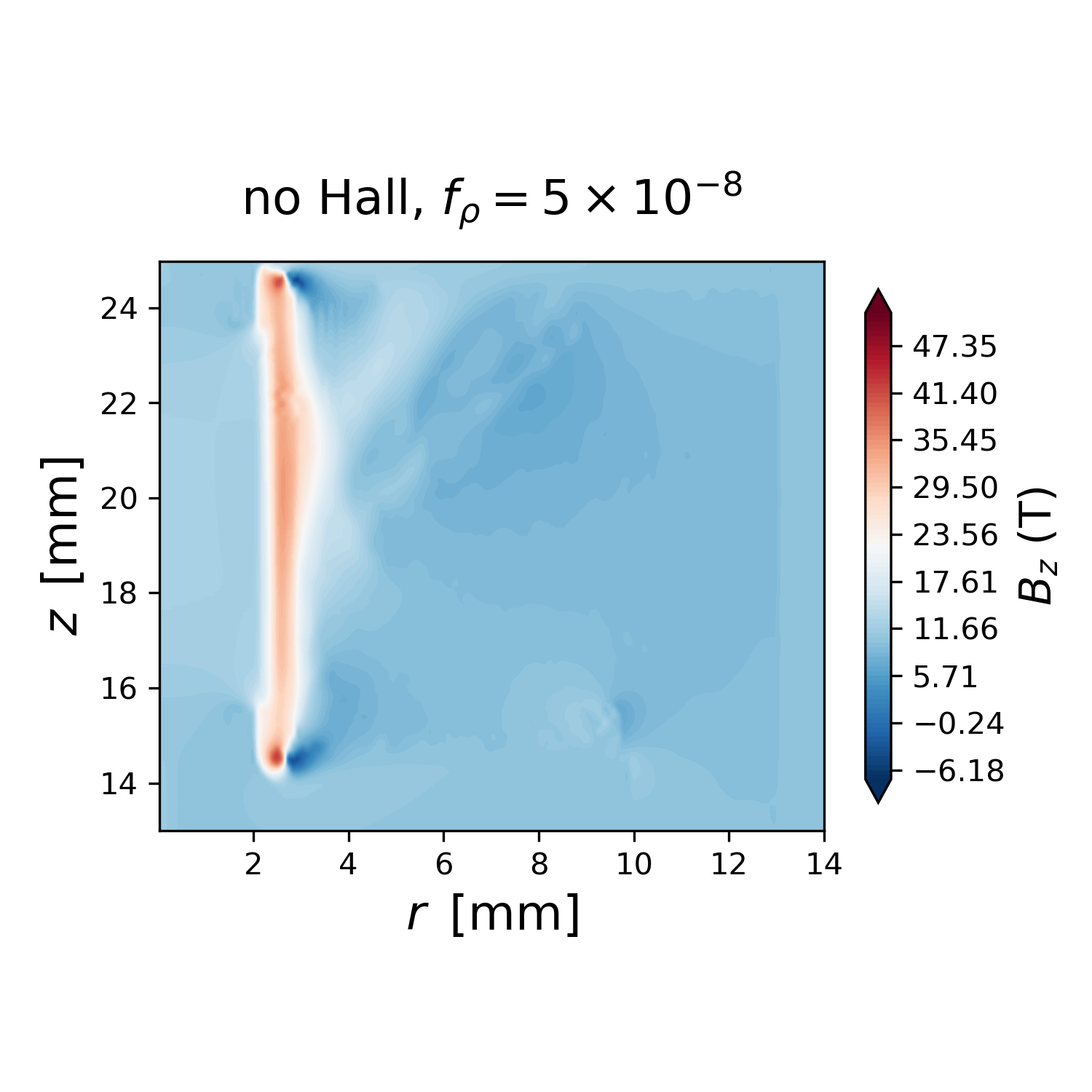}\\
\caption{Displays of axial $B$-field at 80 ns at $\rho_{\rm floor} = f_{\rho}\rho_{\rm solid}$ with (i) Hall included, $f_{\rho} = 10^{-11}$, (ii) Hall omitted, $f_{\rho} = 10^{-11}$, (iii) Hall included, $f_{\rho} = 10^{-10}$, (iv) Hall omitted, $f_{\rho} = 10^{-10}$, (v) Hall included, $f_{\rho} = 5\times 10^{-8}$, (vi) Hall omitted, $f_{\rho} = 5\times 10^{-8}$.  $\rho_{\rm  multiplier} = 1.01$.  Conclusions from Fig. \ref{fig:rz_rho_hall_0hall_80} also apply to axial flux compression against the liner.}
\label{fig:rz_bz_hall_0hall_80}
\end{figure*}

We begin by examining convergence with respect to density floor, and how this convergence is affected by Hall physics.  Figures \ref{fig:rz_rho_hall_0hall_80} display density at 80 ns, comparing 6 cases: with and without Hall modeled at floors of $10^{-11}\rho_{\rm solid}$, $10^{-10}\rho_{\rm solid}$ and $5\times 10^{-8}\rho_{\rm solid}$.  Figures \ref{fig:rz_bz_hall_0hall_80} compare the axial $B$-field for these same 6 cases.   From comparing the $10^{-10}\rho_{\rm solid}$-floor cases to the $5\times 10^{-8}\rho_{\rm solid}$- floor case, the use of a lower density floor allows modeling of vacuum plasma not captured at the higher floor, even though the density of this vacuum plasma is several orders of magnitude higher than the $5\times 10^{-8}\rho_{\rm solid}$ floor value.  

The $B_z$ displays at a $10^{-10}\rho_{\rm solid}$ floor show that this vacuum plasma carries substantial amounts of current, in particular azimuthal current, that compresses the axial $B$-field against the liner, resulting in considerable amplification of $B_z$ compared to the corresponding $B_z$ display for the $5\times10^{-8}$ floor.  The modeling of vacuum plasma, and consequent axial flux compression, requires the use of density floors which are much less than those typically used in MHD codes.  A similar conclusion was reached in Ref.~\onlinecite{seyl18} in which an analytic EOS was used; the present study shows that the same is true for the use of a tabular EOS.  Also, the vacuum plasma density is $\sim 10^{17}$ cm$^{-3}$, which is consistent with the values of $10^{15}$ to $10^{17}$ cm$^{-3}$ reported in Bennett et al Ref.~\onlinecite{benn21}.

At sufficiently low density floors ($< 5\times 10^8$ of solid density), Figs. \ref{fig:rz_rho_hall_0hall_80} and \ref{fig:rz_bz_hall_0hall_80} show, by comparing corresponding figures with versus without Hall included, that Hall physics noticeably increases the amount of low-density plasma entering the simulation volume and being swept against the liner, which increases axial flux compression against the liner.  Recall that in this geometry, the presence of an in-plane $B$-field enables a Hall dynamo mechanism that proceeds as follows: $B_z$ and $J_r$ produce an $E_{\theta}$ that models electron $E\times B$ drift, which generates a $J_{\theta}$ from Ampere's law \eqref{ampere}, which in turn amplifies the existing $B_z$, again from Ampere's law.  This mechanism tends toward the generation of a force-free current in the low-density plasma, and is also described in Refs. such as~\onlinecite{seyl18} and~\onlinecite{wool23}.

We therefore have two minimum (though not sufficient) requirements in order to more accurately model the interaction of low-density plasma with the $B$-field and resulting current coupling onto the liner. Namely, (i) a sufficiently low density floor must be used, and (ii) Hall physics must be modeled.  A fully correct model requires additional physics, as discussed in the Conclusion.   

%\begin{widetext}
%\end{widetext}

\begin{figure}[!h]
(i) \includegraphics[scale=0.5]{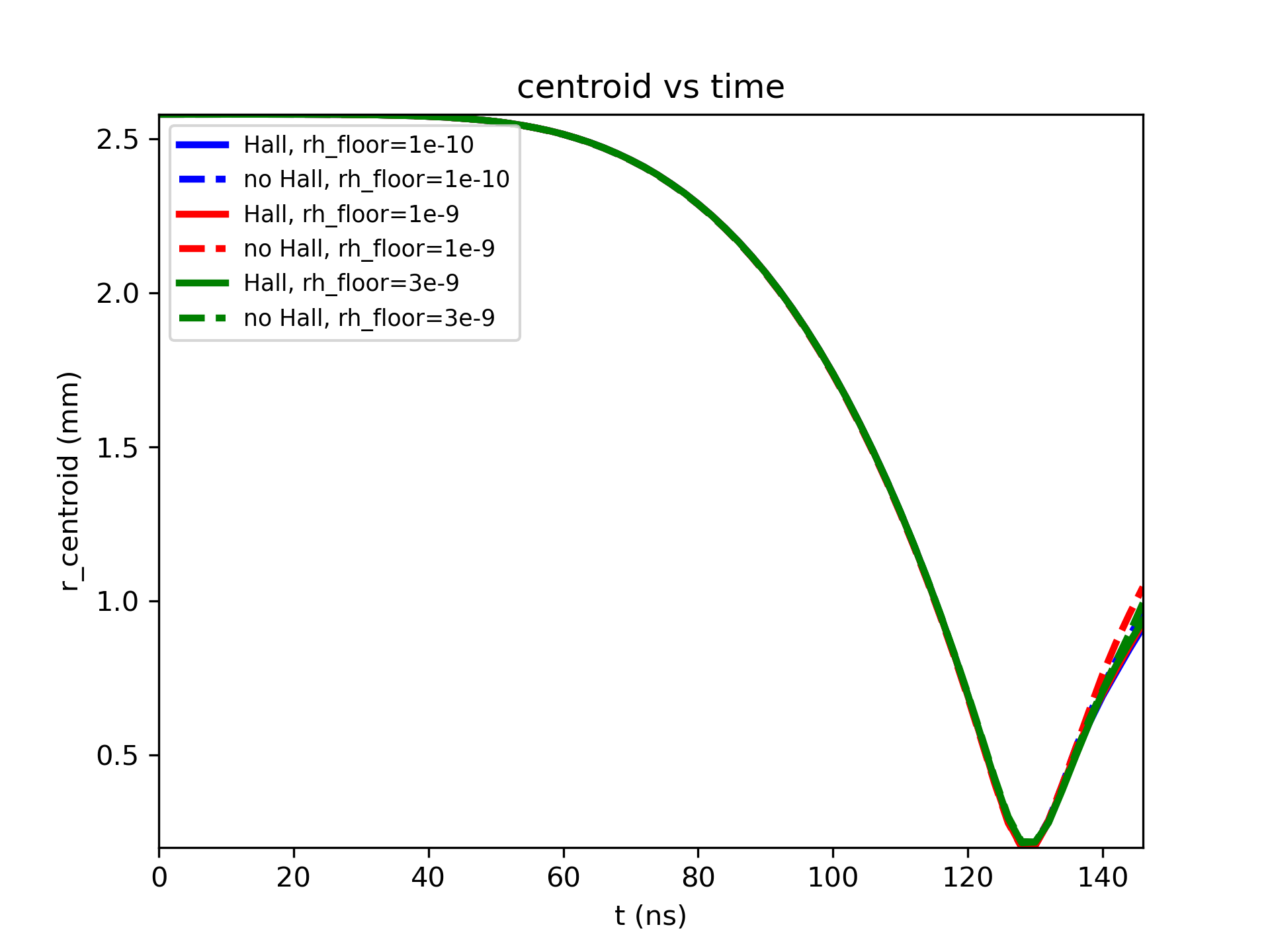}\\
(ii) \includegraphics[scale=0.5]{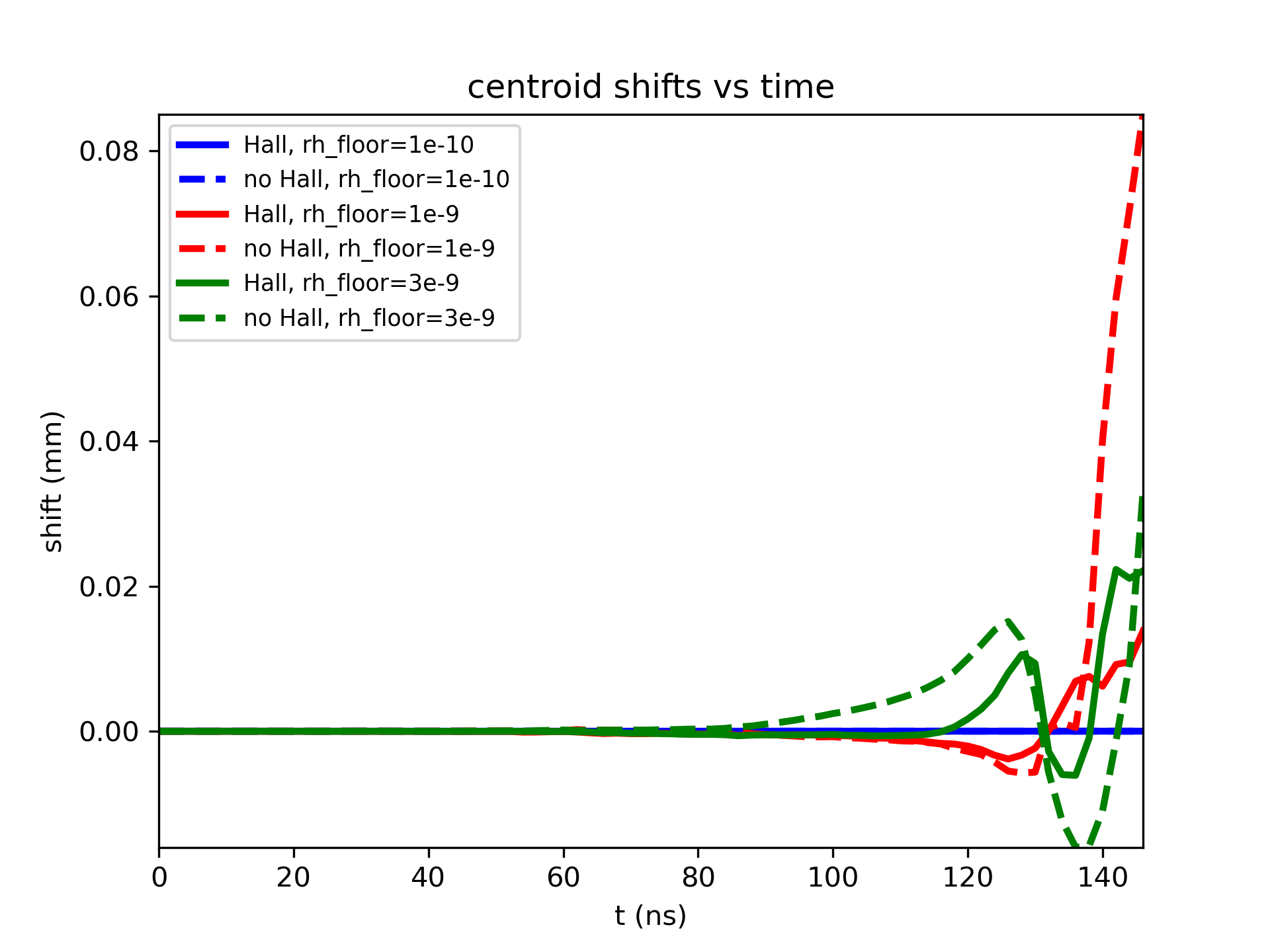}\\
(iii) \includegraphics[scale=0.49]{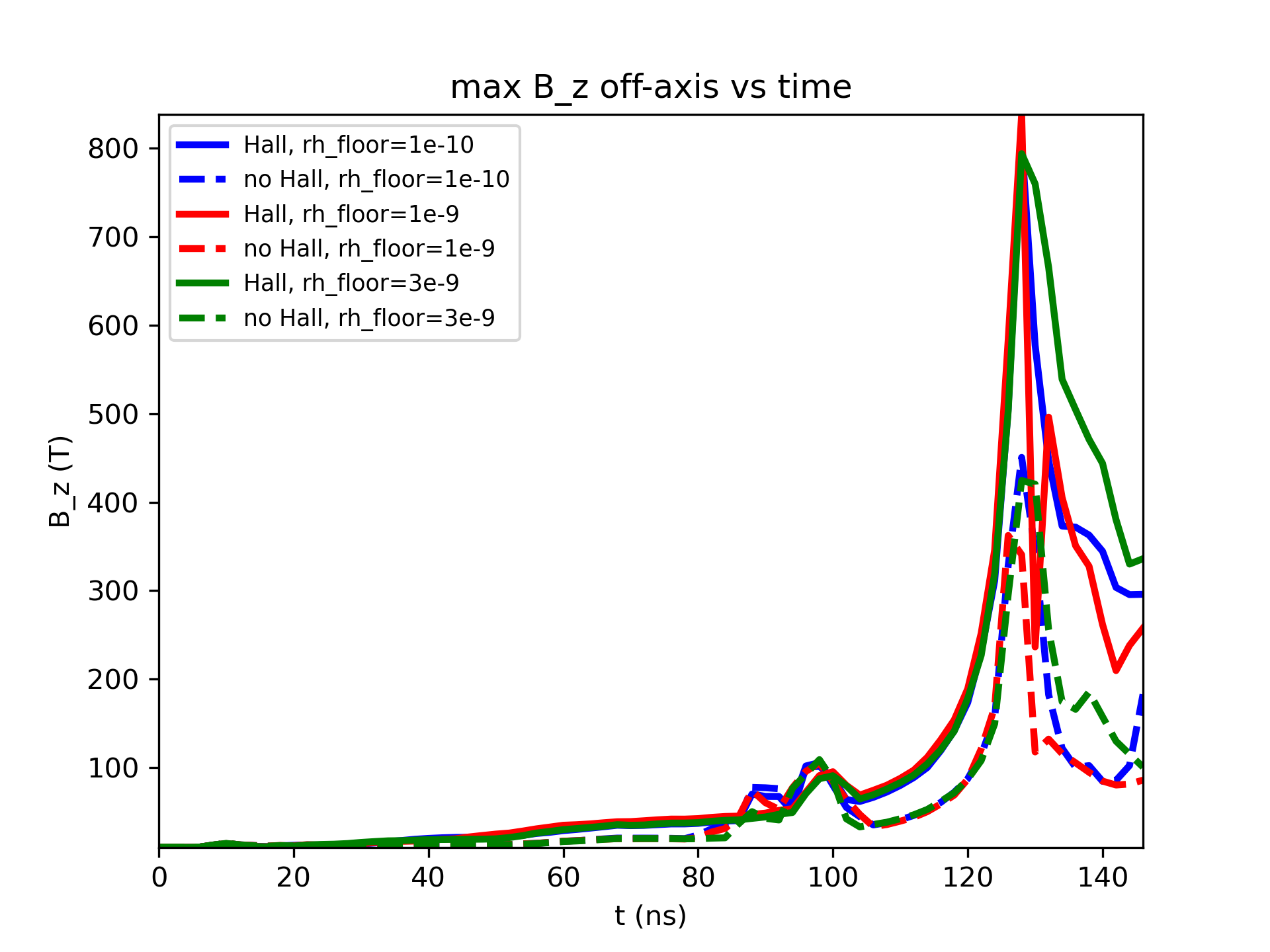}
\caption{Time-dependence of (i) centroid of liner, (ii) shift in liner centroid relative to the corresponding $10^{-10}\rho_{\rm solid}$ floor case, and (iii) maximum axial $B$-field just outside liner, each comparing Hall (solid curves) and no-Hall (dashed curves) cases at $3\times 10^{-9}\rho_{\rm solid}$, $10^{-9}\rho_{\rm solid}$, and $10^{-10}\rho_{\rm solid}$ density floors. Vacuum temperature is 116 K, and $\rho_{\rm  multiplier} = 1.01$.  Hall physics provides faster floor convergence of centroid position.}
\label{fig:rz_hall_0hall_cent_Bz}
\end{figure}

Hall physics more accurately captures current coupling onto the liner.  We now examine this impact more quantitatively, including the effect on bulk implosion dynamics.  Figures \ref{fig:rz_hall_0hall_cent_Bz} show the time-dependence of (i) the centroid of the liner, (ii) the shift in the liner centroid relative to the corresponding $10^{-10}\rho_{\rm solid}$ floor case, and (iii) the maximum axial $B$-field just outside the imploding liner, where each graph compares Hall and no-Hall cases at $3\times 10^{-9}\rho_{\rm solid}$, $10^{-9}\rho_{\rm solid}$, and $10^{-10}\rho_{\rm solid}$ density floors.  Each of these quantities are computed along a radial lineout through the middle of the liner, using Eq. \eqref{centroid} with $r_{\rm max} = 5$ mm.  From comparing solid to dashed curves in Fig. \ref{fig:rz_hall_0hall_cent_Bz} (iii), the Hall term further compresses and amplifies $B_z$ by a factor of about 2 to 5, depending on simulation time.   

%Note the close correlation between the amount of axial flux compression, i.e. maximum $B_z$, and the shift in the centroid position, which shifts downward, corresponding to an earlier implosion, as less axial flux is compressed against the liner.  

Note the more rapid floor convergence during the late implosion, from green to red to blue, of the solid curves (with Hall modeled) compared to the dashed curves (no Hall term modeled), indicating that the Hall term is accelerating convergence with respect to the density floor.  

Note also that during the implosion phase, the centroid shifts are about 2 orders of magnitude larger than the centroid shifts plotted in Sec. \ref{ssec:1Dfloor} for the 1-D liner implosion (before stagnation), indicating a higher vacuum modeling sensitivity of the bulk implosion dynamics in this $r-z$ axisymmetric problem compared to the 1-D geometry. 

The accelerated density-floor convergence provided by Hall can be understood in terms of conductivity; see the discussion in Sec. \ref{ssec:1Dfloor}.  The floor convergence study was repeated with the vacuum temperature initialized at 300 K instead of 116 K, and the above trends were again observed.

These convergence properties are not the result of a particular material model, e.g. tabular, or numerical artifacts.  In particular, Sec. III B 2 of Ref.~\onlinecite{haml19} examines a similar liner implosion in 2-D $r-z$ axisymmetric geometry, but using analytic EOS and conductivity models that differ from the tabular models used in the present study.  However, Ref.~\onlinecite{haml19} finds that, as with the present study, (i) the results converge (become insensitive to density floor) at sufficiently low floors, and (ii) Hall physics facilitates this convergence.  

\begin{figure} %[!h]
(i) \includegraphics[scale=0.5]{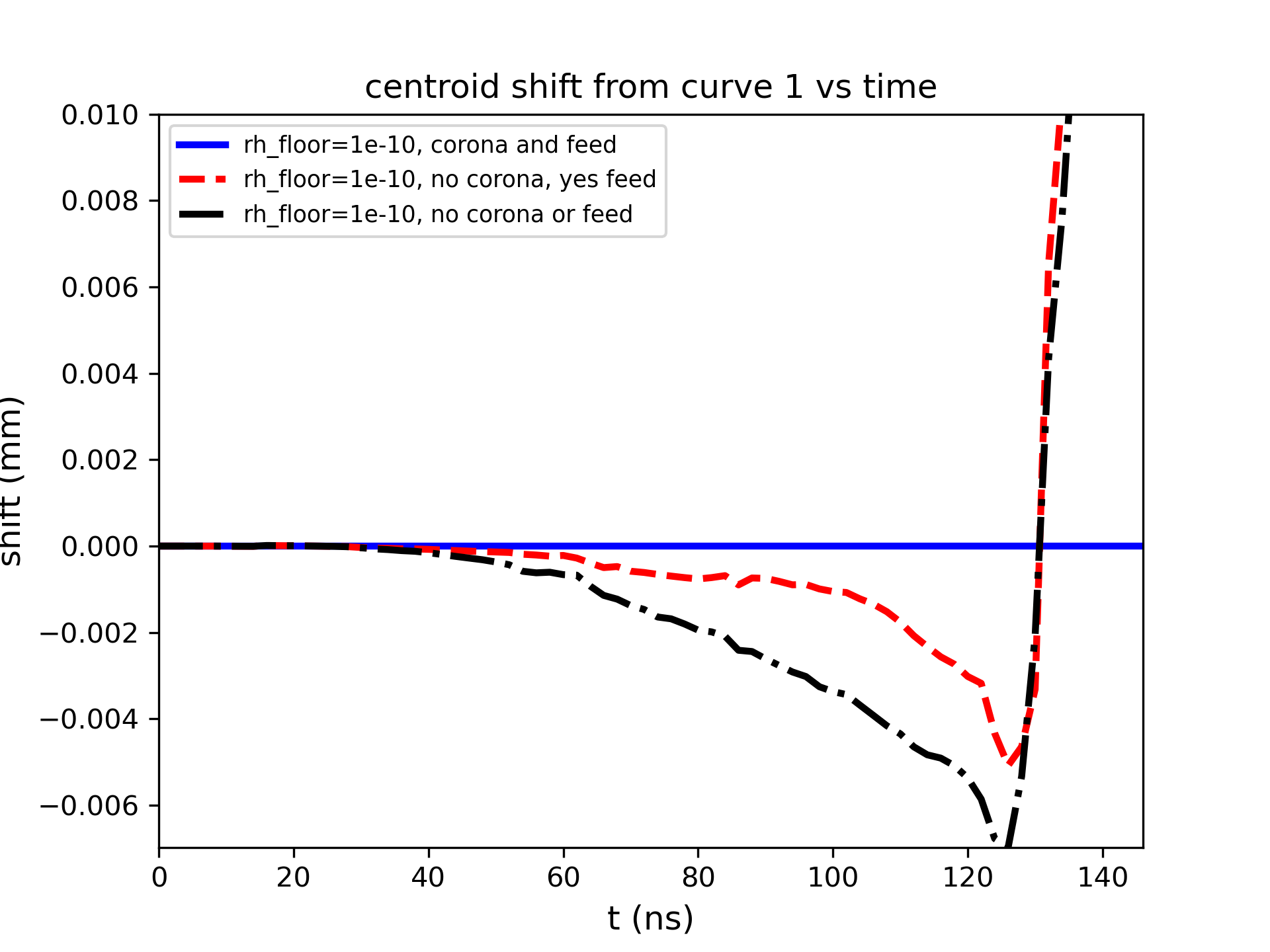}\\
(ii) \includegraphics[scale=0.5]{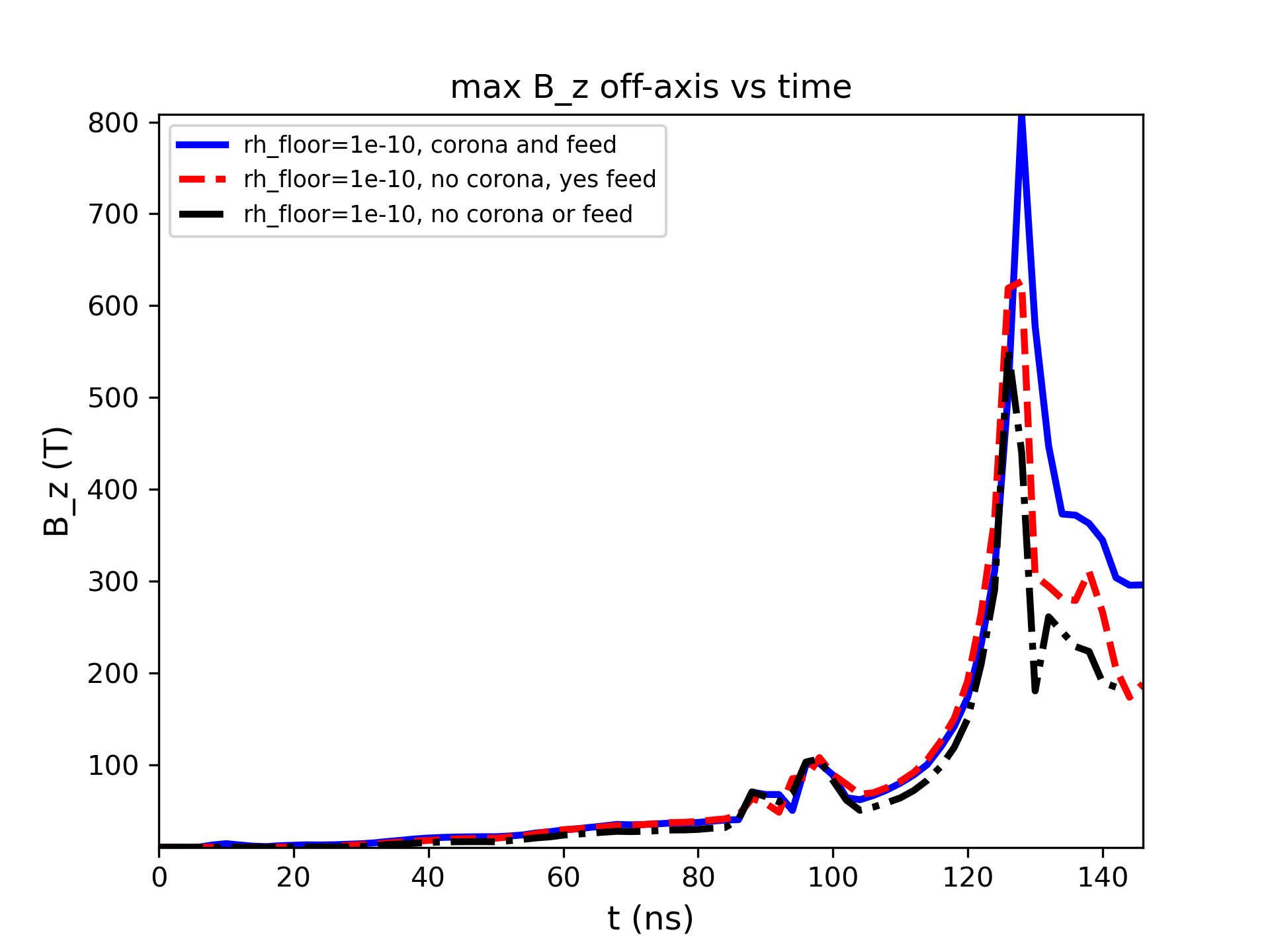}
\caption{Influence of coronal and feed plasma: Time-dependence of (i) shift in liner centroid relative to first case, and (ii) maximum axial $B$-field just outside liner, each comparing initializations of (i) coronal and feed plasma, (ii) feed plasma, but no coronal plasma, and (iii) neither coronal plasma nor feed plasma.  Hall is modeled at $10^{-10}\rho_{\rm solid}$ floor and $\rho_{\rm  multiplier} = 1.01$.  During the implosion, the coronal plasma and feed plasma have comparable influence on axial flux compression and liner implosion.}
\label{fig:rz_hall_0cor_0feed_cent_Bz}
\end{figure}

Figures \ref{fig:rz_hall_0cor_0feed_cent_Bz} show, for the case with Hall physics and $10^{-10}\rho_{\rm solid}$ floor, the influence of the coronal and feed plasma on the axial flux compression and consequent implosion delay.  Axial flux compression is influenced comparably by both the $10^{16}$ cm$^{-3}$, 5-cell-thick layer of coronal plasma just outside the liner and the $10^{19}$ cm$^{-3}$, 1-cell-thick layers of feed plasma lining the anode and cathode.  That is, the addition of feed plasma (from black to red curve) and the addition of coronal plasma (from red to blue curve) both result in comparable shifts in maximum $B_z$ and comparable delays of implosion timing (upward shift of centroid position due to additional axial flux amplification).  Note that even in the absence of initial coronal and feed plasma, a small amount of plasma is still sourced from the liner and electrode surfaces in the liner region.

\subsubsection{Sensitivity to density buffer}
\label{ssec:rzbuffer}

\begin{figure} %[!h]
(i) \includegraphics[scale=0.45]{Hall4_0hall_jmax_2e6_h10_density_log_80ns.png}\\
(ii) \includegraphics[scale=0.45]{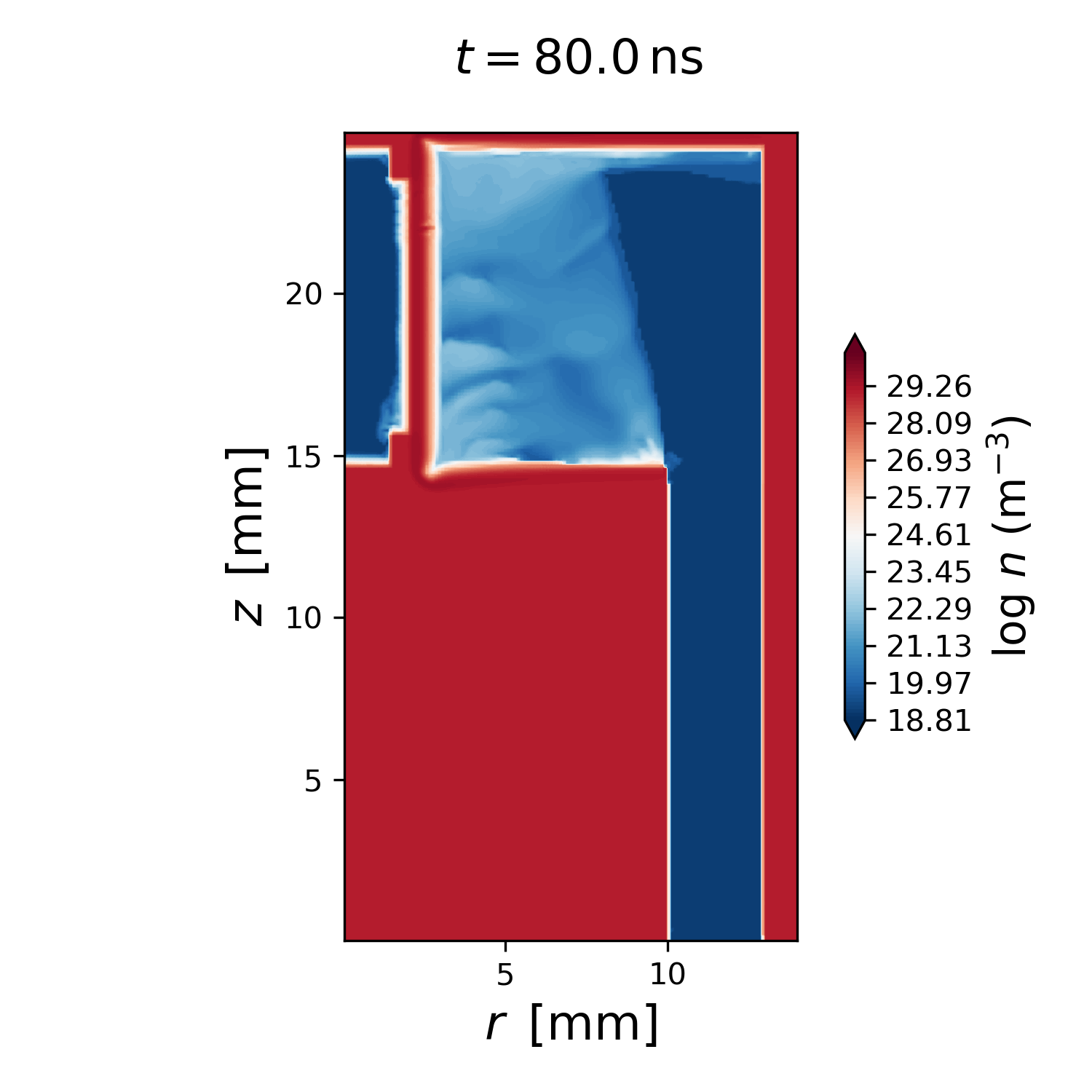}\\
(iii) \includegraphics[scale=0.45]{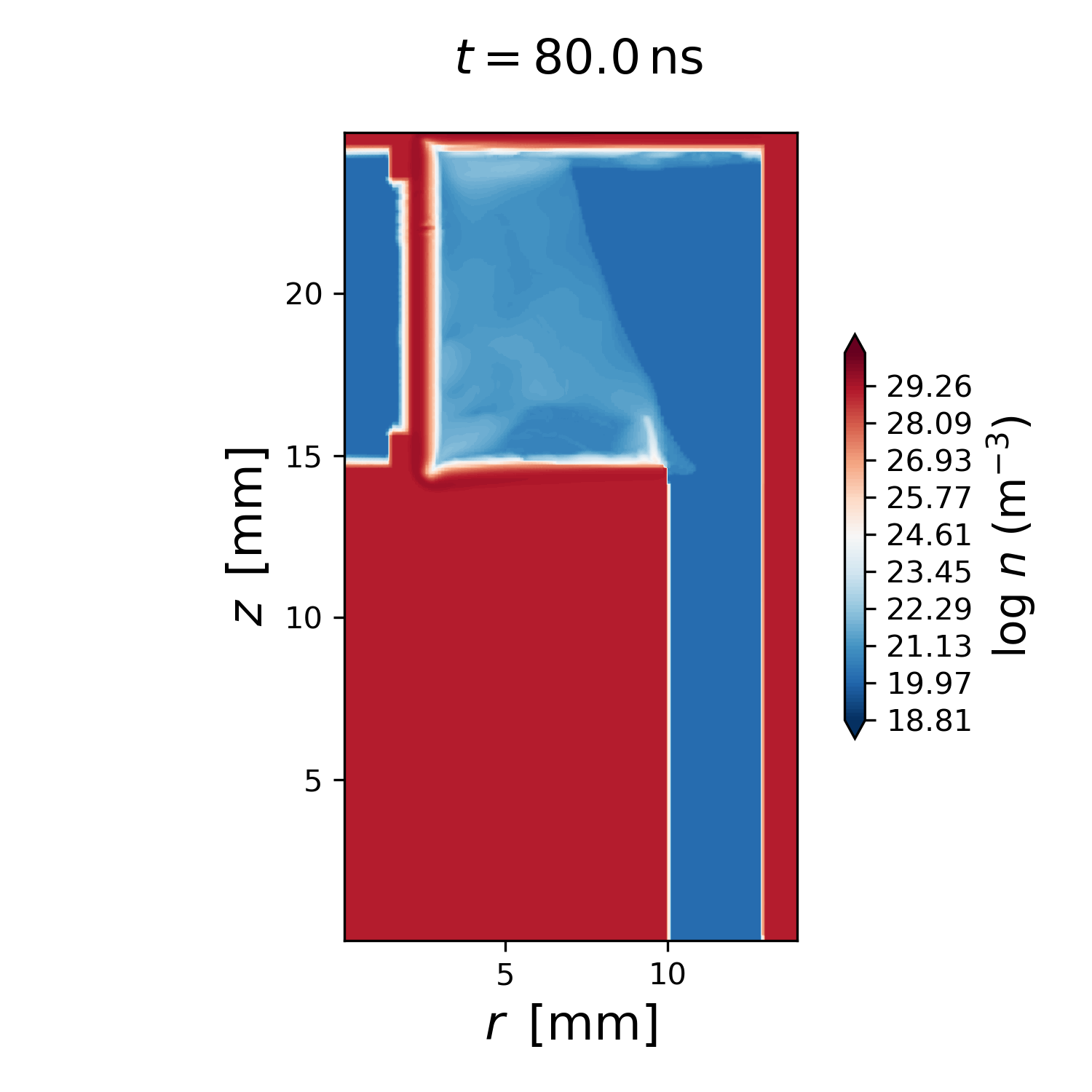}
\caption{Displays of base-10 log density at 80 ns comparing $(\rho_{\rm floor}/\rho_{\rm solid},\rho_{\rm  multiplier}) = $ (i) $(10^{-10}, 1.01)$, (ii) $(10^{-10}, 4)$, and (iii)  $(10^{-9}, 4)$.  With Hall term included.  A higher floor multiplier artificially reduces mass of vacuum plasma and therefore flux compression against the liner.}
\label{fig:rz_rho_hall_mult4_100}
\end{figure}

\begin{figure} %[!h]
(i) \includegraphics[scale=0.45]{Hall4_0hall_jmax_2e6_h10_Bz_80ns.png}\\
(ii) \includegraphics[scale=0.45]{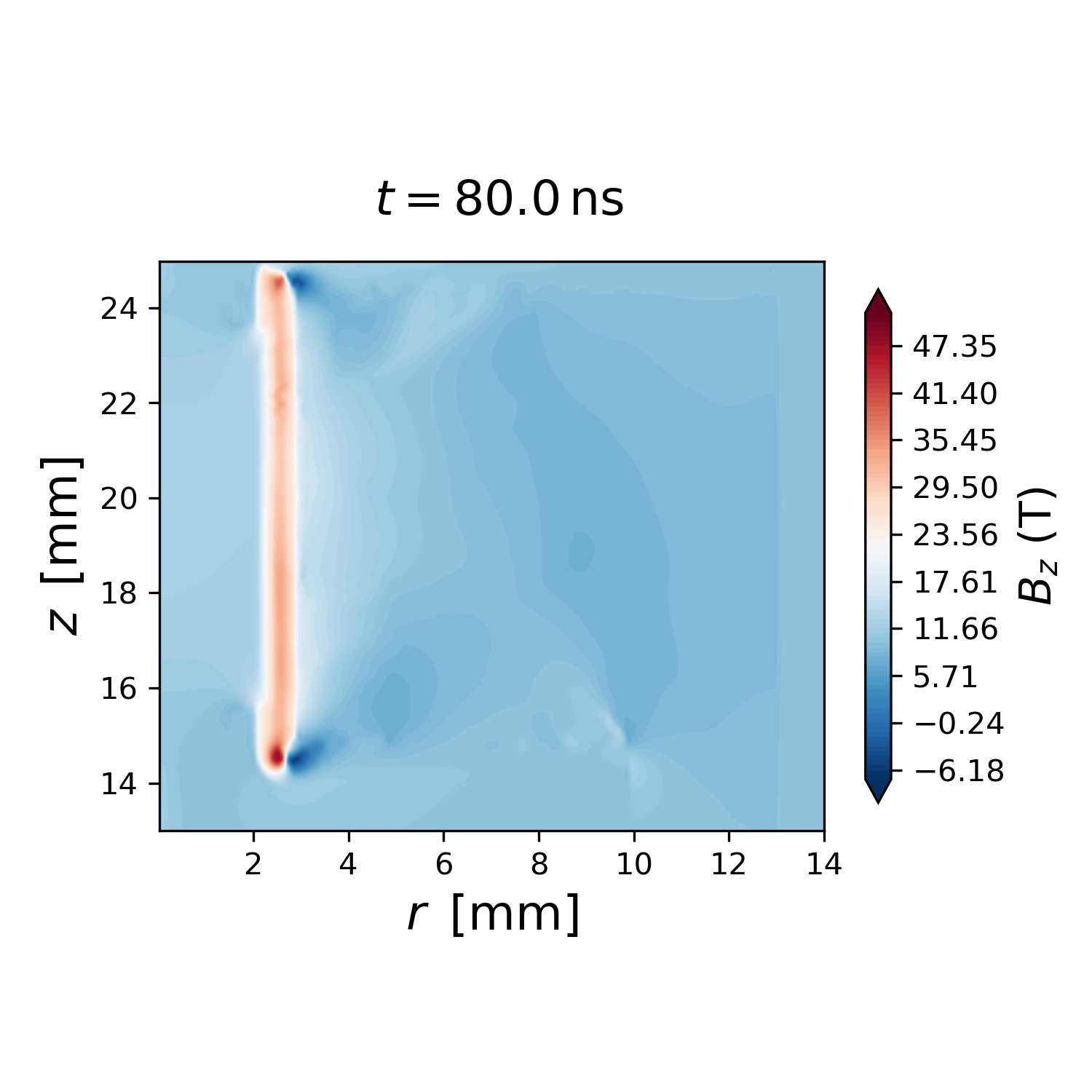}\\
(iii) \includegraphics[scale=0.45]{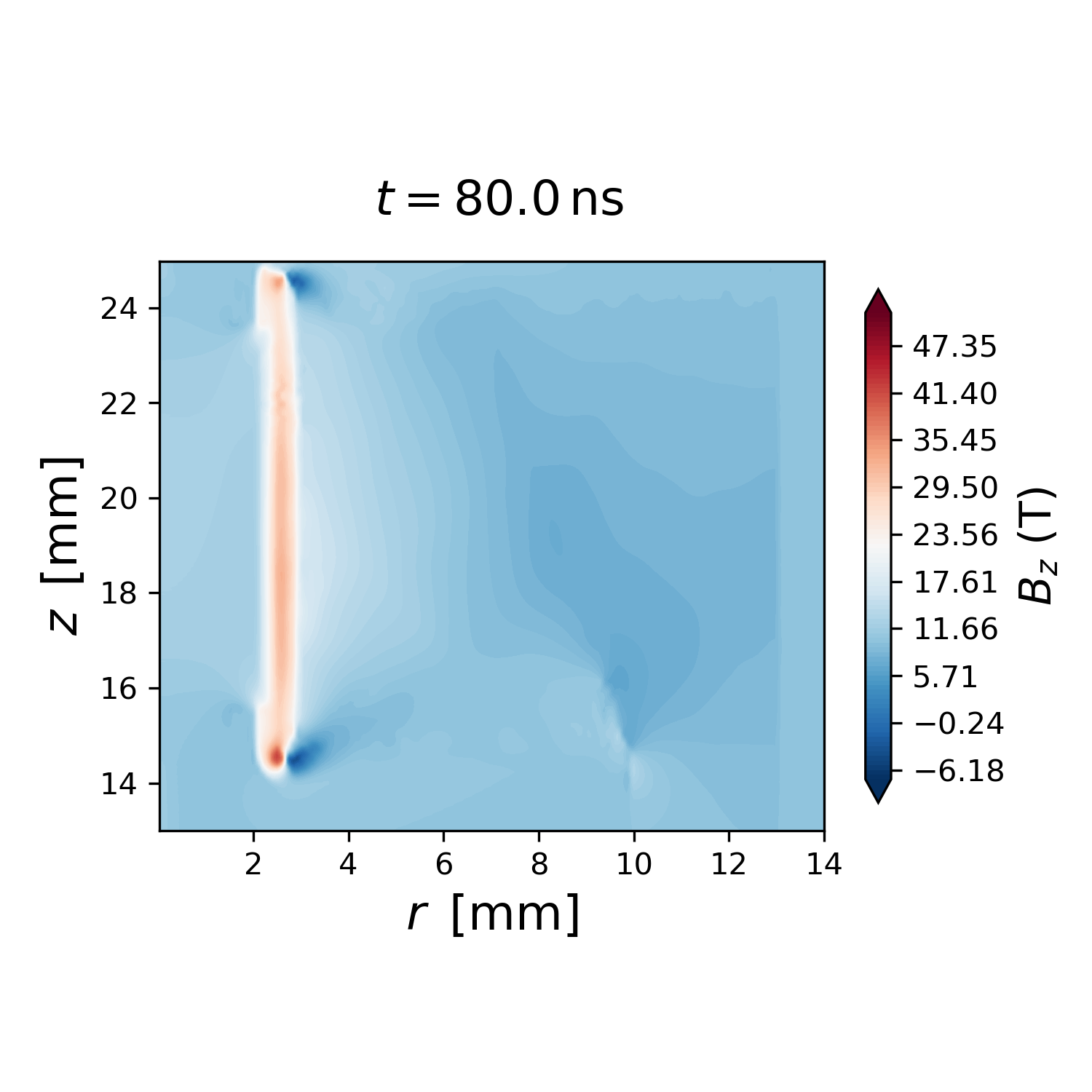}
\caption{Displays of axial $B$-field at 80 ns comparing $(\rho_{\rm floor}/\rho_{\rm solid},\rho_{\rm  multiplier}) = $ (i) $(10^{-10}, 1.01)$, (ii) $(10^{-10}, 4)$, and (iii)  $(10^{-9}, 4)$.  With Hall term included.  A higher floor multiplier artificially reduces mass of vacuum plasma and therefore flux compression against the liner.}
\label{fig:rz_Bz_hall_mult4_100}
\end{figure}

\begin{figure} %[!h]
(i) \includegraphics[scale=0.5]{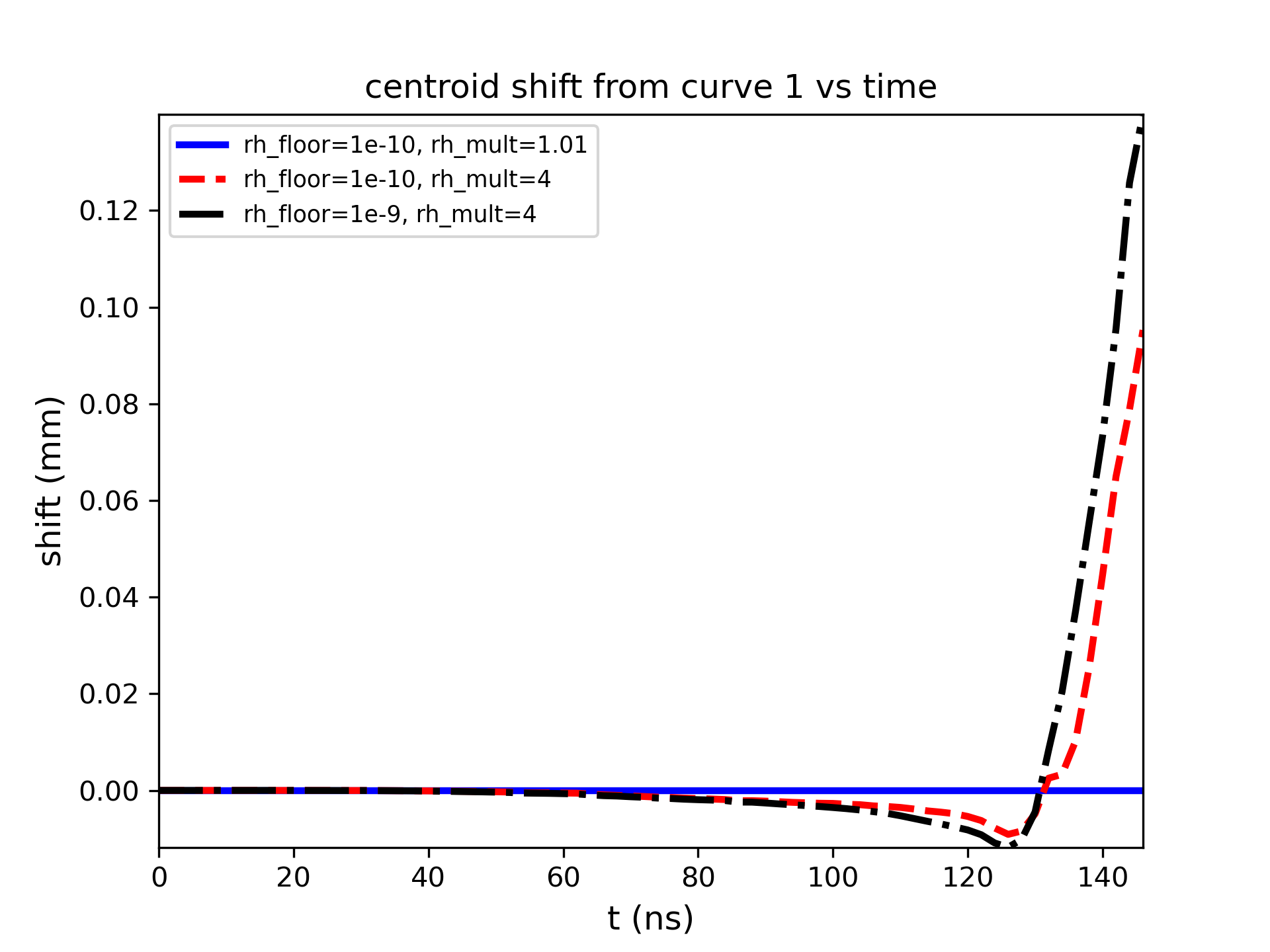}\\
(ii) \includegraphics[scale=0.5]{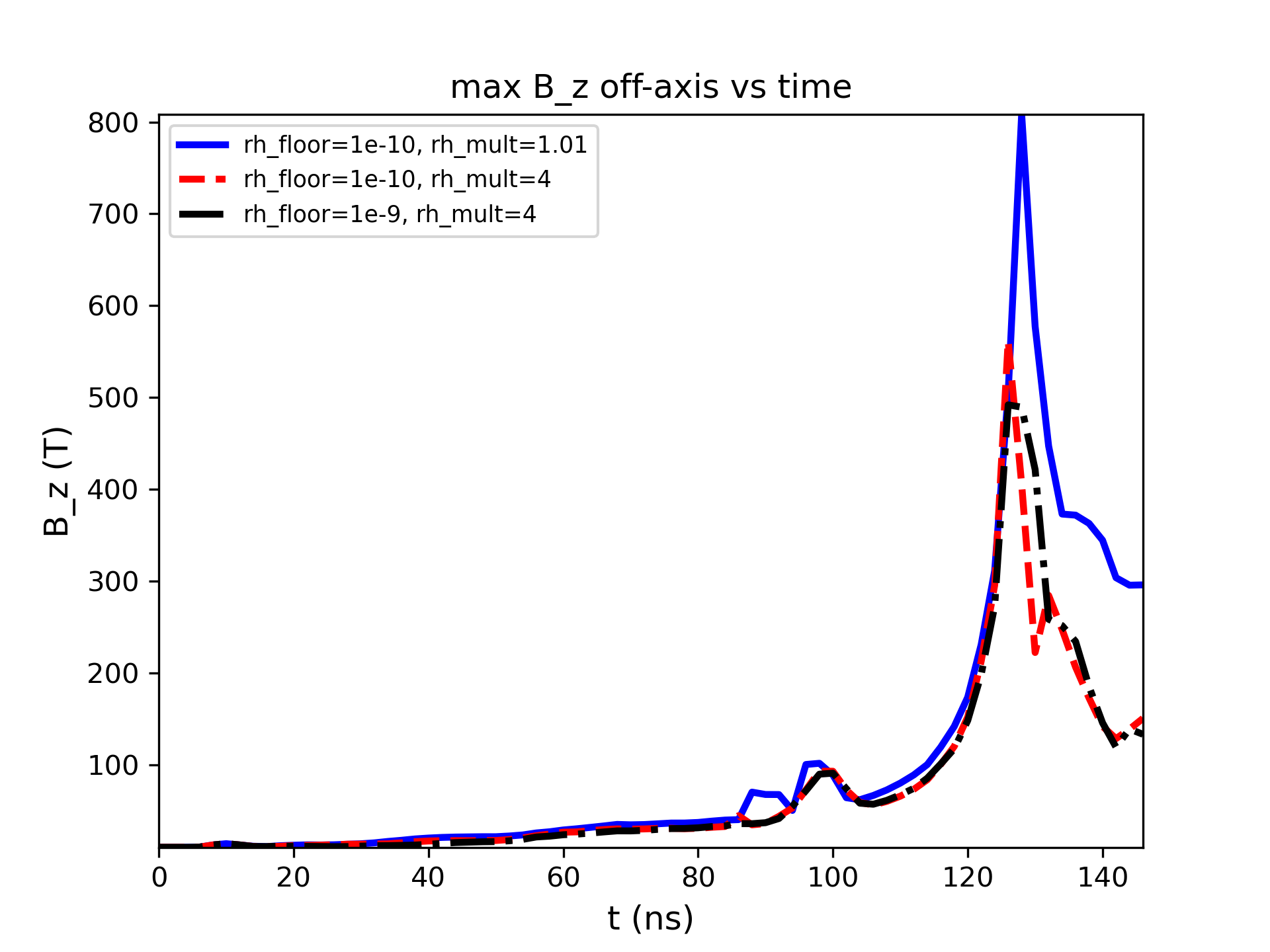}
\caption{Time-dependence of (i) shift in liner centroid relative to first case plotted, and (ii) maximum axial $B$-field just outside liner, each comparing $(\rho_{\rm floor},\rho_{\rm  multiplier}) = $ (i) $(10^{-10}, 1.01)$, (ii) $(10^{-10}, 4)$, and (iii)  $(10^{-9}, 4)$. The earlier implosion at a higher floor multiplier is artificial, as the floor multiplier removes physical vacuum plasma and axial flux compression.  Hall physics is modeled.  Vacuum temperature is 116 K.}
\label{fig:rz_hall_mult_cent_Bz}
\end{figure}

The $r-z$ liner simulations thus far have used a floor multiplier of 1.01 in the density buffer, where recall from Sec. \ref{ssec:limiting} that current density, momentum, and energy are reset to floor values for $\rho \le \rho_{\rm floor}\rho_{\rm multiplier}$.  Figures \ref{fig:rz_rho_hall_mult4_100} compare the densities, and Figs. \ref{fig:rz_Bz_hall_mult4_100} compare the axial $B$-field between (i) the floor multiplier of 1.01 used in the results thus far in this section, and (ii) a floor multiplier of 4.  The higher floor multiplier has a small but noticeable effect on the implosion dynamics, but clearly reduces both the amount of plasma in the vacuum and therefore the axial flux compression against the liner.  As seen from the graphs of centroid shift and maximum $B_z$ versus time in Figs. \ref{fig:rz_hall_mult_cent_Bz}, the reduction in $B_z$ amplification artificially shifts the implosion earlier in time, by an amount which is comparable at both $10^{-10}\rho_{\rm solid}$ and $10^{-9}\rho_{\rm solid}$ floors, with and without the Hall term modeled.  This suggests that at such a high floor multiplier, there simply isn't enough current-carrying plasma outside the liner for appreciable axial flux compression.

Recall that PERSEUS requires little to no density buffer, while MHD codes often require floor multipliers of at least 4 for numerical stability, which, based on these results, likely reduces the axial flux compression modeled by these codes, and affects the timing of the liner implosion.  Recall from Sec. \ref{ssec:limiting} that a density buffer is often used by MHD codes to mitigate instability that would otherwise be introduced by abruptly switching on electromagnetics and hydrodynamics for any $\rho > \rho_{\rm floor}$.   By including extended-MHD physics, a code is able to more physically capture the transition to vacuum conditions at low densities, including the transition to zero conductivity, and is therefore less susceptible to instability associated with an abrupt transition. 

\subsubsection{Summary}

The salient results of this section are summarized below, regarding liner implosions in $r-z$ axisymmetric geometry.

\begin{enumerate}
\item The vacuum modeling sensitivities are primarily in the dynamics of low-density vacuum plasma, and have less of an effect on the implosion of the liner itself, though this effect is greater (larger centroid shift) than for the 1-D test problem.
%\item In order to stabilize the system against (numerical) thermal runaway, and prevent Hall velocities exceeding the numerical speed of light, current must be limited at a maximum Hall velocity.
\item At lower density floors, the vacuum plasma carries more current, and compresses more axial flux.  Density floors less than $5\times 10^{-8}$ of solid density are required to model this effect.
\item At sufficiently low floors, Hall physics enhances axial flux compression, likely through Hall dynamo in a force-free low-density plasma.  
\item The above two items lead to the following minimum requirements in order to capture the correct low-density dynamics and current-coupling onto the liner: (i) the density floor must be sufficiently low, and (ii) Hall physics must be modeled.
\item These mechanisms introduce greater vacuum-modeling sensitivity of the bulk implosion dynamics in the $r-z$ axisymmetric geometry with axial inflow BCs, compared to the 1-D radially convergent geometry driven at the outer radial boundary.
\item Whether or not Hall physics is modeled, a density buffer artificially reduces vacuum plasma production and resulting axial flux compression, even at low density floors, when using floor multipliers typical of MHD codes.  This artificially affects the timing of the liner implosion.
%\item The available Beryllium EOS models do not permit both the vacuum temperature and pressure to be simultaneously initialized at physical values of $\sim 300$ K and $\sim 10^{-5}$ Torr, respectively.  Unless otherwise stated, the simulations used $(T_{\rm solid},T_{\rm vacuum},P_{\rm vacuum}) \sim (300\ {\rm K},116\ {\rm K},10^{-50}\ {\rm Torr})$.  When $(1160\ {\rm K},1160\ {\rm K},10^{-5}\ {\rm Torr})$ was tried, there was a noticeable delay in the liner implosion, compared to the 300-K, 116-K initialization.
\item A clear correlation exists between coupling of the Hall term into $E$ and $B$-field generation, and the amount by which the Hall term reduces vacuum sensitivity of bulk implosion dynamics, both of which are greater for this $r-z$ axisymmetric problem than for the 1-D radially convergent geometry or a 2-D $r-\theta$ geometry.  This correlation can likely be understood in terms of the conductivity modeled by Hall physics, which is a 3x3 tensor relating $E$ and $J$, obtained by solving for $J$ in terms of $E$ in the GOL \eqref{gol}; see Appendix \ref{app:XMHD} for an expression for this tensor.  As the Hall term plays a stronger role in the electromagnetics, there emerges a larger difference between the Hall conductivity tensor and the scalar conductivity used in the absence of Hall, and in particular this tensor plays a larger role in modeling a more physical transition to zero current perpendicular to the $B$-field at vacuum densities.
\end{enumerate}

\subsection{2-D MRT development}
\label{sec:vac2dmrt}

Vacuum modeling sensitivity has important consequences for numerical predictions about instability development, as we demonstrate in this section in the context of the magneto-Rayleigh-Taylor (MRT) instability.  MRT is seeded on a Copper liner in $r-z$ axisymmetric cylindrical geometry, using SESAME EOS and conductivity tables for Copper, where the EOS table uses Maxwell constructions in the vapor dome.  MRT is an instability that arises in imploding ICF targets when the radially inward ${\bf J}\times {\bf B}$ force acts on a nonuniform surface, which can then cause the amplitudes of those nonuniformities to grow and thereby limit target performance, including the generation of fusion energy.  MRT has been studied extensively in references such as \onlinecite{sina10,sina11,lau11,weis15}.  An azimuthal $B$-field is driven at the outer radius using the current profile \eqref{icurr} described in Sec. \ref{ssec:vac1d}.  The domain spans $3.84\times 2.4$ mm, extending from $r = 0$ to $r = 3.84$ mm, and the resolution is 20 $\mu$m, corresponding to $192\times 120$ cells.  The liner has an outer radius of 2.88 mm and inner radius of 2.79 mm.  The MRT is seeded with a 1$\%$ sinusoidal temperature perturbation in the outer layer of cells on the liner surface, with a wavelength of 200 $\mu$m, so that the domain spans 12 wavelengths axially (typical of experimentally observed MRT) with 10 cells per wavelength.  

\begin{figure}
\includegraphics[scale=0.4]{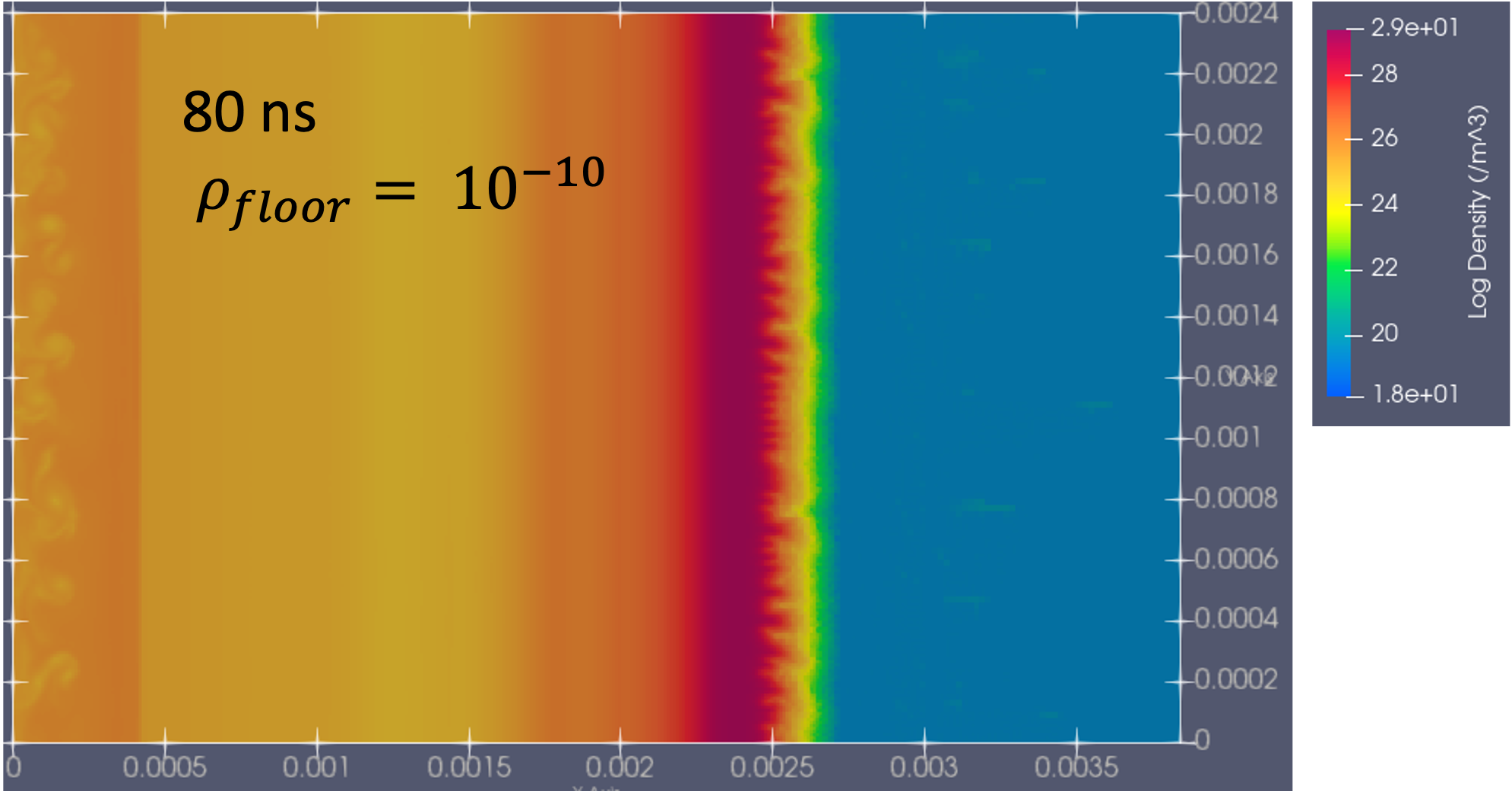}
\includegraphics[scale=0.4]{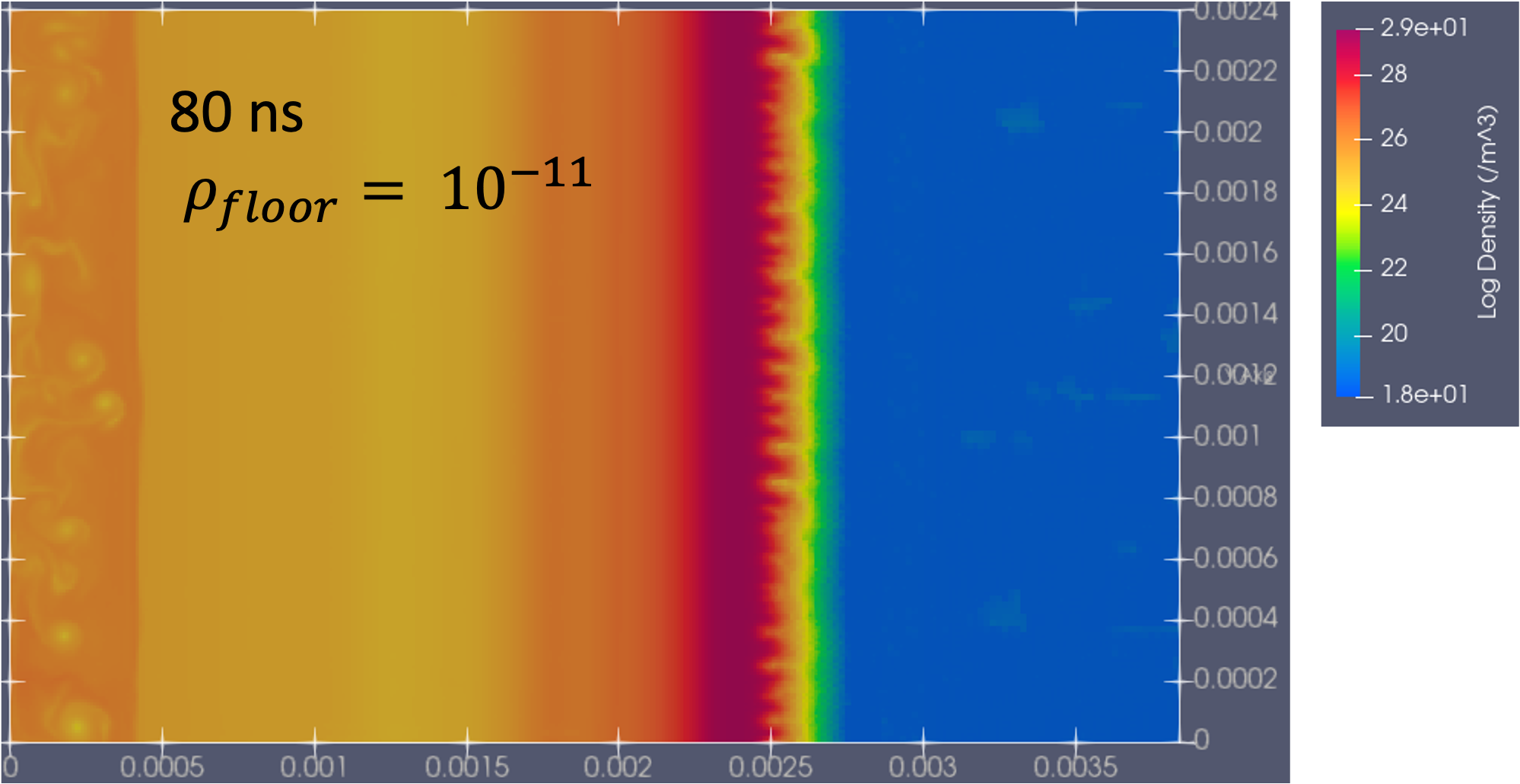}
\includegraphics[scale=0.4]{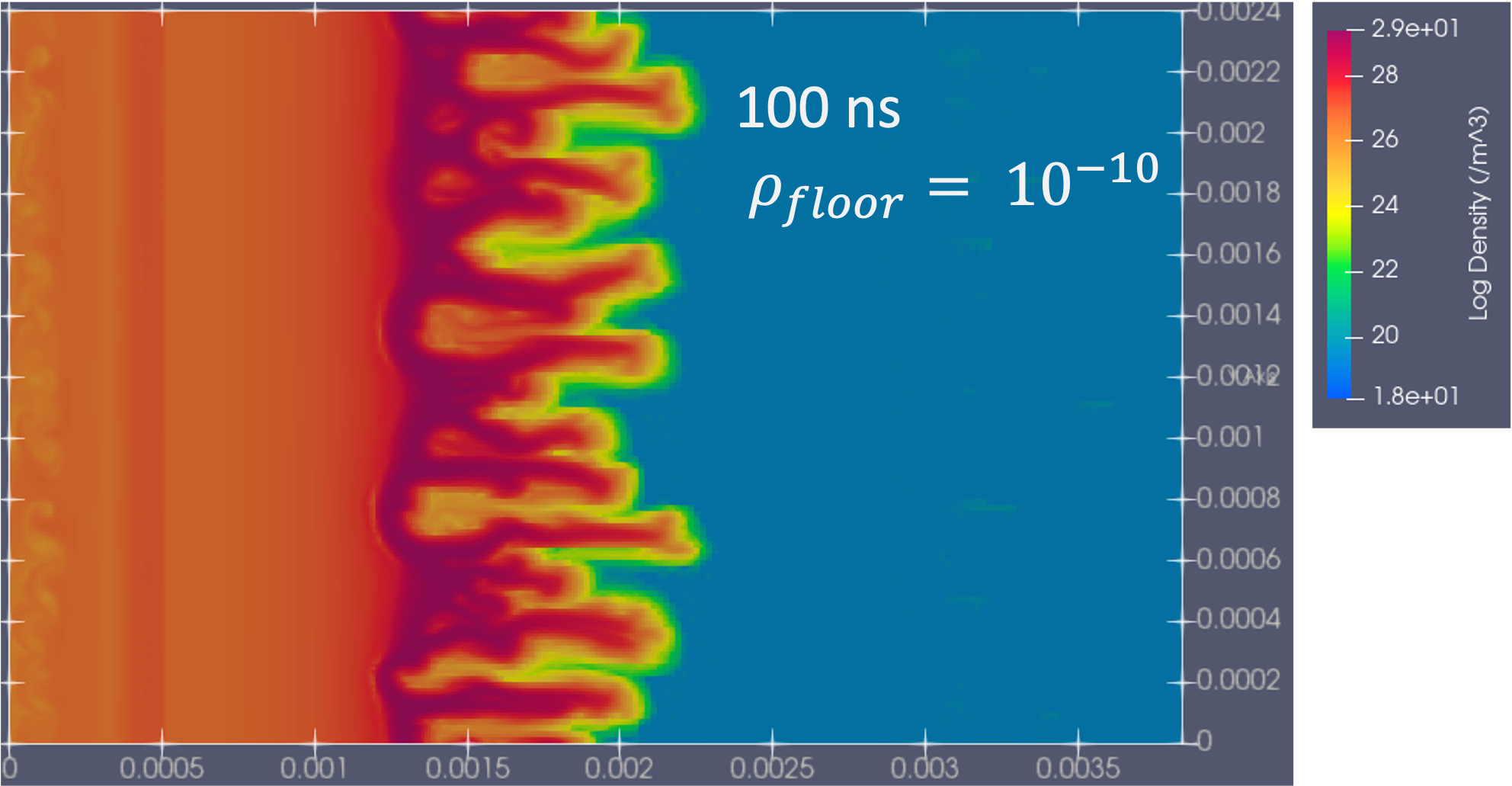}
\includegraphics[scale=0.4]{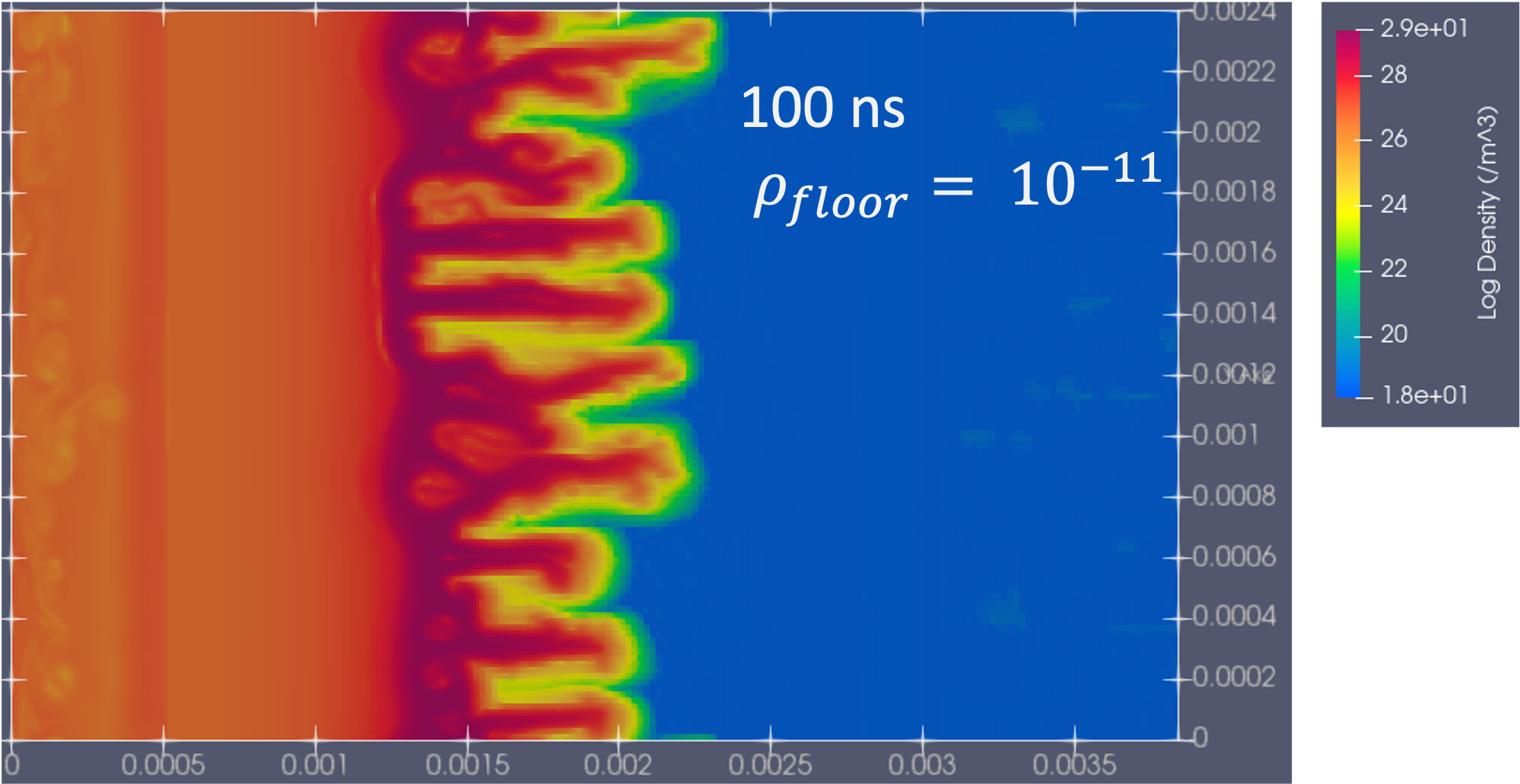}
\caption{Base-10 log density displays at 80 ns and 100 ns using density floors of $10^{-10}$ and $10^{-11}$ of solid Copper}
\label{fig:2-Dmrt1}
\end{figure}

Figures \ref{fig:2-Dmrt1} show base-10 log density displays of the imploding liner at density floors of $10^{-10}$ and $10^{-11}$ of solid Copper.  The displays at 80 ns show a much higher spatial frequency of MRT than the 200-$\mu$m wavelength that was seeded, due to the excitation of secondary modes likely associated with the electrothermal instability (ETI).  By 100 ns, these modes have merged to a wavelength of $\sim 200\ \mu$m, indicating that the dominant MRT mode has a wavelength close to the seeded value.

\begin{figure}
\includegraphics[scale=0.4]{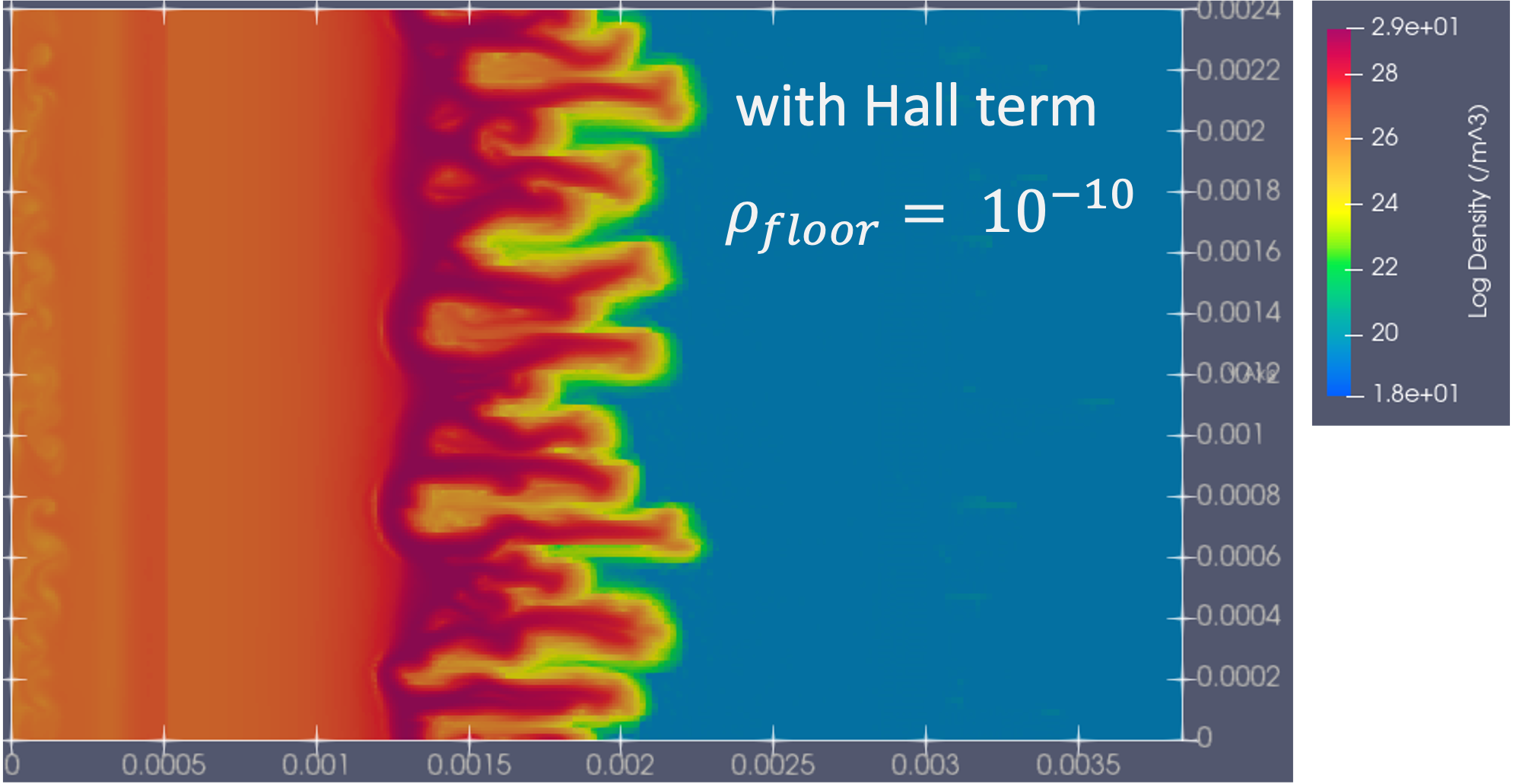}
\includegraphics[scale=0.4]{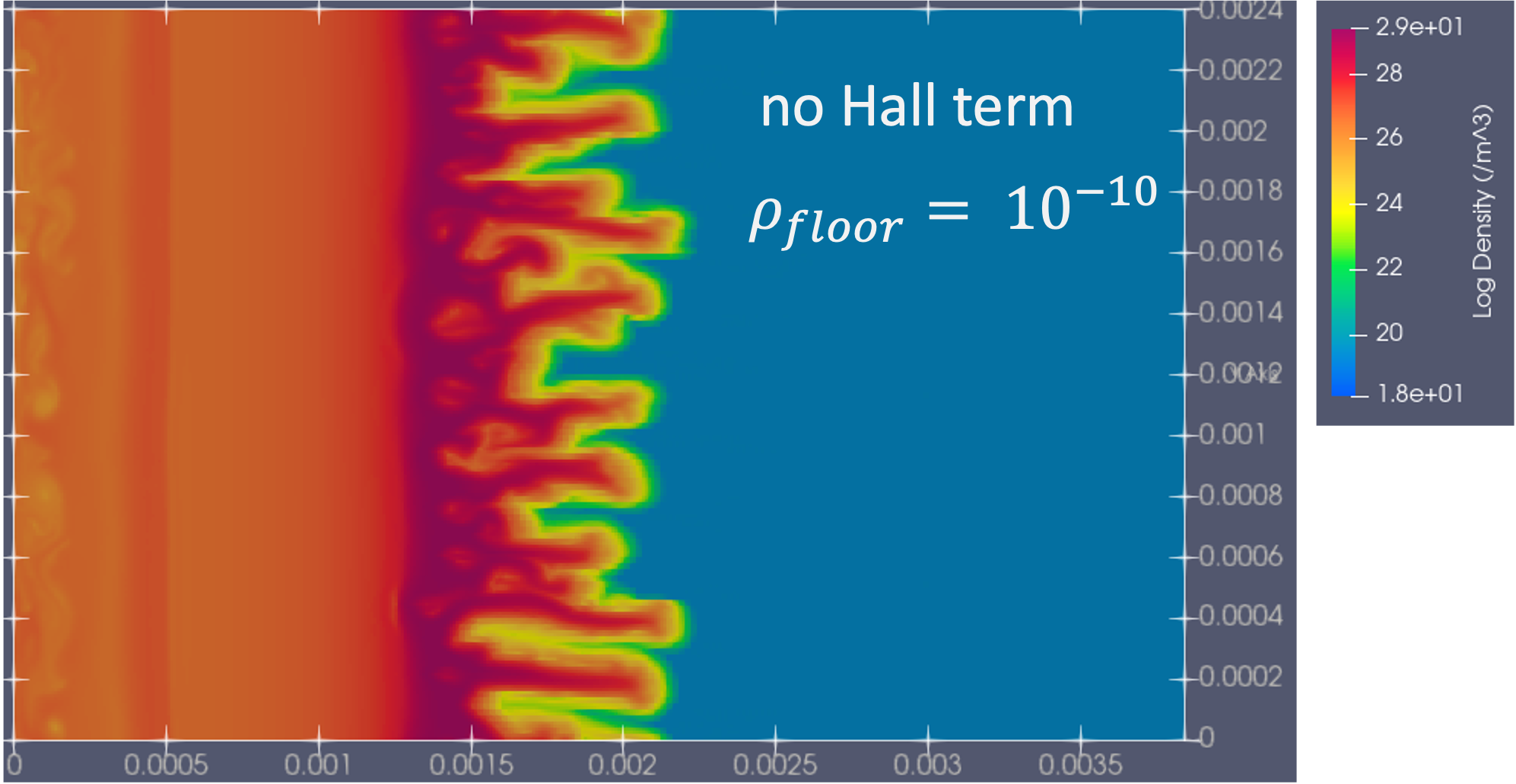}
\caption{Base-10 log density displays at 100 ns with versus without the Hall term being modeled, using a density floor of $10^{-10}$ of solid Copper}
\label{fig:2-Dmrt2}
\end{figure}

At each of the times displayed, there is no qualitative difference between the two density floors, either in the backside ablation or MRT development.  Figures \ref{fig:2-Dmrt2} show a comparison of log density with versus without the Hall term being modeled.  For this problem, the Hall term does not seem to have a noticeable impact.  The use of Hall physics has previously been observed to introduce a ${\bf B}\times \nabla n$ axially-propagating drift mode in MRT development \cite{huba98}.  This was not observed in the present study, likely because the MRT wavelength exceeds the relevant ion inertial lengths by several orders of magnitude. 

At these low density floors, the results have become insensitive to other aspects of vacuum modeling.  As one example, these simulations were run both with and without resets to pressure and temperature floors, and the results were qualitatively unaffected.  

\begin{figure}
\includegraphics[scale=0.4]{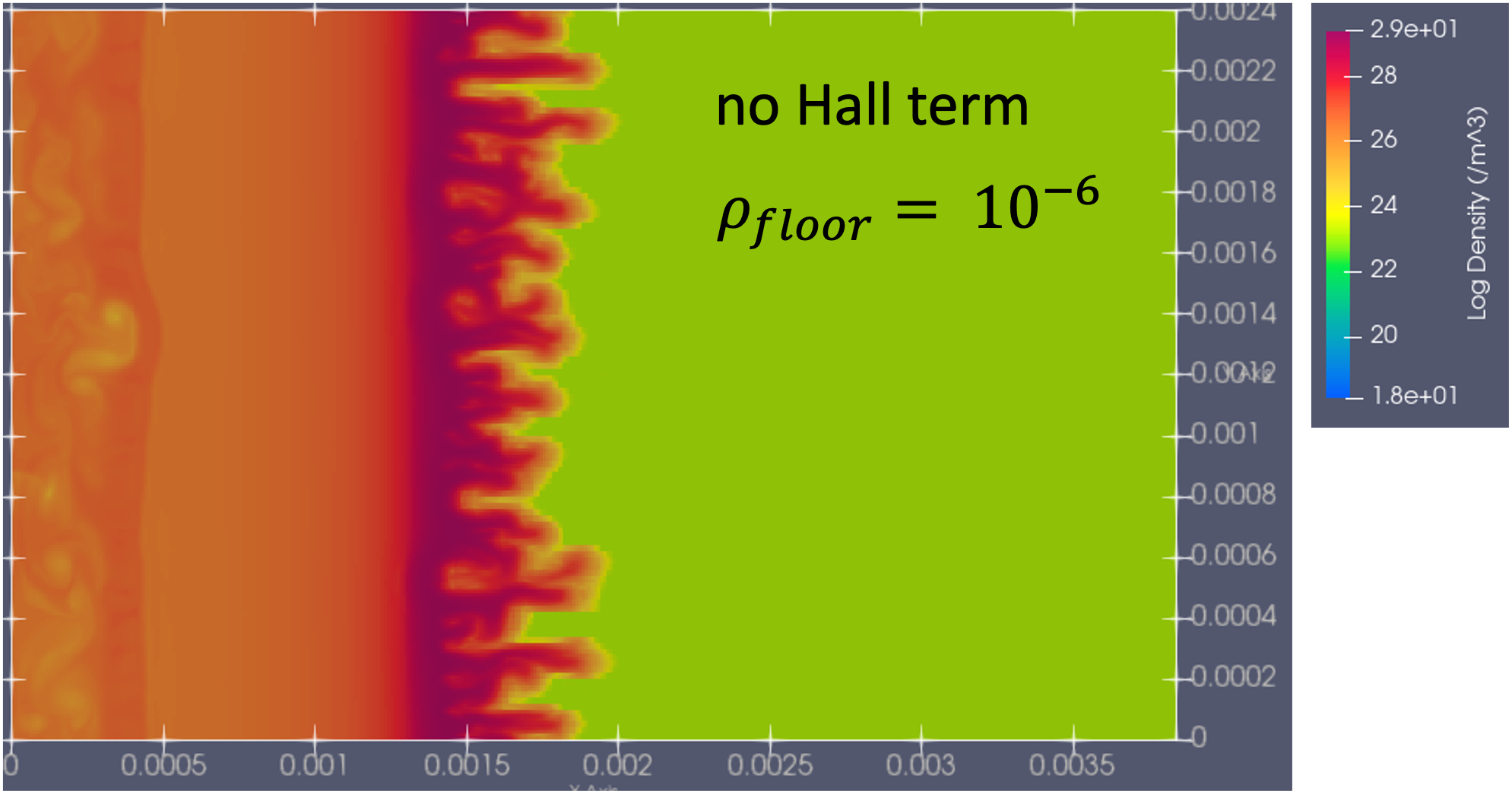}
\caption{Base-10 log density displays at 100 ns using a density floor of $10^{-6}$ of solid Copper}
\label{fig:2-Dmrt3}
\end{figure}

Figure \ref{fig:2-Dmrt3} shows the corresponding log density display when a density floor of $10^{-6}$ of solid Copper is used.  Here, the MRT spikes are noticeably shorter than at the lower density floors, indicating that convergence with respect to vacuum has not been achieved at this density floor, which is typical of values used in many MHD codes.  This lack of convergence evidently impacts numerical predictions about MRT growth rates, and more broadly, about instability development and its impact on target performance.  As with the other floors, the Hall term did not noticeably impact MRT growth at the $10^{-6}$ floor, and therefore does not appear to be accelerating density floor convergence for this problem.

%\subsection{Verification of Linear MRT Growth}
%\label{ssec:linmrt}

%\subsection{Cross-Validation of Tabular Interface}
%\label{ssec:valtab}

\section{Conclusion}

To predict the performance of an ICF target in a pulsed-power device, e.g. the fusion yield from a MagLIF target, requires modeling densities that span 12 orders of magnitude from solid-density electrodes down to vacuum density.   In particular, much of the current coupled onto the target is carried by plasma whose densities are lower than can be modeled by resistive MHD codes, whose timesteps are often restricted by an Alfven speed which is unbounded as the density drops. The extended-MHD code PERSEUS, which has now been interfaced with tabular material models, provides several important advances toward a predictive model of a physical plasma-vacuum interface in a pulsed-power device, and therefore a more predictive model of ICF target performance.  These advances are as follows: 

\begin{enumerate}
\item The inclusion of displacement current limits the Alfven speed by the numerical speed of light, and thereby limits the reduction in timestep at low densities.  
\item The extended-MHD model, in particular the use of Hall physics, models a more physical transition from plasma to vacuum, including a conductivity tensor for which the current perpendicular to the $B$-field tends to zero as the density tends to vacuum levels.  This substantially reduces the need for a density buffer to artificially model a transition from plasma to vacuum conditions.  
\item The semi-implicit time advance circumvents the need to resolve electron cyclotron and electron plasma frequencies, while also being implicit only in the source terms, and therefore circumventing the need for a computationally expensive global matrix solve.
\end{enumerate}

From the 1-D and 2-D numerical results presented, we see that these advances have noticeably reduced the sensitivity, to parameters characterizing the numerical vacuum, of the bulk and low-density dynamics.  That is, for a given grid resolution, we have a single-material model which is minimally dependent on numerical vacuum parameters.  This is particularly true of the 1-D implosion of a Copper liner.  In general, in each geometry that was tested, there was a drop in sensitivity to vacuum modeling when (i) a lower density floor was used, and (ii) Hall physics was modeled.

Somewhat greater vacuum sensitivity was found in the 2-D $r-z$ axisymmetric implosion of a Beryllium liner with applied axial $B$-field and axial Poynting inflow.  Recall that, of the problems tested, this is the setup for which Hall physics is expected to have the most influence, and indeed, was observed to enhance axial flux compression at higher density floors, and reduce sensitivity to density floor.  Compared to the 1-D setup, increased sensitivity to the density buffer was observed for this 2-D setup, which in particular remained even at floors of $10^{-10}\rho_{\rm solid}$.  However, this test also demonstrates that (i) PERSEUS does not require the use of such buffers, and (ii) the results can be substantially altered by using floor multipliers typical of MHD codes, often required to model an artificial transition to vacuum conditions when a more physical extended-MHD model is not included.    

In both 1-D and 2-D tests, convergence of the results with respect to density floor consistently required the use of floors which are orders of magnitude less than the values $\sim 10^{-6}\rho_{\rm solid}$ typical of resistive MHD codes.  Moreover, in each test, Hall physics accelerated convergence of the results with respect to density floor, which is consistent with the ability of Hall physics to model a more physical transition to zero current perpendicular to the $B$-field at vacuum densities, by modeling the Hall conductivity tensor described in Appendix \ref{app:XMHD}, which is not captured by a resistive MHD model.

Compared to the 1-D liner implosion, the 2-D $r-z$ implosion shows higher coupling of Hall physics into the liner dynamics through $E$ and $B$-field generation, which was found to be correlated with an increase in potential vacuum modeling sensitivity of the liner implosion dynamics, and a larger decrease in this sensitivity when the Hall term is modeled.
The $r-z$ liner implosion also exhibits higher vacuum modeling sensitivity of the bulk implosion dynamics, showing greater floor sensitivity, in liner centroid position and axial flux compression, than the 1-D radially convergent test.  This is also the test for which the Hall term is more influential, owing likely to the Hall dynamo generation of complementary $E$ and $B$-field components enabled by the axial Poynting inflow BCs and applied in-plane $B$-field.

Moreover, unlike the 1-D test, in the 2-D test the Hall term introduced differences in low-density plasma dynamics regardless of the density floor setting, showing greater vacuum plasma production resulting in enhanced axial flux amplification against the liner.  This shows that in complex 2-D and 3-D geometries, in order to more accurately capture coupling of current onto an imploding target, not only does the density floor need to be significantly less than the density of current-carrying vacuum plasma, but Hall physics also needs to be modeled (along with other physics; see below).

Clearly, there is a correlation between coupling of the Hall term into $E$ and $B$-field generation, and the amount by which the Hall term reduces vacuum sensitivity of bulk implosion dynamics.  This correlation can likely be understood in terms of the 3x3 conductivity tensor modeled by Hall physics.  As the Hall term plays a stronger role in the electromagnetics, a larger difference emerges between the Hall conductivity tensor and the scalar conductivity used in the absence of Hall, and in particular this tensor plays more of a role in modeling more physical behavior of current-carrying plasma at vacuum densities; see Appendix \ref{app:XMHD}.

As shown in Sec. \ref{sec:vac2dmrt}, a reduction in vacuum modeling sensitivity increases the reliability of numerical predictions about instability growth, including MRT, and the resulting impact on target performance.

%In order to stabilize the 2-D $r-z$ setup against thermal runaway in the low-density plasma, limiting was needed on the maximum current density per unit mass.  This was found to be more of a function of the available tabular material models, as opposed to the numerical model in PERSEUS.  For this reason, the thermal runaway instability, along with much of the observed vacuum sensitivity, will likely be mitigated once (i) thermal conduction is incorporated, and (ii) the vacuum plasma is modeled as a realistic separate material, e.g. hydrogen or deuterium, and not a vaporized version of the solid material.  These capabilities are under development.

Further capabilities, which are currently under development in FLEXO (Flux-Limited Extended-MHD Ohm's Law; under development at Sandia, and using a similar extended-MHD formulation to the one in PERSEUS), will fully enable a physical solution of the plasma-vacuum interface problem under vacuum conditions relevant to pulsed-power systems.  These include (i) a multi-material model, (ii) adaptive mesh refinement (AMR) for improved resolution, and spatial convergence, of stagnation conditions and micro-scale surface physics, and (iii) other magnetically anisotropic transport, including radiation, thermal conduction, Nernst, Ettinghausen, viscosity, and replacement of the scalar resistivity $\eta$ with a magnetically anisotropic conductivity tensor, e.g. following the Lee-More-Desjarlais model.  It should be noted that this tensor is distinct from the Hall conductivity tensor discussed elsewhere in this study, which arises from solving Eq. \eqref{gol} for ${\bf J}$ in terms of the field ${\bf E} + {\bf u}\times {\bf B}$ in the co-moving frame, and is given in Appendix \ref{app:XMHD}.

A more immediate next step, to be included in a forthcoming publication, is a fully integrated 3-D simulation for validation against a suitable experiment being run on Sandia's Z Accelerator.  The simulation will be performed both at experimentally-relevant floor settings, and at floor settings typical of MHD codes, to demonstrate the improvements in experimental validation provided by minimizing vacuum sensitivity.

By minimizing sensitivity to numerical parameters, the vacuum model in PERSEUS has become more readily portable between different experimental setups, which will enhance its reliability in upcoming validation against various experimental setups.  That is, there are fewer knobs that need to be re-adjusted each time validation is performed against data from a new experiment.  The reliability of each validation exercise is greatly enhanced if they all use the same vacuum model, and do not require separate settings of various numerical knobs.

In this way, the PERSEUS extended-MHD model enables validation against a broader range of pulsed-power design concepts extending to higher current densities, larger timescales, and smaller spatial scales.

\begin{acknowledgments}
The authors gratefully acknowledge valuable advice provided by Tom Gardiner and Allen Robinson.
Sandia National Laboratories is a multi-mission laboratory managed
and operated by National Technology and Engineering Solutions of
Sandia LLC, a wholly owned subsidiary of Honeywell International
Inc. for the U.S. Department of Energy’s National Nuclear Security
Administration under contract DE-NA0003525.  The data that support the findings of this study are available from 
the corresponding author upon reasonable request.
\end{acknowledgments}

\appendix

%\section{Appendixes}

\section{PERSEUS extended-MHD model of current in low-density plasma}
\label{app:XMHD}

We seek to understand how the PERSEUS extended-MHD formulation provides a more physical description of the behavior of current in a near-vacuum plasma.  This derivation is similar to the one presented in Sec. IV A of Ref.~\onlinecite{haml18b}, but also explicitly includes the GOL ``relaxation" density discussed in Sec. \ref{ssec:model}.  The GOL used by PERSEUS is as follows, where $n_e$ is physical electron density and $n_r$ is a density intended to enforce relaxation of the GOL to a Hall MHD regime (negligible electron inertia).

\begin{align}
\frac{\partial {\bf J}}{\partial t} &\ =\ \frac{n_re^2}{m_e}\left({\bf E} + {\bf u}\times {\bf B} - \frac{\bf J}{n_ee}\times {\bf B} - \eta {\bf J}\right)\label{golq}
\end{align}

\medno Denote the $E$-field in the co-moving frame of the fluid by

\begin{align}
{\bf E}' &\ =\ {\bf E} + {\bf u}\times {\bf B}
\end{align}

For purposes of analytically representing the contribution of the $\partial {\bf J}/\partial t$ term, let ${\bf E}$, ${\bf B}$, and ${\bf J}$ have the linearized form

\begin{align}
{\bf B} &\ =\ {\bf B}_0 + {\bf B}_1\exp[i({\bf k}\cdot {\bf r} - \omega t)]\nonumber\\
{\bf E} &\ =\ {\bf E}_0 + {\bf E}_1\exp[i({\bf k}\cdot {\bf r} - \omega t)]\nonumber\\
{\bf J} &\ =\ {\bf J}_0 + {\bf J}_1\exp[i({\bf k}\cdot {\bf r} - \omega t)]\label{bejlin}
\end{align}

\medno Then the linearized version of GOL \eqref{golq} is

\begin{align}
-i\omega  {\bf J}_1 &\ =\ \frac{n_re^2}{m_e}\left({\bf E}'_1 - \frac{{\bf J}_1}{n_ee}\times {\bf B}_0  - \frac{{\bf J}_0}{n_ee}\times {\bf B}_1 - \eta {\bf J}_1\right)
\end{align}

\medno which can also be written

\begin{align}
{\bf E}'_1 - \frac{{\bf J}_1}{n_ee}\times {\bf B}_0 - \frac{{\bf J}_0}{n_ee}\times {\bf B}_1 - \eta' {\bf J}_1 &\ =\ 0
\end{align}

\medno where

\begin{align}
\eta' &\ =\ \eta + i\omega \frac{m_e}{n_re^2}
\end{align}

\medno Invoking the linearization of Faraday's law, ${\bf B}_1 = {\bf k}/\omega\times {\bf E}_1$, we have

\begin{align}
\frac{{\bf J}_0}{n_ee}\times {\bf B}_1  &\ =\ \frac{{\bf J}_0}{n_ee}\times \left(\frac{\bf k}{\omega}\times {\bf E}_1\right)\nonumber\\
&\ =\ \frac{1}{n_ee\omega}\left[{\bf k}({\bf J}_0\cdot{\bf E}_1) - ({\bf k}\cdot{\bf J}_0){\bf E}_1\right]
\end{align}

\medno We finally have 

\begin{align}
{\bf E}''_1 - \frac{{\bf J}_1}{n_ee}\times {\bf B}_0 - \eta' {\bf J}_1 &\ =\ 0,\qquad {\rm where}\nonumber\\
{\bf E}''_1 &\ \equiv\ {\bf E}'_1 - \frac{1}{n_ee\omega}\left[{\bf k}({\bf J}_0\cdot{\bf E}_1) - ({\bf k}\cdot{\bf J}_0){\bf E}_1\right]
\end{align}

\medno Solving for ${\bf J}_1$ in terms of ${\bf E}_1$, we have

\begin{align}
{\bf J}_1 &\ =\ \frac{1}{\eta'}\left[{\bf E}''_{1,\parallel} + \frac{{\bf E}''_{1,\perp}}{1 + \Omega^2\tau^2} + \frac{\Omega\tau (\hat{b}\times {\bf E}''_1)}{1 + \Omega^2\tau^2}\right]\label{j1}
\end{align}

\medno where $\Omega\tau = B_0/(n_ee\eta')$ is the resistive Hall, or magnetization, parameter, and 

\begin{align}
{\bf E}''_{1,\parallel} &\ =\ ({\bf E}''_1\cdot \hat{b})\hat{b}\\
{\bf E}''_{1,\perp} &\ =\ {\bf E}''_1 - ({\bf E}''_1\cdot \hat{b})\hat{b}\\
\end{align}

\medno The $3\times3$ Hall conductivity tensor ${\bf \sigma}$ is obtained from writing Eq. \eqref{j1} as

\begin{align}
{\bf J}_1 &\ =\ {\bf \sigma}\cdot {\bf E}''_1\label{j1sigma}
\end{align}

For the problems in the present study, the ``relaxation" density $n_r$ is chosen so that electron inertial physics is negligible (typically $n_r \sim 10^{-3}n_{\rm solid}$), corresponding to $\eta' \approx \eta$, which means that

\begin{align}
\omega \frac{m_e}{n_re^2} \ll \eta
\end{align}

\medno In this limit, $\Omega\tau \approx B_0/(n_ee\eta)$, and thus the conductivity tensor is uninfluenced by electron inertial physics.  In particular, the density $n_r$ only influences the relative importance of electron inertial physics, and has no influence on the importance of the Hall term relative to the $\eta J$ or $E + u\times B$ terms in the GOL \eqref{golq}.  Note that in the limit of negligible electron inertial physics, $\eta' \approx \eta$, Eq. \eqref{j1} is valid for any $J$, $E$, and $B$, i.e. no linearization is required, and the ``1" and ``0" subscripts can be removed from these quantities.

In the strongly magnetized limit $\Omega\tau \gg 1$, the ${\bf E}''_{1,\perp}$ term can be dropped, so that Eq. \eqref{j1} becomes

\begin{align}
{\bf J}_1 &\ \rightarrow\ \frac{1}{\eta'}\left({\bf E}''_{1,\parallel} + \frac{\hat{b}\times {\bf E}''_1}{\Omega\tau}\right)\label{j1b}
\end{align}

\medno or

\begin{align}
{\bf J}_1 &\ \rightarrow\ \frac{{\bf E}''_{1,\parallel}}{\eta'} + \frac{n_ee}{B_0}(\hat{b}\times {\bf E}''_1)\label{j1c}
\end{align}

\medno The strongly magnetized limit therefore corresponds to a superposition of field-aligned current ${\bf E}''_{1,\parallel}/{\eta'}$, also called force-free current, and electron $E\times B$ drift current $({n_ee}/B_0)(\hat{b}\times {\bf E}''_1)$.  This is often the case with a low-density current-carrying plasma in vacuum, for which the collisionality is low. 

\medno In particular, when the system is strongly magnetized enough, and in the presence of a low-density plasma with a dynamo mechanism of field generation, field-aligned current dominates electron $E\times B$ drift current, often corresponding to a helical force-free configuration of the $B$-field and current.  This phenomenon is also discussed in Ref.~\onlinecite{seyl18}.

This behavior of the current at low densities (field-aligned configurations and electron $E\times B$ drift) is not captured unless the Hall term is modeled.

%\nocite{*}
\bibliography{perseus_rev} % Produces the bibliography via BibTeX.

\end{document}